\documentclass[twoside,11pt]{article}

\usepackage{jmlr2e}
\usepackage{natbib}
\usepackage{float}
\usepackage{lineno,hyperref}
\modulolinenumbers[5]
\usepackage[linesnumbered,ruled,vlined]{algorithm2e}
\usepackage{cite}
\usepackage{amsmath,amssymb,amsfonts}
\usepackage{algorithmic}
\usepackage{graphicx}
\usepackage{textcomp}
\usepackage{wrapfig}
\usepackage{bm}
\usepackage{color,soul}
\usepackage{epstopdf}
\usepackage{subfigure}
\usepackage{amssymb}
\usepackage{booktabs}
\usepackage{multirow}
\usepackage{caption}
\usepackage{array}
\usepackage[usestackEOL]{stackengine}
\usepackage{enumerate}
\makeatletter
\renewcommand*\env@matrix[1][*\c@MaxMatrixCols c]{%
 \hskip -\arraycolsep
 \let\@ifnextchar\new@ifnextchar
 \array{#1}}
\makeatother
\usepackage[marginal]{footmisc}   
\usepackage{framed,enumitem} 
\usepackage{adjustbox}
\usepackage{setspace}
\usepackage{mathtools}
\newcommand{\defeq}{\vcentcolon=}

\jmlrheading{1}{2023}{1-48}{4/00}{10/00}{Zhu \textit{et al.}}

\ShortHeadings{SCOREH+: A High-Order Node Proximity Spectral Clustering}{Zhu et al.,}
\firstpageno{1}

\begin{document}

\title{SCOREH+: A High-Order Node Proximity Spectral Clustering on Ratios-of-Eigenvectors Algorithm for Community Detection}

\author{\name Yanhui~Zhu \email yanhui@iastate.edu \\
       \addr Department of Mathematics and Statistics\\
       University of West Florida\\
       Pensacola, FL 32514, USA\\
       \addr Department of Computer Science\\
       Iowa State University\\
       Ames, IA 50011, USA
       \AND
       \name Fang Hu \email naomifang@ieee.org \\
       \addr Department of Mathematics and Statistics\\
       University of West Florida\\
       Pensacola, FL 32514, USA\\
       \addr College of Information Engineering\\
       Hubei University of Chinese Medicine\\
       Wuhan 430065, P.R. China.
       \AND
       \name Lei~Hsin~Kuo \email lkuo@uwf.edu \\
       \addr Department of Mathematics and Statistics\\
       University of West Florida\\
       Pensacola, FL 32514, USA
       \AND
       \name Jia Liu \email jliu@uwf.edu \\
       \addr Department of Mathematics and Statistics\\
       University of West Florida\\
       Pensacola, FL 32514, USA\\
       }


\maketitle

\begin{abstract}
The research on complex networks has achieved significant progress in revealing the mesoscopic features of networks. Community detection is an important aspect of understanding real-world complex systems. We present in this paper a High-order node proximity Spectral Clustering on Ratios-of-Eigenvectors (SCOREH+) algorithm for locating communities in complex networks. The algorithm improves SCORE and SCORE+ and preserves high-order transitivity information of the network affinity matrix. We optimize the high-order proximity matrix from the initial affinity matrix using the Radial Basis Functions (RBFs) and Katz index. In addition to the optimization of the Laplacian matrix, we implement a procedure that joins an additional eigenvector (the $(k+1)^{th}$ leading eigenvector) to the spectrum domain for clustering if the network is considered to be a "weak signal" graph. The algorithm has been successfully applied to both real-world and synthetic data sets. The proposed algorithm is compared with state-of-art algorithms, such as ASE, Louvain, Fast-Greedy, Spectral Clustering (SC), SCORE, and SCORE+. To demonstrate the high efficacy of the proposed method, we conducted comparison experiments on eleven real-world networks and a number of synthetic networks with noise. The experimental results in most of these networks demonstrate that SCOREH+ outperforms the baseline methods. Moreover, by tuning the RBFs and their shaping parameters, we may generate state-of-the-art community structures on all real-world networks and even on noisy synthetic networks.
\end{abstract}


\section{Introduction}
\label{sec:intro}
Complex networks model many entities and their relations as nodes and edges in real-world scenarios, which are ubiquitous and can be applied to any data as long as pair-wise interactions exist among the objects. Non-trivial topological features preserved in the network structure have attracted researchers from various fields, for example, biology \citep{guerrero2017adaptive, berahmand2021spectral}, climate \citep{boers2019complex}, epidemiology~\citep{renteria2018epidemiology,kinsley2020multilayer}, etc. In complex networks, community detection discovers clusters in the network model to mine latent information among the objects. The nodes are densely connected by edges within clusters while sparsely connected between clusters. Researchers have developed various algorithms to discover community structures, for example, Walktrap \citep{pons2005computing}, Infomap \citep{rosvall2007information}, Louvain algorithm \citep{blondel2008fast}, deep learning-based algorithms \citep{yang2016modularity, jin2021survey}, spectral-based methods \citep{ng2002spectral, jin2015fast, hu2019computing}, diffusion-based algorithms \citep{roghani2022fast}, algorithms on overlapping networks \citep{kumar2017upper, gupta2020overlapping}, to name a few. On the other hand, hiding nodes from community algorithms for privacy purposes \citep{liu2022protect}, and vulnerability assessment \citep{li2021measuring} have also become hot spots.

Spectral clustering, rooted in graph theory, is one of the state-of-the-art algorithms for detecting communities in complex networks. A spectral clustering algorithm consists of three procedures: (1) regularization of a suitable adjacency or Laplacian matrix; (2) a form of spectral truncation; and (3) a k-means algorithm on the reduced spectral domain \citep{zhou2019analysis}. For the first step, the formation and selection of the graph proximity method and Laplacian are significant. A similarity matrix models the local neighborhood relationships of pair-wise data points. Researchers typically use the network affinity matrix as the node similarity representation or implement the similarity measures to construct a new similarity matrix. Radial Basis Functions (RBFs) are commonly used kernels in constructing those similarity matrices \citep{law2017deep, park2018spectral}. In the third step, the number of clusters is a prerequisite. Nonetheless, by decomposing the Laplacian matrix, the first large gap between two eigenvalues generally indicates the number of clusters. That is to say, the number of eigenvalues before this gap is the number of clusters \citep{NIPS2001_a5a61717}. However, this approach lacks a theoretical justification \citep{NIPS2004_40173ea4}.            

The challenges and drawbacks of existing spectral-based and community detection are presented as follows:

\begin{itemize}
    \item The affinity matrix is insufficient to capture graph local information. It only captures the direct neighbors' information between nodes, while the higher-order motif information is natural and essential for forming a community \citep{shang2022local}.
    \item Given the number of clusters $k$, researchers generally preserve the exact top $k$ eigenvectors for posting k-means clustering. However, for some networks, other eigenvectors may also carry information for clustering \citep{jin2018score+}. Thus, an eigen-selection strategy is essential.
    \item Most graph affinity matrices are large, sparse matrices with huge condition numbers (can be approximated by the ratio of the largest and smallest eigenvalues or singular values). If the linear system is ill-conditioned with huge condition numbers, we may have a large error in the numerical results from the eigenvalue and eigenvector methods. Therefore, optimization of the affinity matrix is a necessary step. 
\end{itemize}

This paper proposes an improved community detection algorithm to address those challenges. First, it utilizes RBFs to approximate the similarity matrix from the original affinity matrix and considers its high-order proximity. The optimization of the RBFs helps reduce the condition number of the original affinity matrix. The resulting linear system will be less ill-conditioned so that the performance of the eigenvalue decomposition will be stable and robust. We select the Katz index to obtain the high-order proximity on the resulting matrix from the RBF transformation. The Katz index is easy to implement and has a relatively lower time complexity than other methods, such as Common neighbors, Propagation, and Eigenvector Centrality. We analyze the eigenvalue distribution and determine if one more eigenvector is necessary.
Moreover, given the substantial impact of an appropriate RBF shaping parameter~\citep{noorizadegan2022effective}, we conduct a comprehensive analysis encompassing a range of optimal RBFs and their shaping parameters.

We tested the proposed algorithm and demonstrated its effectiveness on eleven real-world networks spanning various areas. In addition, we generated benchmark networks using a criterion presented by Lancichinetti, Fortunato, and Radicchi \citep{lancichinetti2008benchmark}, in short, the LFR benchmark. Taking advantage of these data sets, we analyzed the multi-view results concerning the number of clusters and mixing parameters.

Overall, our paper makes the following contributions:
\begin{itemize}
\item \textbf{Well-conditioned network data}. We apply the RBF transformation to the affinity matrix of the network data. This procedure can reduce the condition number of the similarity matrix from the network affinity, resulting in a well-conditioned matrix, which is important for data clustering.
\item \textbf{High-order proximity}. Compared to the local direct relationship between nodes, high-order information is essential for preserving the network. We utilize the Katz index to compute the high-order proximity of nodes, allowing us to preserve more node-local information for more accurate clustering purposes.
\item \textbf{RBF shaping parameter optimization process}. We analyze the selection of RBFs and their shaping parameters. Extensive experiments show that some RBFs have an ``optimal" shaping parameter domain where near-optimal results can be obtained.
\end{itemize}

The remainder of this paper is organized as follows: In Section \ref{sec:related}, an overview of related work is presented, encompassing topics such as spectral clustering, SCORE/SCORE+ algorithms, high-order proximities, and Radial Basis Function (RBF) applications. Section \ref{sec:the_algorithm} offers an in-depth exploration of our algorithm, delving into its design, stepwise execution, pseudocodes and complementing these with a comprehensive flowchart illustration. Moving forward, Section \ref{sec:experimental_evaluation} outlines our experimental setup, dataset selection, evaluation metrics, baseline algorithms, and more. Section \ref{sec:experiment-analyses} is dedicated to the presentation of our experimental outcomes, followed by their thorough analyses. Finally, we draw our paper to a close with a concise summary of our findings and a glimpse into future research prospects, detailed in Section \ref{sec:conclusion}.

\section{Related Work}
\label{sec:related}
%
In this section, we present the related work concerned with the community detection algorithms, SCORE, and SCORE+ algorithms, higher-order proximities, and RBF applications.

\subsection{Community Detection Algorithms}
Community detection models are basic tools that enable us to discover the organizational principles in the network. In the last two decades, researchers have developed various efficient and effective models, for example, the famous Louvain algorithm \citep{blondel2008fast}, and the fast greedy algorithm \citep{clauset2004finding}, to name a few. The abovementioned algorithm can only solve simple static networks; however, in real applications, network data is changing and evolving~\citep{zhang2019dynamic}. Therefore, Li et al. introduced a novel dynamic fuzzy community detection method~\citep{li2022characterizing} and a new highly efficient belief dynamics model~\citep{li2022fast}. 

\medskip
\noindent
\textbf{Spectral Methods.} Spectral clustering is one of the common community detection algorithms. It uses information from the eigenvalues of the Laplacian of an affinity matrix and maps the nodes to a low-dimensional space where data is more separable, enabling us to perform eigen-decomposition and form clusters \citep{ng2002spectral}. Spectral clustering has wide applications in other datasets, such as attributed network~\citep{berahmand2021spectral}, multi-layer network~\citep{dabbaghjamanesh2019novel}, multi-view network~\citep{huang2019multi}, etc. In particular, if one considers the network's topology structures as well as node attribute information, the algorithm should apply to attributed networks. Berahmand et al. proposed a spectral method for attributed graphs showing that the identified communities have structural cohesiveness and attribute homogeneity~\citep{berahmand2022novel}. To reveal the underlying mechanism of biological processes, a novel spectral-based algorithm was proposed for attributed protein-protein interaction networks~\citep{berahmand2021spectral}.

\medskip
\noindent
\textbf{SCORE and its Variants.} Jin et al.\citep{jin2015fast} first proposed the SCORE algorithm, which uses the entry-wise ratios between eigenvectors for clustering to improve the effectiveness of spectral clustering. SCORE effectively removes the effect of degree heterogeneity by taking entry-wise ratios between the first leading eigenvector and each of the other eigenvectors. The SCORE provides novel ideas about computing communities in networks and can be extended in various directions. 

Jin et al. \citep{jin2018score+} applied two normalizations and eigen selections to improve SCORE's performance. The new algorithm SCORE+ demonstrated the rationality of Laplacian regularization as a pre-PCA normalization and retained an additional eigenvector as a post-PCA normalization. SCORE+ has two tuning parameters, but each is easy to set and not sensitive. Therefore, SCORE is fast, and SCORE+ is slightly slower. Their experimental results showed that the clustering error rate was reduced dramatically compared to SCORE on testing networks. Researchers have borrowed ideas from SCORE and SCORE+ for the community detection field \citep{gao2018community, duan2019state}.

\subsection{Higher-Order Proximities} 
In networks, to measure the similarity of every pair of nodes, the adjacency matrix and Laplacian matrix represent the first-order proximity, which simulates the local pair-wise proximity between vertices. Cosine similarity, Euclidean similarity, and Jaccard similarity are also popularly used. However, these similar methods can only preserve local information by using its connectivity to its neighbors. They are not sufficient to fully simulate the pair-wise proximity between nodes. How to preserve high-order proximity, therefore, has become a hot topic recently. People have also explored higher-order similarities to simulate the strength between two nodes \citep{cao2015grarep, tang2015line}.

Three commonly used high-order proximities are Common Neighbors and Propagation \citep{liben2007link}, Katz Proximity \citep{katz1953new}, and Eigenvector Centrality \citep{bonacich2007some}. The Katz index was proposed by Katz \citep{katz1953new} to compute the similarity of two nodes in a heterogeneous network by computing the walks between two nodes. We selected the Katz index in our paper for two reasons. First, it has been popularly used in related areas, for example, graph embedding \citep{ou2016asymmetric} \citep{zhang2018arbitrary}, complex networks \citep{lu2009similarity}, and relationship prediction in networks \citep{chen2015katzlda, zhang2017katzlgo}. Second, compared to Common Neighbors and Propagation \citep{liben2007link} and Eigenvector Centrality \citep{bonacich2007some}, the Katz index has lower complexity in implementation since Eigenvector Centrality \citep{bonacich2007some} requires the eigenvector computations, which is known to be time-consuming. Ou et al. \citep{ou2016asymmetric} proposed applying multiple high-order proximity measurements, e.g., Katz index \citep{katz1953new} on the graph embedding task. This work has attracted much attention. Plenty of proximity measurements have emerged in the last century. The Katz index is a widely used measurement considering the total number of walks between two nodes rather than the shortest one.

\subsection{RBF applications} 
The Gaussian similarity function $\exp(-r^2/(2c^2))$ is one of the most common Radial Basis Functions (RBFs) or similarity functions in the neural networks, where $r$ is the distance between the two nodes and $c$ is the shaping parameter. This function is equivalent to the Gaussian RBF in Table \ref{Tab:RBFs}. The Multiquadric (MQ) RBF is effective in geographical data sets, and the density of the local dataset determines the shaping parameter $c$. The selection of RBFs used for the interpolation matrix is strongly problem-dependent. On the other hand, the interpolation matrix, the same weight matrix we use in the complex network, is highly ill-conditioned. Thus, the selection of RBFs and the parameters are significant. Zhang et al. \citep{zhang2021graph} proposed a framework that integrates the attention mechanism and auto-kernel learning. The hyperparameter tuning for kernels largely facilitated improving the performance of graph convolutional networks.

In network sciences, researchers consider undirected graphs where the weighted adjacency matrix $\mathbf{W} = \mathbf{W}^T$ is symmetric. The structures of the networks could be studied by exploring the structures of the matrix $\mathbf{W}$ \citep{newman2013community}. The pattern of the vertices and edges of the adjacency graph or the corresponding adjacency matrix may reveal information about the network's divisions, clusters, and communities. The first step is to transform the given data set into a graph called a ``similarity graph". The goal of constructing the similarity graph is to model the local neighborhood relationships from the network data. 

A well-condition matrix is important and has better properties \citep{sharma2015greedy}. The condition number of a matrix can be approximated by the ratio of the largest eigenvalue and smallest eigenvalue (in absolute value). Therefore, to reduce the condition number, we need the absolute eigenvalue to be bounded away from 0. RBFs are such methods that bound the eigenvalues with at least $\Omega(\delta/\sqrt{d})$ \citep{ball1992eigenvalues}, where $\delta$ is the minimum separation and $d$ is the dimension of the data. 

\section{The High-order Proximity Preserved Spectral Clustering}
\label{sec:the_algorithm}
In this section, we present a fast community detection algorithm that uses the ratios of eigenvectors and the Eigen selection strategy while preserving higher-order proximities in the networks using the RBF and Katz index to capture more node-local information. 

Before introducing the algorithm, we clarify the symbols and definitions used. Table \ref{t:notations} summarizes the notations used in this paper. The following sections will formally define these notations when we introduce our network model and technical details.

\begin{table}[!t]
\renewcommand{\arraystretch}{1.2}
\caption{Important Notations and Naming Conventions}
\label{t:notations}
\centering
\resizebox{0.7\textwidth}{!}{
\begin{tabular}{|c|l|}
 \hline {Notation} & {Description} \\ \hline
$\mathbf{G= \{V,\ E\}}$ & Graph with set of nodes and edges \\
$n, m$ & \Longunderstack{number of nodes, number of edges in graph $\mathbf{G}$ \\ {\em (italic math style letter represents a single variable)}}   \\
$\Phi(\mathbf{v})$ & \Longunderstack{Radial basis
function value of node vector $\mathbf{v}$\\{\em (bold math style letter represents a vector/matrix)}}  \\
$d_{max}$ & maximal node degree in a graph\\
$\mathbf{A}$ & Graph affinity matrix from $\mathbf{G}$\\
$\mathbf{D}$ & Degree matrix from $\mathbf{G}$\\
$\mathbf{W}$ & Weighted matrix\\
$\mathbf{K}$ & High-proximity Katz matrix \\
$\mathbf{I}$ & Identity matrix with size $n \times n$\\
$\beta$ & decay parameter of Katz index\\
$\sigma$ & ridge regularization parameter\\
$\mu$ & mixing parameter of LFR network\\
$c$ & shaping parameter of RBFs\\
$k$ & number of clusters\\
\hline
\end{tabular}
}
\end{table}

\subsection{The Algorithm}
\label{subsec:algorithm}
The algorithm constructs the high-order proximity matrix while preserving the high-order transitivity information from the original affinity matrix using the RBF technique and Katz index. From the high-order proximity matrix, we obtain the normalized graph Laplacian. Next, we normalize the $k$ leading eigenvectors of the proximity matrix by dividing the leading eigenvectors into an additional $(k+1)^{th}$ eigenvector, which will be used for clustering if the network is considered to be a ``weak signal". 

\subsubsection{Radial Basis Functions}
\label{subsec:rbf}
We begin by outlining the methodology for creating an RBF matrix from the graph's original ill-conditioned affinity matrix.

For a given node vector $\left\{\mathbf{v}_i\right\}_{i=1}^n \in \mathbf{A}$ where $\mathbf{A}$ is an $n$-dimensional affinity matrix. 
The approximation that utilizes RBF is an unknown function $f$, which can be expressed as linear combinations of data norms.
\begin{equation}\label{rbfequation}
f\left( \mathbf{v}\right) \approx \tilde{f}\left( \mathbf{v} \right) = \sum_{i=1}^n \alpha_i\Phi\left(  \left\|\mathbf{v}-\mathbf{v}_i \right\| \right), \quad \mathbf{v}\in\Omega
\end{equation}
where $\left\|\cdot\right\|$ represents the Euclidean norm on $\mathbb{R}^d$, $\alpha_i$ is the coefficients, and $\Phi: \mathbb{R}^{d} \rightarrow \mathbb{R}$ is called a Radial Basis Function (RBF)~\citep{fasshauer2007meshfree} if,
\begin{align}
\Phi\left(\mathbf{v}\right) = \Phi\left(\mathbf{u}\right), \quad \text{whenever} \quad \|\mathbf{v}\| = \|\mathbf{u}\|, \quad \mathbf{v},\mathbf{u} \in \mathbb{R}^{d}
\end{align}
In Table~\ref{Tab:RBFs}, we list the most common RBFs widely used in neural networks and the numerical approximations. The symbol $c$ is a shaping parameter, and the symbol $r$ in the table denotes the Euclidean distance of $\mathbf{v}\in \mathbb{R}^d$ from the original point, $r = \left\| \mathbf{v} \right\|_2 = \sqrt{\sum_{i=1}^d x_i^2}$

\begin{table}[htpb]
\begin{center}
\caption{Some common choices of RBFs}
\label{Tab:RBFs}
\setlength{\parskip}{0.8\baselineskip}
\resizebox{80mm}{12mm}{
\begin{tabular}{@{}ll@{}}\toprule
Choice of RBF  & Definition \\\midrule
Multiquadric (MQ) & $\Phi(r,c) = \sqrt{c^2 + r^2}$ \\
Inverse Multiquadric (IMQ) & $\Phi(r,c) = 1/ \sqrt{c^2 + r^2}$ \\
Gaussian &  $\Phi(r,c) = \exp\left({r^2/c^2}\right) $ \\ 
\bottomrule
\end{tabular}
}
\end{center}
\end{table}
Follows Equation~\eqref{rbfequation} with the collocation scheme,
\begin{equation}
\tilde{f}\left( \mathbf{v}_i \right) = f\left( \mathbf{v}_i \right), \quad i=1,2,\cdots,n
\end{equation}
leads to a system of linear equations,
\begin{equation}
\mathbf{W} \boldsymbol{\alpha} = \mathbf{f}
\end{equation}
where $\boldsymbol{\alpha} = \left[ \alpha_1,\cdots,\alpha_n \right]^T$, $\mathbf{f} = \left[  f\left( \mathbf{v}_1 \right),\cdots,f\left( \mathbf{v}_n \right) \right]^T$, and a matrix $\mathbf{W}_{ij} = \Phi\left( r_{ij} \right)\in \mathbb{R}^{n\times n}$. The distance matrix, $r_{ij}$, contained within $\mathbf{W}$ can be expressed as follows,
\begin{equation}\label{distancematrix}
r_{ij} = \left[ \begin{array}{ccc}
\left\|\mathbf\!{u}_{1}\! - \!\mathbf\!{v}_{1}\!\right\|_{2} & \cdots & \left\|\mathbf\!{u}_{1}\! -\! \mathbf\!{v}_{n}\!\right\|_{2}\\
\vdots & \ddots & \vdots\\
\left\|\mathbf\!{u}_{n}\! - \!v_{1}\!\right\|_{2} & \cdots & \left\|\mathbf\!{u}_{n}\! - \! \mathbf\!{v}_{n}\!\right\|_{2}
\end{array}\right],
\quad i,j\!=\!1,\!\cdots\!, n
\end{equation}

After applying RBF to the data sets, we obtain the similarity graph and the weighted matrix $\mathbf{W}_{ij}$, with entries $\mathbf{W}_{ij} = \Phi (\|\mathbf{v}_{i} - \mathbf{v}_{j} \|_{2})$, $i = 1, \cdots, n$, also the interpolation matrix. $\mathbf{W}_{ij}$ consists of the functions serving as the basis of the approximation space. For distinct data points in the data sets and a constant shape parameter $c$, $\mathbf{W}_{ij}$ is a nonsingular matrix. Both the choice of the RBF and its corresponding shaping parameter play an important role in calculating the final interpolations and partitions in the resulting graph. 

\subsubsection{Higher-Order Proximities}
\label{subsec:HOP}
We compute the high-order similarity matrix with the RBF transformation from Section \ref{subsec:rbf} using the Katz index. 

The Katz index \citep{katz1953new, ou2016asymmetric} computes the relative influence of a node within a network. We call nodes that are directly connected to a node as immediate neighbors. Therefore, the Katz index measures the number of immediate neighbors and the immediate neighbors of its immediate neighbors. It is a weighted summation of the path node-set between two nodes. The weight of a path is an exponential function of its length (the number of nodes on this path). We formularize the {\em Katz high-order proximity matrix} $\mathbf{K}$ as:

\begin{equation}
    \label{equ:katz}
    \mathbf{K}=(\mathbf{I}-\beta \cdot \mathbf{W})^{-1}\cdot \beta \cdot \mathbf{W}
\end{equation}

where $\mathbf{W}$ is the weighted matrix acquired from the RBF transformation, $ \beta$ is a decay parameter, which determines the weight of a path decay speed as the length of the path grows. $\beta$ should be properly set to preserve the series convergence. In practice, the decay parameter $\beta$ must be smaller than the spectral radius of the weighted matrix $\mathbf{W}$. Conventionally, in this paper, we set $\beta$ to $0.0025$. 

The pseudocode for computing the high-order proximity of an affinity matrix using Gaussian RBF is shown in Algorithm \ref{algo:HOP}. This algorithm first generates a list of $n$ shaping parameters $\mathbf{c}$ (Line 2). Then, iteratively find an optimal shaping parameter (Line 3-4) where GaussianRBF(·,·) computes the distance using the Gaussian RBF. We can compute MQ and iMQ RBFs using the procedures analogous to Algorithm \ref{algo:HOP} by substituting Line 4 with the respective RBF distances.

\begin{algorithm}

\SetKwData{Left}{left}
\SetKwData{This}{this}
\SetKwData{Up}{up}
\SetKwFunction{condition}{condition}
\SetKwFunction{GaussianRBF}{GaussianRBF}
\SetKwFunction{linspace}{linspace}
\SetKwFunction{Katz}{Katz}
\SetKwFunction{DMatrix}{DMatrix}
\SetKwFunction{NESSC}{NESSC}
\SetKwInOut{Input}{Input}
\SetKwInOut{Output}{Output}
\caption{\textsc{High-order Proximity (HOP)}}
\label{algo:HOP}
\Input{
Affinity matrix: $\mathbf{A} \in \mathbb{R}^{n \times n}$ \\ }
\Output{High-order matrix: $\mathbf{K}$}
\BlankLine
$\mathbf{x} \leftarrow \linspace(0.001,\ 1,\ n)$\;
$\mathbf{c} \leftarrow \linspace(0.001,\ 1,\ 100)$\;
\For{$i \leftarrow 1 \; \KwTo \; n$}{
    $\mathbf{B} \leftarrow \GaussianRBF(\mathbf{c}_i, \DMatrix(\mathbf{x}^T, \mathbf{x}^T))$\;
}
$\mathbf{C} \leftarrow$ optimal $\mathbf{B}$\;
$\hat{\mathbf{C}} \leftarrow \mathbf{C} \cdot \mathbf{A}$\;
$\mathbf{K} \leftarrow \Katz(\hat{\mathbf{C}})$\;

\end{algorithm}

\subsubsection{Normalized Eigens}
\label{subsec:eigens}
We have obtained the high-order similarity matrix $\mathbf{K}$ in Section \ref{subsec:rbf} and \ref{subsec:HOP}. Then, we obtain the eigen features to prepare for the post-clustering procedure.

The diagonal matrix $\mathbf{D} \defeq \mathrm{diag}(\mathbf{K})$, where the diagonal value $\mathbf{D}_{ii}$ is the degree of the row $i$ of $\mathbf{K}$ and the off-diagonal elements are $0$. The normalized Laplacian matrix $\mathbf{L}_\sigma$ with ridge regularization $\sigma$ can then be formed.
\begin{equation}
\label{equ:L}
    \mathbf{L}_\sigma = (\mathbf{D}+ \sigma \cdot d_{max} \cdot \mathbf{I})^{-\frac{1}{2}} \mathbf{K} (\mathbf{D}+ \sigma \cdot d_{max} \cdot \mathbf{I})^{-\frac{1}{2}} 
\end{equation}
where $d_{max}$ is the maximum node degree of the network. The empirical setting of $\sigma$ is $0.1$.

Next, we compute $k+1$ largest eigenvalues $\bm{\mathbf{\hat{\lambda}}}$ and their corresponding eigenvectors $\bm{\mathbf{\hat{\Xi}}}$, a.k.a. $k+1$ leading eigenvectors, and sort them in non-descending order by $\bm{\mathbf{\hat{\lambda}}}$. Consequently, the feature matrix's dimension is reduced from $n \times n$ to $n \times (k+1)$. The reduced feature matrix $\hat{\bm{\Theta}}$ is computed by:
\begin{equation}
    \label{equ:theta}
    \hat{\bm{\Theta}} \defeq \bm{\hat{\Xi}} \cdot Diag(\bm{\mathbf{\hat{\lambda}}})
\end{equation}

where $Diag(\bm{\mathbf{\hat{\lambda}}})$ forms a diagonal matrix from a tuple of eigenvalues $\hat{\lambda}$. Thus, the feature matrix can be expressed as the matrix dot product of $\bm{\hat{\Xi}}$ and $ Diag(\bm{\mathbf{\hat{\lambda}}})$

\subsubsection{Eigen-selection and Clustering}
\label{subsec:eigen}
We now select a subset of eigen features from $\hat{\bm{\Theta}}$ obtained from Section \ref{subsec:eigens} for clustering.

Suppose the high-order matrix $\mathbf{K}$ contains ``signal" and ``noise" network information. A network with a ``strong signal" has a large gap between the $k^{th}$ and the $(k + 1)^{th}$ eigenvectors of its Laplacian matrix. We determine whether a network has a ``weak" signal profile by setting a threshold $t>0$,
\begin{equation}
    \label{equ:t}
    \delta_{k+1}=\frac{\bm{\mathbf{\hat{\lambda}}}_{k+1}}{\bm{\mathbf{\hat{\lambda}}}_{k}} \ge 1-t
\end{equation}
If the above Equation (\ref{equ:t}) is satisfied, then we say that a network is of ``weak signal" profile. In that case, therefore, the $(k + 1)^{th}$ eigenvector contains useful information as the $k^{th}$ eigenvector. Consequently, we consider it as one more feature contributing to label clustering. Finally, we apply the k-means algorithm to the new normalized feature matrix with $k+1$ dimensions to compute the communities. 

\begin{figure*}
\centering
  \includegraphics[width=0.98\textwidth]{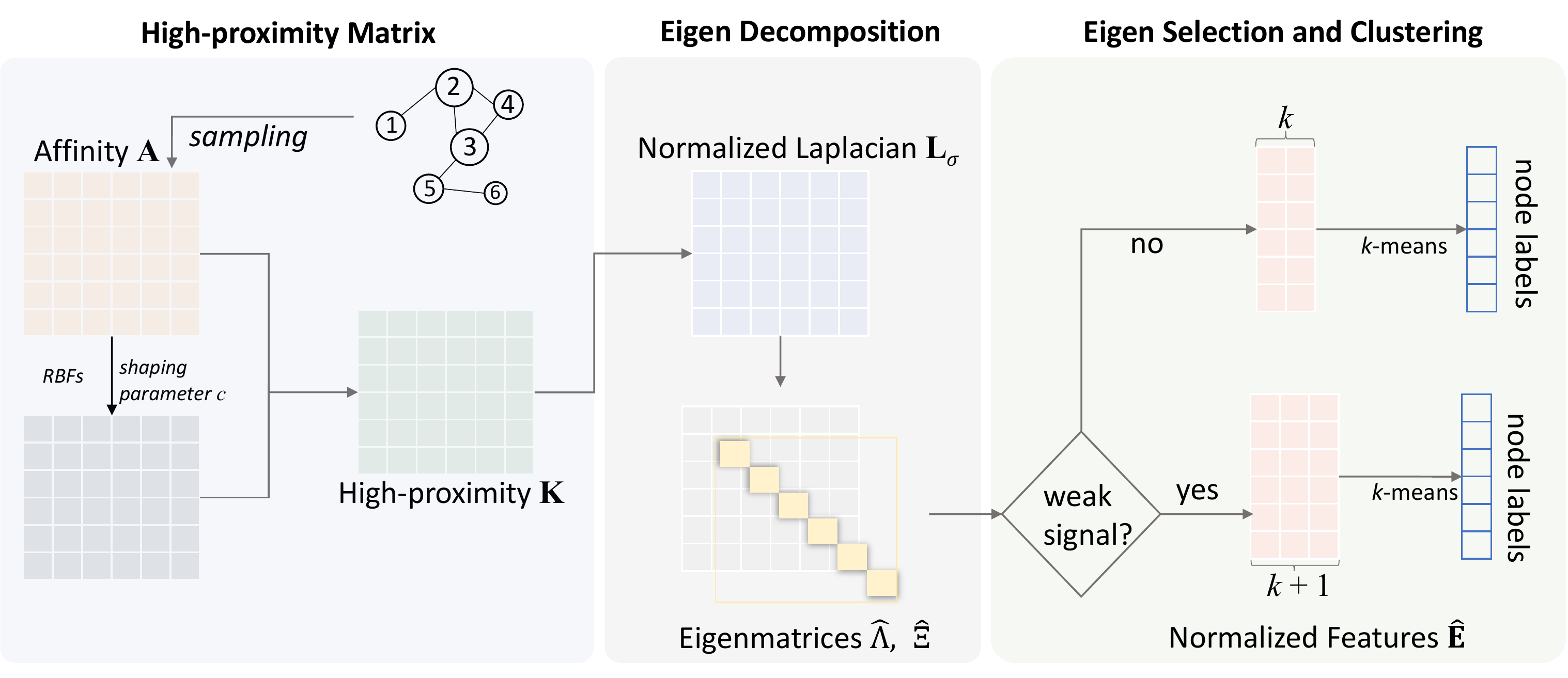}
  \caption{Flowchart of SCOREH+ Model. This model consists of three phases: i) the high-order proximity matrix extraction from the original graph using RBF and Katz; ii)  eigen-decomposition from the normalized Laplacian from the high-proximity matrix; iii) eigen-selection~(lines 8 - 9 of Algorithm~\ref{algo:SCOREH+}) and clustering. A step-by-step computation of this toy example is included in Appendix \ref{app:example}.}
  \label{flowchart}
\end{figure*}

The pseudocode is presented in Algorithm \ref{algo:SCOREH+}. Line $1$ computes the high-order proximity of a network from the original affinity matrix using Algorithm \ref{algo:HOP} and returns a matrix with the same dimension as its input. Then, we collect eigenpairs (eigenvalue, eigenvector) of $\mathbf{K}$ and arrange them in decreasing order by eigenvalues (Lines $2$ to $5$). Lines $8$ - $9$ determine if the network has a strong or weak signal profile and assign $k'$ accordingly. The algorithm returns a list of clustered node labels. A flowchart is illustrated in Fig.~\ref{flowchart} for easier access to our model framework.

\begin{algorithm}
\SetKwData{Left}{left}
\SetKwData{This}{this}
\SetKwData{Up}{up}
\SetKwFunction{HOP}{HOP}
\SetKwFunction{Kmeans}{k-means}
\SetKwFunction{Diag}{Diag}
\SetKwFunction{eigsh}{eigsh}
\SetKwInOut{Input}{Input}
\SetKwInOut{Output}{Output}
\caption{SCOREH+}
\label{algo:SCOREH+}

\Input{ $\mathbf{A} \in \mathbb{R}^{n \times n},\sigma >0$, $t\in (0,1), k\in \mathbb{N}_{> 1}$\\ }
\Output{Node labels: $\hat{\mathbf{y}}$}
\BlankLine
$\mathbf{K} \leftarrow \HOP(\mathbf{A})$\;
$\mathbf{D} \leftarrow \mathrm{Diag}(\mathbf{K})$\; 
$\mathbf{L}_\sigma \leftarrow (\mathbf{D}+ \sigma \cdot d_{max} \cdot \mathbf{I})^{-\frac{1}{2}} \mathbf{K} (\mathbf{D}+ \sigma \cdot d_{max} \cdot \mathbf{I})^{-\frac{1}{2}} $\;
$k' \leftarrow k+1$\; 
$\bm{\mathbf{\hat{\lambda}},\  \mathbf{\hat{\Xi}}} \leftarrow  \eigsh(\mathbf{L}_\sigma,\ k')$\; 
$\mathbf{\hat{\Lambda}} \leftarrow \Diag(\bm{\mathbf{\hat{\lambda}}}) $ \;
$\hat{\bm{\Theta}} \leftarrow \bm{\hat{\Xi}} \cdot \hat{\Lambda}$\;

\If{$\frac{\hat{\lambda}_{k'}}{\hat{\lambda}_{k}} < 1-t$}
{$k' \leftarrow k$\; }
$\hat{\mathbf{E}} \leftarrow \{  \frac{\hat{\bm{\Theta}}_{h}}{\hat{\bm{\Theta}}_{1}},\cdots,  \frac{\hat{\bm{\Theta}}_{k'}}{\hat{\bm{\Theta}}_{1}} \}$;

$\hat{\mathbf{y}} \leftarrow \Kmeans(\hat{\mathbf{E}},\  k)$;
\end{algorithm}

\subsection{Time Complexity}
The time complexity of the classic spectral clustering (SC) is $O(n^3)$, which consists of the construction of similarity matrix ($O(n^2)$), computing the $k$ leading eigenvectors ($O(n^3)$) and the post-k-means clustering ($O(nk^2 \ell)$), where $\ell$ is the number of iterations of convergence at time. Our SCOREH+ algorithm enjoys the same time complexity bound as SC. Algorithm \ref{algo:HOP} takes at most $O(n^3)$ to find the optimal RBF and calculate the Katz index of the resulting RBF matrix. Algorithm \ref{algo:SCOREH+} also requires $O(n^3)$ to get the leading eigenpairs. Note that the post-k-means clustering may need $O(n(k+1)^2 \ell)$ since we consider the $(k+1)^{th}$ eigenvector as a feature for clustering. However, the overall time complexity is still $O(n^3)$. In comparison, the time complexity of FastGreedy \citep{clauset2004finding} and Louvain \citep{blondel2008fast} algorithms are $O(nm^2)$ and $O(n \log n)$, respectively.

The most expensive step of spectral-based clustering is the computation of the eigenvalues/eigenvectors of the Laplacian matrix. However, eigenvalues/eigenvectors can be efficiently computed with approximation algorithms \citep{boutsidis2015spectral}. Moreover, the complexity of k-means can also be reduced to $O(k^2 \log^2 k)$ \citep{tremblay2016compressive}. In social networks/complex networks, the network is sparse due to the small-world effect. Therefore, the matrix computations in social networks are efficient in practice.

\section{Experimental Settings}
\label{sec:experimental_evaluation}
We first give an overview of the real-world network and synthetic networks. Then, we also briefly present the baseline algorithms for comparisons. Subsequently, we introduce two widely used metrics (modularity and NMI) in community detection tasks. Finally, we gave the detailed experimental design, including the programming language, experiment platform, online resources, etc.

\subsection{Datasets}
\subsubsection{Real-world Networks}
We tested our algorithm and others (ASE, Louvain, fast-Greedy, SC, SCORE, and SCORE+) on $11$ public real-world network datasets originating from diverse domains such as social sciences and political sciences. The specific statistical details for each dataset are provided in Table \ref{t:stat}. Note that the Polbooks network is accessible to the public through http://www.orgnet.com/divided.html.

\begin{table}[ht] 
 \centering 
 \footnotesize 
 \caption{Statistics of the 11 real-world datasets. The number of nodes and edges are $n$ and $m$, respectively; the number of clusters is $k$, and the minimum and maximum of the node degrees are $\min(d)$ and $\max(d)$, respectively.} \label{t:stat} 
 \setlength{\parskip}{0.5\baselineskip} 
  \resizebox{140mm}{25mm}{
 \begin{tabular}{clccccccc}
 \hline {No.} & {Dataset} & {Reference} & {$n$} & {$m$} & {$k$} & {$\min(d)$} & {$\max(d)$}  \\ \hline
$1$ & $\textbf{Les~Mis{\'e}rable}$ & \citep{knuth1993stanford} & $77$ & $254$ & $11$ & $1$ & $36$  \\
$2$ & $\textbf{Caltech}$ & \citep{traud2011comparing} & $590$ & $12,822$ & $172$ & $1$ & $179$   \\
$3$ & $\textbf{Dolphins}$ &\citep{lusseau2003bottlenose} & $62$ & $159$ & $2$ & $1$ & $12$   \\
$4$ & $\textbf{Football}$ & \citep{girvan2002community} & $110$ & $568$ & $11$ & $7$ & $12$   \\
$5$ & $\textbf{Karate}$ &\citep{zachary1977information} & $34$ & $78$ & $2$ & $1$ & $17$   \\
$6$ & $\textbf{Polbooks}$ & \citep{politicalblogs2008} & $92$ & $374$ & $2$ & $1$ & $24$   \\
$7$ & $\textbf{Blog}$ &\citep{adamic2005political} & $1,222$ & $16,714$ & $2$ & $1$ & $351$  \\
$8$ & $\textbf{Simmons}$ & \citep{traud2012social} & $1,137$ & $24,257$ & $4$ & $1$ & $293$   \\
$9$ & $\textbf{UKfaculty}$ & \citep{nepusz2008fuzzy}& $79$ & $552$ & $3$ & $2$ & $39$   \\
$10$ & $\textbf{Github}$ & \citep{rozemberczki2021multi}& $37,700$ & $289,003$ & $2$ & $1$ & $9,458$  \\
$11$ & $\textbf{Facebook}$ & \citep{rozemberczki2021multi}& $22,470$ & $171,002$ & $4$ & $1$ & $709$  \\
 \hline 
 \end{tabular} 
 }
\end{table}
\subsubsection{Synthetic Networks}

We generate a series of synthetic networks using the LFR criteria, first proposed by Lancichinetti, Fortunato, and Radicchi \citep{lancichinetti2008benchmark}. A network can be generated in terms of the given parameters: the power-law exponent for the degree distribution $\tau_1$, the power-law exponent for the community size distribution $\tau_2$, the number of nodes $n$, the average degree $ \langle d_{ave}\rangle$, the minimum of communities  $c_{min}$, and the mixing parameter $\mu$. Most importantly, $\mu$ controls the fraction of edges between communities. Thus, it reflects the amount of noise in the network. When $\mu =0$, all links are within community links; when $\mu =1$, all links are between nodes from different communities.

We generate networks by setting the number of nodes ranging from $150$ to $10,000$ and the key parameter $\mu$ (mixing parameter) from $0.15$ to $0.85$.

\subsection{Baseline Algorithms}
We select traditional greedy methods (Louvain \citep{blondel2008fast}, Fast-Greedy \citep{clauset2004finding}), spectral methods (spectral clustering (SC) \citep{ng2002spectral}, Spectral Clustering on Ratios-of-Eigenvectors (SCORE) \citep{jin2015fast}, SCORE+ \citep{jin2018score+}), and a graph embedding based algorithm ASE as baseline algorithms for comparisons.
\begin{itemize}
    \item Louvain~\citep{blondel2008fast}: Louvain is one of the most successful community detection algorithms with a good balance of effectiveness and efficiency.
    \item Fast-Greedy~\citep{clauset2004finding}: A greedy algorithm that iteratively searches for nodes with the largest modularity.
    \item SC~\citep{ng2002spectral}: SC is the classic spectral clustering, wherein the top $k$ eigenvectors are retained to form the feature matrix, followed by using k-means as the post-clustering algorithm.
    \item SCORE~\citep{jin2015fast}: SCORE is a modification to the SC algorithm by incorporating a normalization step that involves the leading eigenvector.
    \item SCORE+~\citep{jin2018score+}: This algorithm extends the SCORE algorithm, and it considers one more eigenvector as the feature matrix for ``weak" signal networks.
    \item Adjacency spectral embedding (ASE)~\citep{sussman2012consistent}: ASE is an approach used to infer the latent positions of a network that is represented as a Random Dot Product Graph. This embedding serves as both dimensionality reduction and fitting a generative model.    
\end{itemize}

\subsection{Evaluation Metrics}
\label{subsec:eva}
\subsubsection{Modularity}
\label{subsec:q}
Newman and Girvan \citep{newman2004finding} proposed modularity $\mathit{Q}$ to assess the quality of the detected community structure. It represents the difference between the actual number of connections and the expected number of connections in random graphs. The equation is as follows:%
\begin{equation}
     Q = \sum\limits_{s=1}^k\left[\dfrac{\ell_s}{\ell}-\left(\dfrac{d_s}{2\ell}\right)^2\right] 
\end{equation}
where $k$ is the number of communities in the network, $\ell$ is the total number of edges, $\ell_s$ is the sum of all edges in the community $s$, and $d_s$ is the sum of the degree of all nodes in $s$. Modularity usually takes value from $[-0.5,1]$, with positive values indicating the possible presence of community structure \citep{newman2006modularity}. Modularity is used when the ground-truth labels of a network are unknown. Usually, higher modularity implies a better community structure.

\subsubsection{Normalized Mutual Information}
\label{subsec:nmi}
Normalized mutual information (NMI) \citep{danon2005comparing} is used to evaluate the community detection algorithms' network partitioning. It is widely used due to its comprehensive meaning and ability to compare community labels $\mathbf{\hat{y}}$ obtained from an algorithm and $\mathbf{y}$, the list of ground-truth labels. Denote $H(\cdot)$ as the entropy function for graph partitioning. Then, the mutual information between the ground truth and the detected labels is:
\begin{equation*}
    \text{MI}(\mathbf{\hat{y}}, \mathbf{y}) = \sum_{i=1}^{|\mathbf{\hat{y}}|} \sum_{j=1}^{|\mathbf{y}|} \frac{|\mathbf{\hat{y}}_i \cap \mathbf{y}_j|}{n}\log\left(n\frac{|\mathbf{\hat{y}}_i \cap \mathbf{y}_j|}{|\mathbf{\hat{y}}_i||\mathbf{y}_j|}\right) 
\end{equation*}
 Then, the normalized mutual information is:
 \begin{equation*}
      \text{NMI}(\mathbf{\hat{y}}, \mathbf{y}) = \frac{\text{MI}(\mathbf{\hat{y}}, \mathbf{y})}{\text{mean}(H(\mathbf{\hat{y}}), H(\mathbf{y}))} 
 \end{equation*}
This metric is independent of the absolute values of the labels and the number of communities. Note that NMI takes value from 0 to 1, inclusive. 
When network labels are known, NMI is a good metric to evaluate the ``accuracy" of the results of a community detection algorithm.

\subsection{Experimental Design}
Our experiments were carried out on a 16.0 GB RAM, 1.90GHz Intel(R) Core(TM) i7-8650U CPU. The implementations of all algorithms are based on Python 3.9 and its packages. Specifically, the classic spectral clustering (SC) was adopted from scikit-learn. We re-implement the MatLab codes of SCORE and SCORE+ from \citep{jin2015fast, jin2018score+} using Python, where we utilized scipy and numpy packages for matrix computations. We called the functions of NetworkX package for the Louvain and FastGreedy algorithms. For ASE, we followed the graspologic package \citep{chung2019graspy} by Johns Hopkins University and Microsoft \footnote{https://microsoft.github.io/graspologic/latest/index.html}. The implementations of our algorithm in Python is publicly accessible \footnote{https://github.com/yz24/RBF-SCORE}.

We tested the proposed algorithm and the baselines on the above-mentioned two types of data sets. For every network, we run each model ten times on spectral-based baselines and SCOREH+, evaluate two metrics, and then report the mean and variance of the results in the form of $mean~(variance)$.

\section{Experimental Results and Analyses}
\label{sec:experiment-analyses}
In this section, we show how to choose the number of clusters and select parameters (RBF function and the corresponding shaping parameter) for RBFs. We also compare our algorithm SCOREH+ with baseline algorithms on real-world and synthetic networks. We report the experimental results by scoring the quality of the discovered communities with modularity and NMI. Moreover, to show the differences among these models, we compare and analyze the topological structures of four small networks: Karate, UKfalculty, Dolphin, and Polbooks.   

\subsection{Experimental Results on Real-world Datasets}
\subsubsection{Number of Clusters}
 \label{subsec:clusters}
We use the method introduced in Section \ref{subsec:eigen} to determine the number of clusters for our algorithm. Table \ref{t:hp-delta} reports the type (weak or strong) of each network by using the value $\delta_{k+1}$ from the high-order matrix computed by Algorithm \ref{algo:HOP}. We defer the analogous results with respect to the original affinity matrix to Appendix \ref{app:eigen}.

\begin{table}[ht] 
 \centering 
 \footnotesize 
 \caption{type of networks from the high-order matrix and the eigenvalue ratios} \label{t:hp-delta} 
 \setlength{\parskip}{0.5\baselineskip} 
  \resizebox{90mm}{25mm}{
 \begin{tabular}{clccccc}
 \hline {No.} & {Dataset} & {Type} & {$\delta_{k+1}$} & {$\delta_{k+2}$} & {$\delta_{k+3}$}\\ \hline
1  & Les~Mis{\'e}rable & Strong & 0.172 & 0.04  & 0.231 \\
2  & Caltech       & Weak   & 0.079 & 0.025 & 0.075 \\
3  & Dolphins      & Strong & 0.486 & 0.187 & 0.076 \\
4  & Football      & Weak   & 0.019 & 0.098 & 0.101 \\
5  & Karate        & Strong & 0.257 & 0.426 & 0.164 \\
6  & Polbooks      & Strong & 0.818 & 0.092 & 0.148 \\
7  & Blog         & Strong & 0.369 & 0.198 & 0.041 \\
8  & Simmons       & Strong & 0.242 & 0.328 & 0.205 \\
9  & Ukfaculty     & Strong & 0.397 & 0.109 & 0.143 \\
10 & Github        & Strong &  0.224  &  0.124  &  0.031 \\
11 & Facebook      &  Strong  &  0.115  & 0.365  & 0.216 \\
 \hline 
 \end{tabular} 
 }
 \end{table}

Algorithm \ref{algo:HOP} alters Football to ``weak" signal while the Simmons becomes ``strong". For each ``weak" network, our algorithm selects one more eigenvector. We report the comparison results in Table \ref{t:r:q} (Modularity) and Table \ref{t:r:nmi} (NMI). In this comparison, we use the default Gaussian RBF with shaping parameter $0.1$. We will analyze the effect of various RBF choices and shaping parameters in Section \ref{subsec:rbfs}.

For the two ``weak" signal networks Caltech and Football, we conducted extensive experiments to show if additional eigenvectors beyond the $(k+1)^{th}$ one are necessary, and the results are shown in Appendix \ref{app:weak}. The results demonstrate that more eigenvectors have a limited contribution to the accuracy of the community detection task. In contrast, additional eigenvectors can increase the time complexity. Thus, the eigen selection method in this paper is sufficient.

\begin{table}[ht] 
 \centering 
 \footnotesize 
 \caption{Numerical results on real-world networks (Modularity)} \label{t:r:q} 
 \setlength{\parskip}{0.5\baselineskip} 
  \resizebox{160mm}{25mm}{
 \begin{tabular}{clccccccc}
 \hline {No.} & {Dataset} & {ASE} & {Louvain} & {Fast-Greedy} & {SC} & {SCORE} & {SCORE+} & {SCOREH+}
 \\ \hline
$1$ & Les~Mis{\'e}rable & 0.381 & $\textbf{0.556}$ & $0.501$ & $-0.019(0.021)$ & $0.386(0.057)$ & $0.239(0.038)$ & $\underline{0.486(0.002)}$\\
$2$ & Caltech & 0.322 & $\textbf{0.412}$ & $0.335$ & $0.39(0.0)$ & $0.372(0.002)$ & $0.373(0.001)$ & $0.368(0.003)$ \\
$3$ & Dolphins & 0.223 &  $\textbf{0.519}$ & $0.495$ & $0.379(0.0)$ & $0.276(0.0)$ & $0.353(0.0)$ & $0.379(0.0)$ \\
$4$ & Football & 0.622 &  $0.622$ & $0.58$ & $\textbf{0.624(0.0)}$ & $0.622(0.021)$ & $\textbf{0.624(0.0)}$ & $0.622(0.01)$ \\
$5$ & Karate & 0.37 &  $\textbf{0.42}$ & $0.381$ & $0.36(0.0)$ & $0.371(0.0)$ & $0.36(0.0)$ & $0.371(0.0)$ \\
$6$ & Polbooks & 0.475 &  $\textbf{0.51}$ & $0.503$ & $0.479(0.0)$ & $0.473(0.0)$ & $0.479(0.0)$ & $0.479(0.0)$\\
$7$ & Blog & 0.25 &  $\textbf{0.427}$ & $\textbf{0.427}$ & $0.0(0.0)$ & $\underline{0.423(0.0)}$ & $0.424(0.0)$ & $0.415(0.001)$ \\
$8$ & Simmons & 0.343 &  $\textbf{0.486}$ & $0.478$ &  $0.482(0.0)$ & $0.462(0.0)$ & $0.46(0.0)$ & $0.447(0.004)$\\
$9$ & UKfaculty & 0.425 &  $\textbf{0.461}$ & $0.457$ & $0.442(0.0)$ & $0.262(0.131)$ & $0.44(0.0)$ & $0.442(0.0)$\\
$10$ & Github & 0.171 &  $0.153$ & $\textbf{0.284}$ &  $0.11(0.0)$ & $0.185(0.01)$ & $0.271(0.0)$ & $0.281(0.0)$\\
$11$ & Facebook & 0.088 &  $0.113$ & $0.173$ &  $0.154(0.0)$ & $0.145(0.0)$ & $0.097(0.0)$ & $\textbf{0.175(0.0)}$\\
 \hline 
 \end{tabular} 
  }
 \end{table}

\begin{table}[ht] 
 \centering 
 \footnotesize 
 \caption{Numerical results on real-world networks (NMI)} \label{t:r:nmi} 
 \setlength{\parskip}{0.5\baselineskip} 
  \resizebox{160mm}{25mm}{
 \begin{tabular}{clccccccc}
 \hline {No.} & {Dataset} & {ASE} & {Louvain} & {Fast-Greedy} & {SC} & {SCORE} & {SCORE+} & {SCOREH+} \\ \hline
$1$ & Les~Mis{\'e}rable & 0.641 &  $\textbf{0.864}$ & $0.668$ & $0.377(0.038)$ & $0.686(0.032)$ & $0.616(0.022)$ & $\underline{0.752(0.008)}$\\
$2$ & Caltech & 0.554 &   $\textbf{0.685}$ & $0.44$ & $0.63(0.001)$ & $0.58(0.003)$ & $0.574(0.003)$ & $\underline{0.637(0.006)}$ \\
$3$ & Dolphins & 0.383 &   $0.484$ & $0.557$ &$\textbf{1.0(0.0)}$ & $0.588(0.0)$ & $0.811(0.0)$ & $\textbf{1.0(0.0)}$ \\
$4$ & Football & 0.944 &   $0.884$ & $0.74$ & $0.934(0.0)$ & $0.946(0.022)$ & $0.934(0.0)$ & $\textbf{0.958(0.012)}$ \\
$5$ & Karate & \textbf{1.0} &   $0.687$ & $0.692$ & $0.836(0.0)$ & $\textbf{1.0(0.0)}$ & $0.836(0.0)$ & $\textbf{1.0(0.0)}$ \\
$6$ & Polbooks & 0.784 &   $0.553$ & $0.701$ & $0.87(0.0)$ & $\textbf{0.924(0.0)}$ & $0.87(0.0)$ & $0.87(0.004)$ \\
$7$ & Blog & 0.176 &   $0.636$ & $0.654$ & $0.006(0.0)$ & $\underline{0.725(0.0)}$ & $\textbf{0.751(0.0)}$ & $0.646(0.001)$\\
$8$ & Simmons & 0.425 &   $\textbf{0.707}$ & $0.693$ & $\underline{0.702(0.001)}$ & $0.621(0.0)$ & $0.615(0.001)$ & $0.658(0.005)$ \\
$9$ & UKfaculty & 0.621 &   $0.834$ & $0.878$ & $\textbf{0.95(0.0)}$ & $0.658(0.187)$ & $0.917(0.0)$ & $\textbf{0.95(0.0)}$\\
$10$ & Github & 0.146 &   $0.0$ & $0.005$ & $0.12(0.0)$ & $0.212(0.01)$ & $0.241(0.0)$ & $\textbf{0.307(0.0)}$\\
$11$ & Facebook & 0.081 &   $0.002$ & $0.006$ & $0.08(0.0)$ & $0.11(0.0)$ & $0.132(0.0)$ & $\textbf{0.146(0.0)}$\\
 \hline 
 \end{tabular} 
 }
 \end{table}

The modularity and NMI values of each algorithm are reported in Table \ref{t:r:q} (Modularity) and Table \ref{t:r:nmi} (NMI). The numerical results show that using the default RBF parameter settings, SCOREH+ achieves the best and second-best on eight out of eleven networks regarding NMI. It is worth mentioning that Fast-Greedy \citep{clauset2004finding} and Louvain \citep{blondel2008fast} are designed to optimize the metric modularity, while spectral-based methods are not. Therefore, Fast-Greedy \citep{clauset2004finding} and Louvain \citep{blondel2008fast} usually achieve high modularity and low NMI. However, our algorithm is based on graph decomposition, neither optimization of modularity nor NMI. The results from our algorithm are more objective and competitive. Regarding computational efficiency, our algorithm demonstrates similar runtime performance to both SCORE and SCORE+. Detailed comparisons of the running times are provided in Appendix \ref{app:time}.  

\subsubsection{RBF selection and tuning of shaping parameters on real-world networks}
\label{subsec:rbfs}

In Section \ref{subsec:rbf}, we discussed three common RBFs and their definitions. Each RBF has a shaping parameter $c$, and it can affect the final graph interpolation matrix, which, as a consequence, can cause a difference in the final clustering result. Therefore, we fine-tune the type of RBF and shaping parameter $c$, and search the optimal parameters where NMI is the criterion.
 
\begin{figure}[ht]
    \centering   \includegraphics[width=0.7\textwidth]{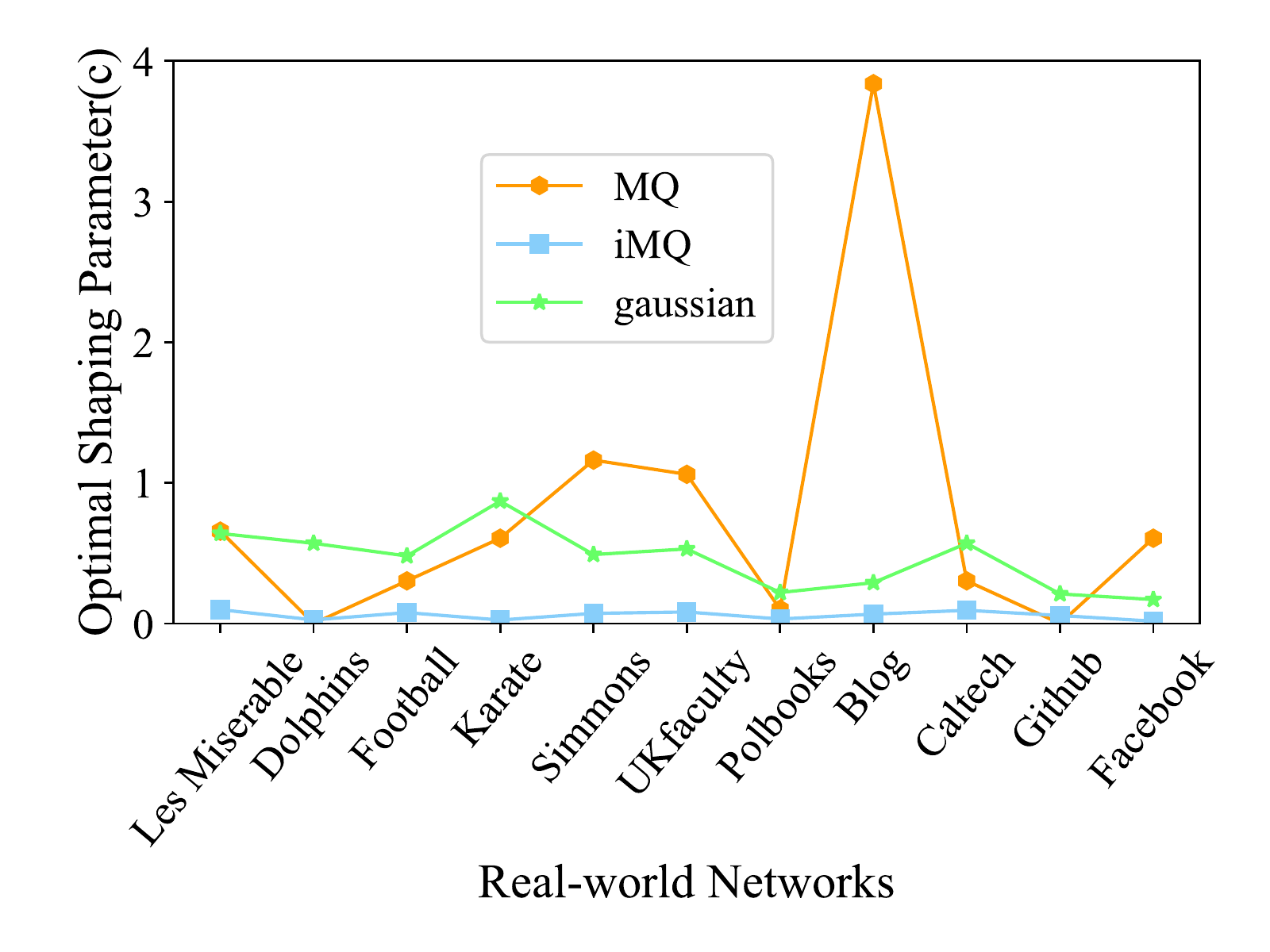}
    \caption{Optimal shaping parameters with RBF choices on real-world networks.}
    \label{fig:rbf}
\end{figure}

Tables \ref{t:r:q} and \ref{t:r:nmi} show that our algorithm has already achieved the best results on Dolphins, Football, Karate, UKfaculty, Facebook, and Github without tuning the RBF shaping parameter. Moreover, we would like to see the domain of optimal shaping parameters for each RBF. Therefore, we experiment on the general range of each RBF and report the results in Fig. \ref{fig:rbf}. From Fig. \ref{fig:rbf}, iMQ RBF is more stable with a constantly small shaping parameter, which means that using iMQ RBF is more likely to achieve optimal results after a small number of iterations. Moreover, although the shaping parameter ranges from $[0,1]$ for iMQ RBF, the empirical experiments from these $11$ real-world networks show that the optimal parameter falls into $[0,0.1]$. This aids in fine-tuning the parameters in real-life applications.

\begin{table}[ht] 
 \centering 
 \footnotesize 
 \caption{Optimal RBF and corresponding shaping parameter} \label{t:optimal-rbf} 
 \setlength{\parskip}{0.5\baselineskip} 
  \resizebox{125mm}{25mm}{
 \begin{tabular}{clccccc}
 \hline {No.} & {Datasets} & {optimal RBF} & {shaping parameter} & {NMI} & {Modularity}\\ \hline

$1$  & Les~Mis{\'e}rable & gaussian & $0.64$ & $0.752$ & $0.48$   \\
$2$   & Caltech  & gaussian & $0.57$ & $0.646$ & $0.4$       \\
$3$   & Dolphins  & MQ & $0$ & $1$   & $0.379$    \\
$4$   & Football  & MQ  & $0.303$  & $0.958$   & $0.62$     \\
$5$   & Karate & iMQ  & $0.0245$ & $1$   & $0.371$ \\
$6$   & Polbooks  & iMQ  & $0.0318$  & $0.924$ & $0.473$    \\
$7$   & Blog & iMQ & $0.0664$   & $0.789$  & $0.415$    \\
$8$   & Simmons & MQ & $1.1616$  & $0.703$  & $0.481$   \\
$9$   & UKfaculty & gaussian   & $0.53$ & $0.95$  & $0.442$\\ 
$10$   & Github & gaussian   & $0.21$ & $0.307$  & $0.281$\\ 
$11$   & Facebook & gaussian   & $0.17$ & $0.146$  & $0.175$\\ 
 \hline 
 \end{tabular} 
 }
 \end{table}

In addition, our algorithm performs better if we fine-tune the shaping parameters. Table \ref{t:optimal-rbf} lists the optimal RBF for each network and the corresponding optimal shaping parameter. We can also interpret that the algorithm has no preferences for any RBFs, and all the common RBFs work well on some networks. The results demonstrate that overall, with the optimal RBF, our algorithm made improvements on Caltech, Polbooks, Blog, and Simmons.  

\subsection{Analyses: Real-World Networks}
\label{sec:real-network-analyses}
In this section, we analyze and visualize the results of spectral-based algorithms on Karate, UKfaculty, Dolphins, and Polbooks. The analyses of the last two networks are deferred to Appendix \ref{app:dolphins-polbooks}. We draw the network structure using the node size to differentiate the node degree and the node color to distinguish the community label. For each smaller network, we plot the topological structure of the results detected by SC, SCORE, SCORE+, and our SCOREH+ to show the differences between those algorithms.

\subsubsection{Analysis of Karate Network}
\begin{figure}[ht]
\centering
\subfigure[Ground-truth network]{
\label{fig:karate:a} 
\includegraphics[width=0.21\textwidth]{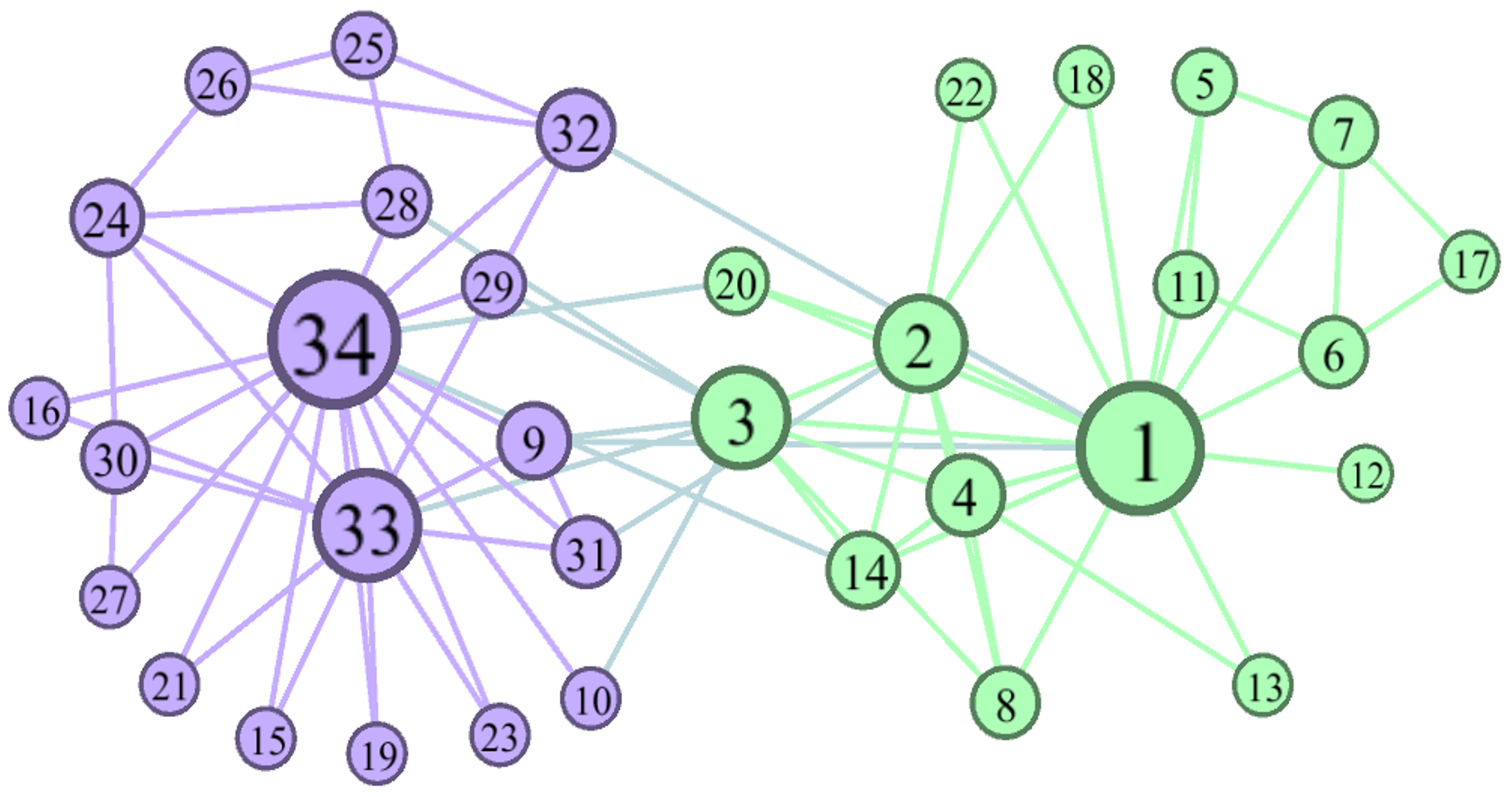}}
~~
\subfigure[Results from SCORE]{
\label{fig:karate:b} 
\includegraphics[width=0.21\textwidth]{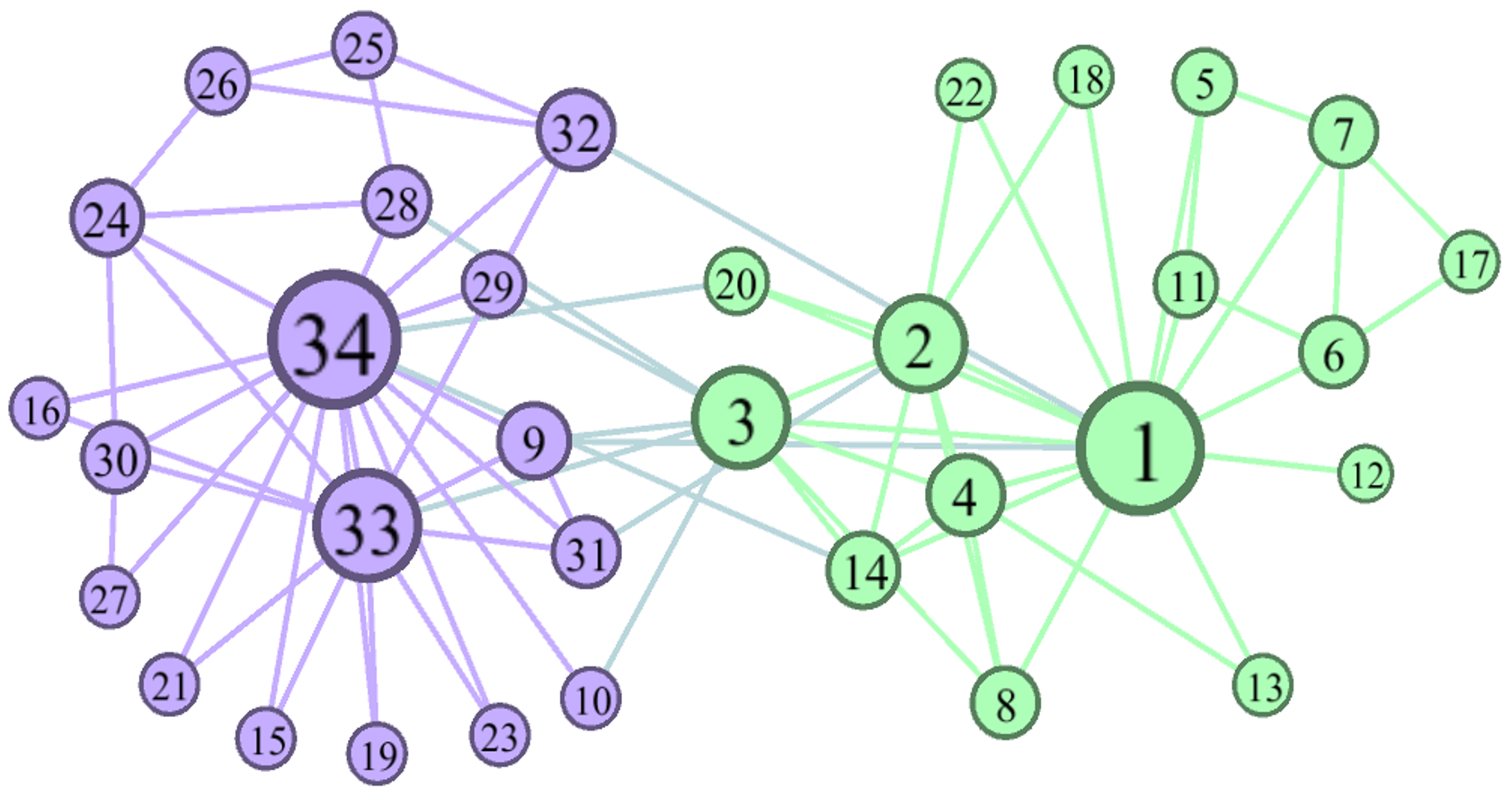}}
~~
\subfigure[Results from SCORE+]{
\label{fig:karate:c} 
\includegraphics[width=0.21\textwidth]{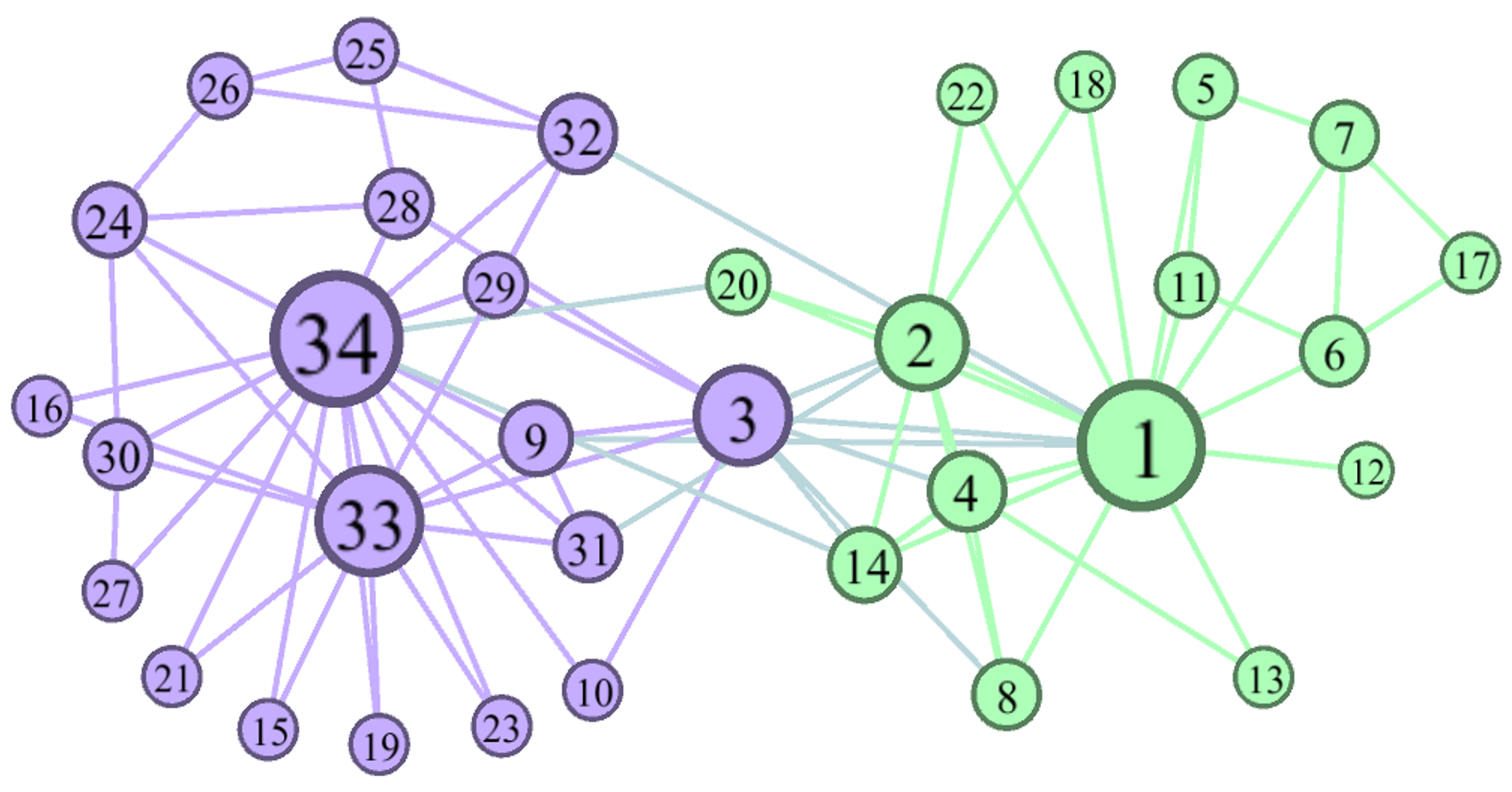}}
~~
\subfigure[Results from SCOREH+]{
\label{fig:karate:d}
\includegraphics[width=0.21\textwidth]{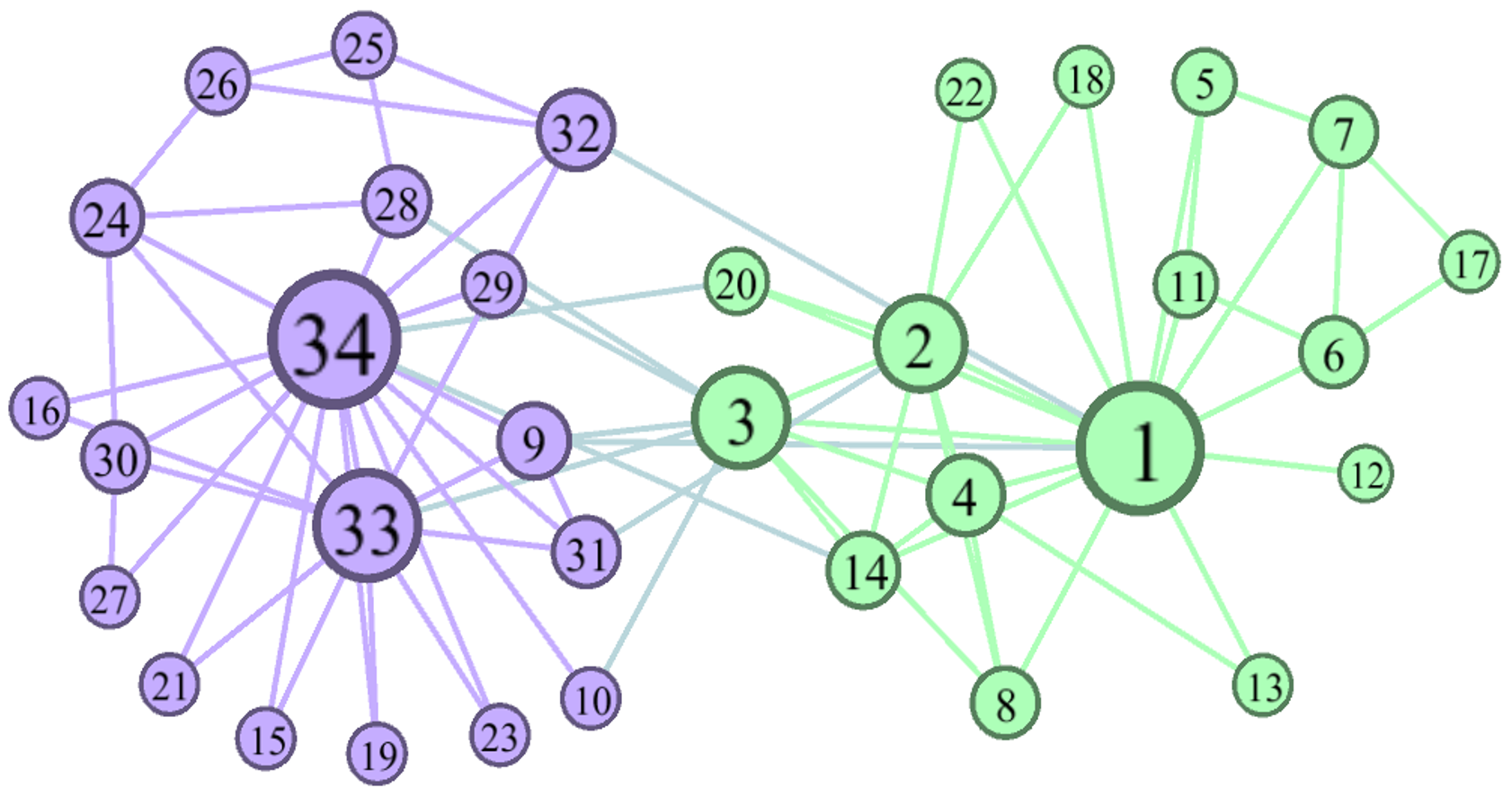}}
\caption{The topological displays for the Karate network from SCORE, SCORE+, and our SCOREH+ algorithms (\ref{fig:karate:a} is plotted from the ground truth of the network for comparison. \ref{fig:karate:b}, \ref{fig:karate:c}) and \ref{fig:karate:d} are from the node labels discovered by SCORE, SCORE+, and our SCOREH+, respectively. The similar settings apply to Fig. \ref{fig:uk}.}
\label{fig:karate} 
\end{figure}

Zachary's karate club network is a social network widely used to test community detection algorithms. This network contains $34$ nodes and $78$ edges, and it was divided into two communities because of the contradictions between the ``president" and the ``instructor" of the karate club. The real topological structure of this network is present in Figure \ref{fig:karate:a}, and the community detected by SCORE, SCORE+, and our SCOREH+ are present in Figure \ref{fig:karate:b}, \ref{fig:karate:c}, and \ref{fig:karate:d}, respectively. Note that the numbering of the subgraphs for the other networks is the same as in Karate.

For this network, SCORE+ misclassified the number $3$ node while SCORE and our SCOREH+ achieved state-of-the-art results.

\subsubsection{Analysis of UKfaculty Network}
The UKfaculty network \citep{nepusz2008fuzzy} is a personal friendship network of UK university faculty, and the school affiliation of each individual is stored as a node label. It comprises $81$ vertices (individuals) and $817$ weighted edges, and three communities.

\begin{figure}[ht]
\centering
\subfigure[Ground-truth network]{
\label{fig:uk:a} 
\includegraphics[width=0.20\textwidth]{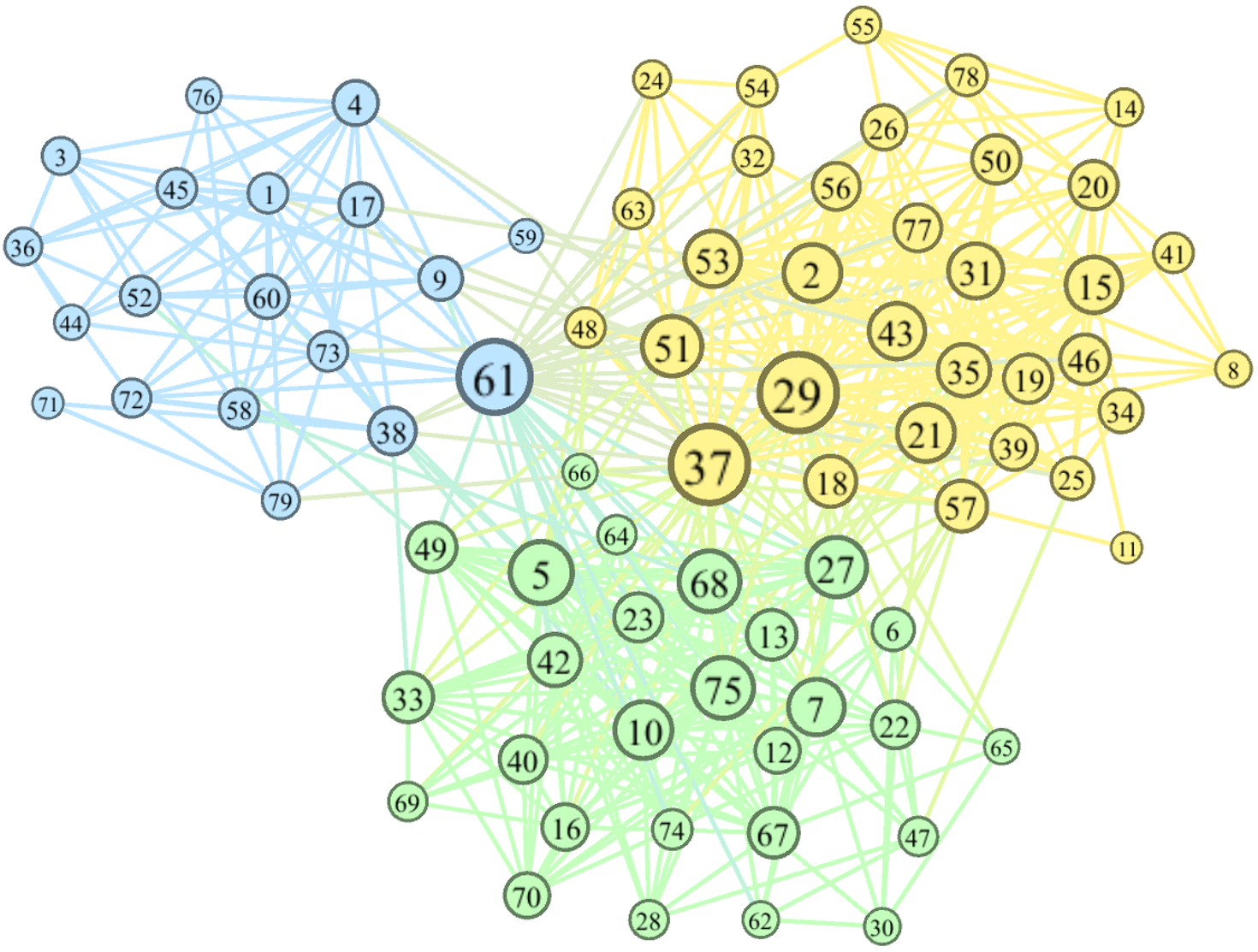}}
~~
\subfigure[Results from SCORE]{
\label{fig:uk:b} 
\includegraphics[width=0.20\textwidth]{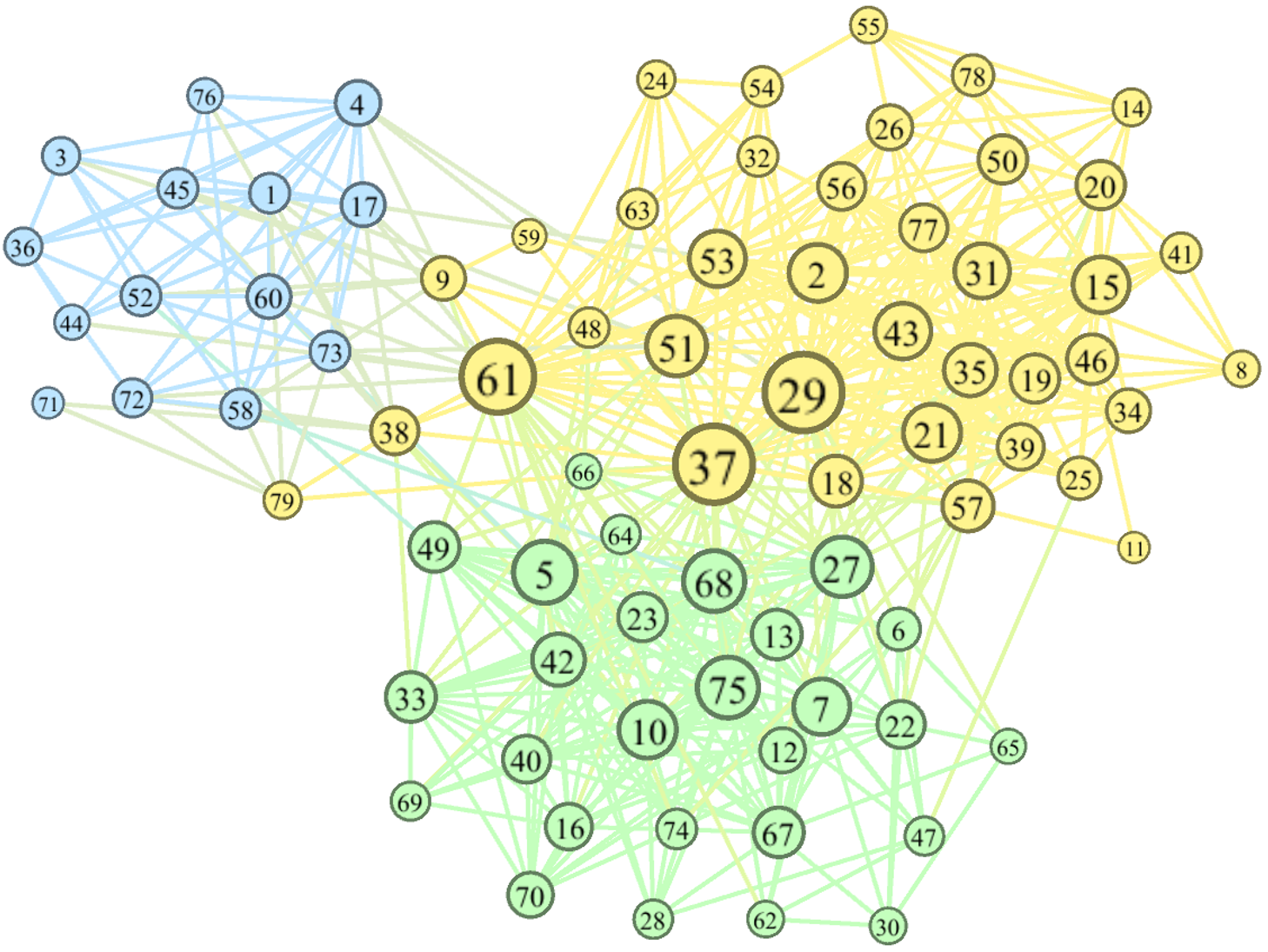}}
~~
\subfigure[Results from SCORE+]{
\label{fig:uk:c} 
\includegraphics[width=0.20\textwidth]{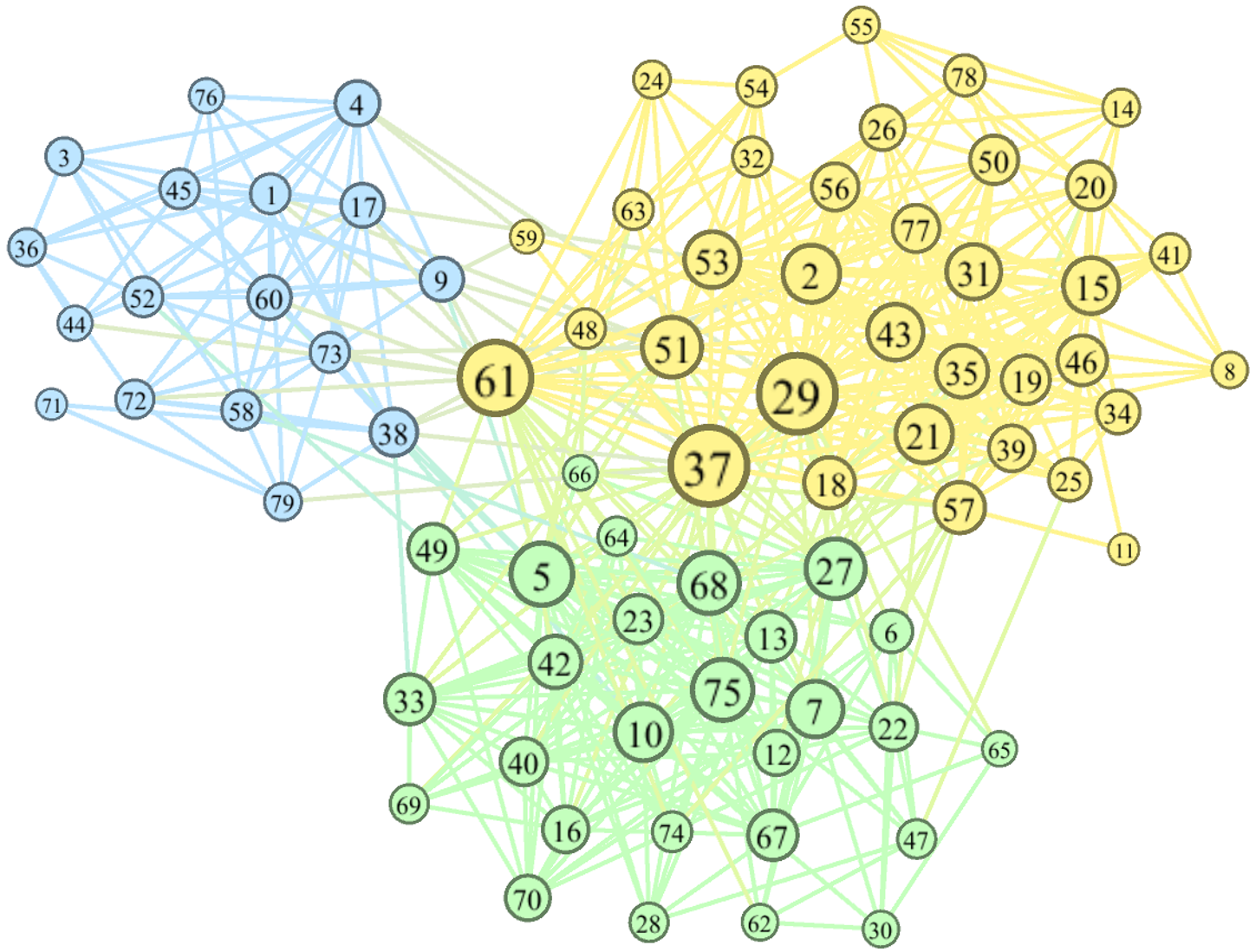}}
~~
\subfigure[Results from SCOREH+]{
\label{fig:uk:d} 
\includegraphics[width=0.20\textwidth]{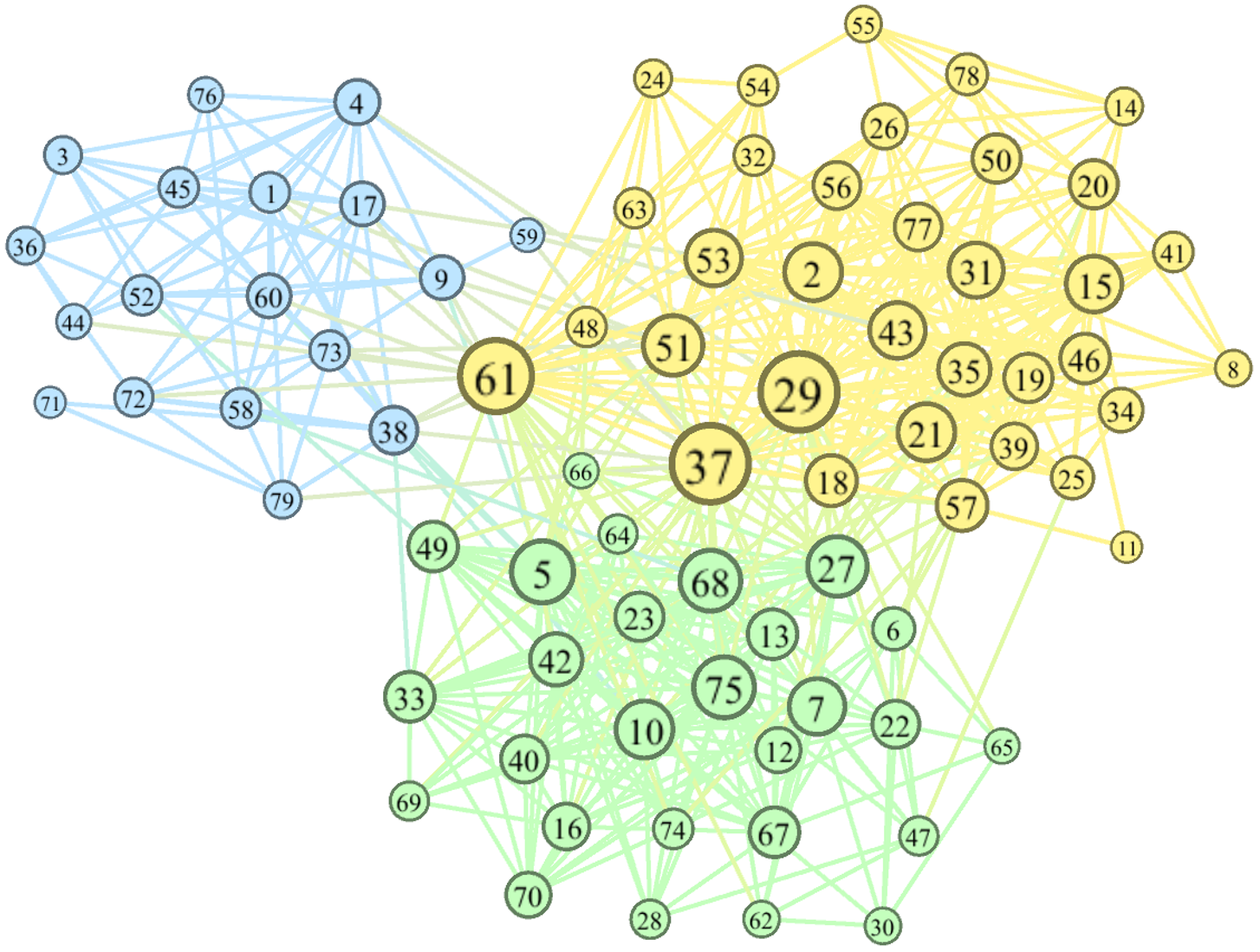}}
\caption{The topological displays for the UKfaculty network from SCORE, SCORE+, and our SCOREH+ algorithms (\ref{fig:uk:a} is from the ground-truth of the network).}
\label{fig:uk} 
\end{figure}

For this network, the experimental comparisons in Fig. \ref{fig:uk} indicate that SCORE misclassified the numbers $9$, $38$, $59$, $61$, $79$ nodes, and SCORE+ made mistakes on the numbers $59$ and $61$ nodes. However, our SCOREH+ only has one misclassified node at node number $61$. 

\subsubsection{Community structures of other Networks}
\begin{figure}
\centering
\subfigure[Caltech]{
\label{fig:real:a} 
\includegraphics[width=0.23\textwidth]{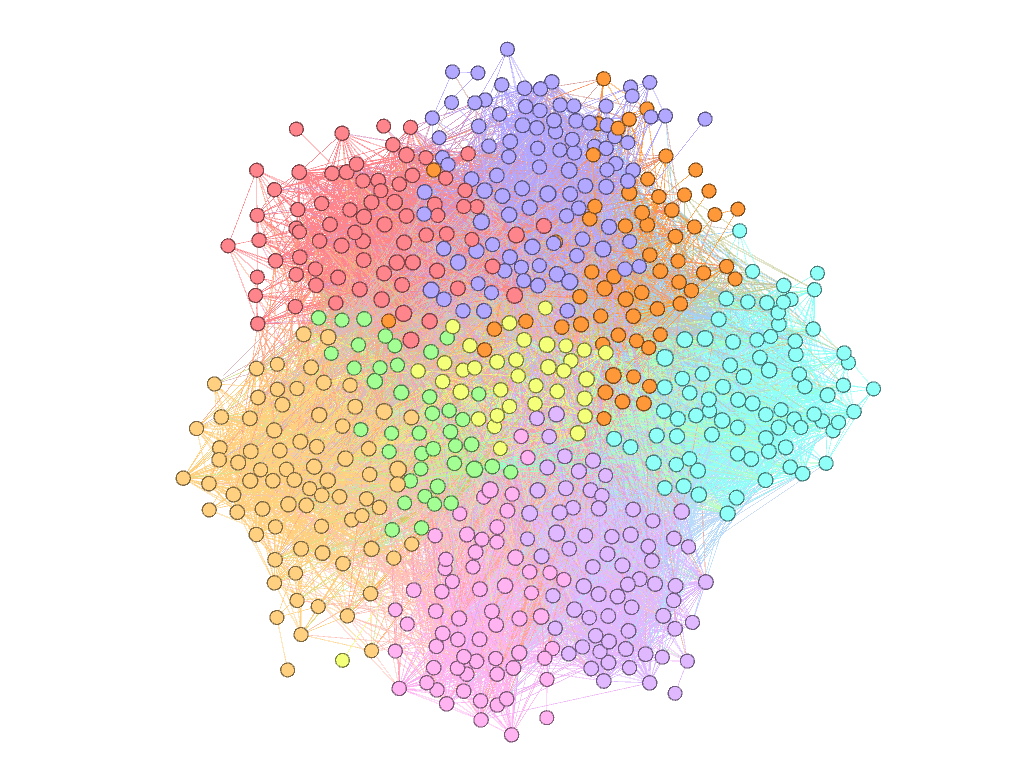}}
\hspace{1em}
\subfigure[Football]{
\label{fig:real:b} 
\includegraphics[width=0.23\textwidth]{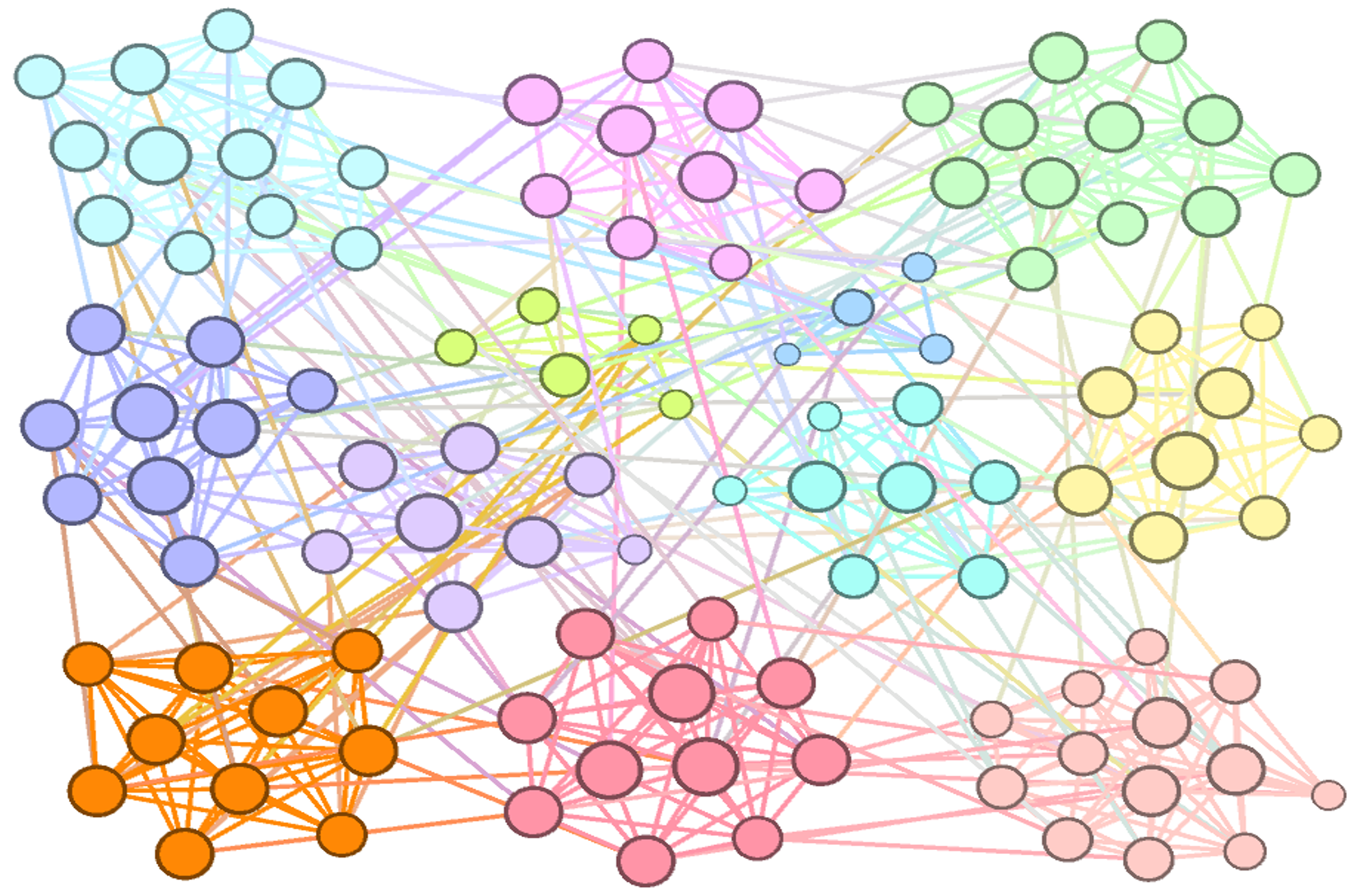}}
\hspace{1em}
\subfigure[Blog]{
\label{fig:real:c} 
\includegraphics[width=0.16\textwidth]{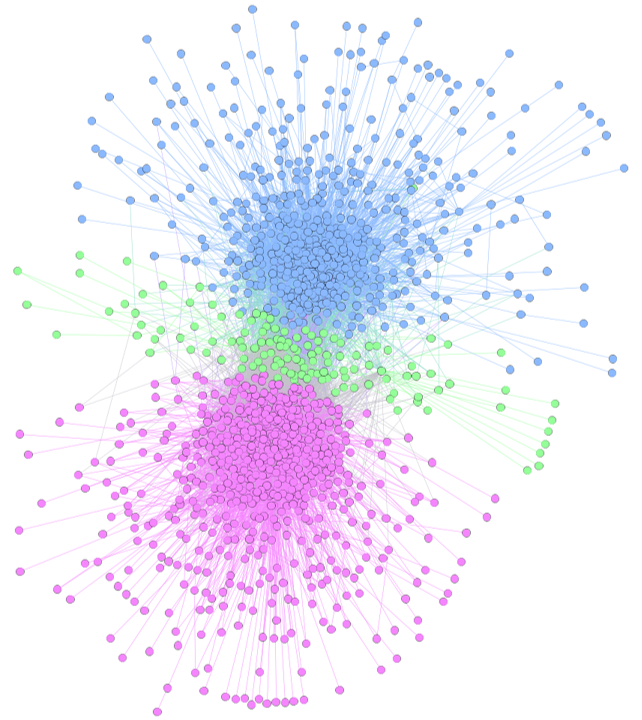}}
\hspace{1em}
\subfigure[Simmons]{
\label{fig:real:d} 
\includegraphics[width=0.16\textwidth]{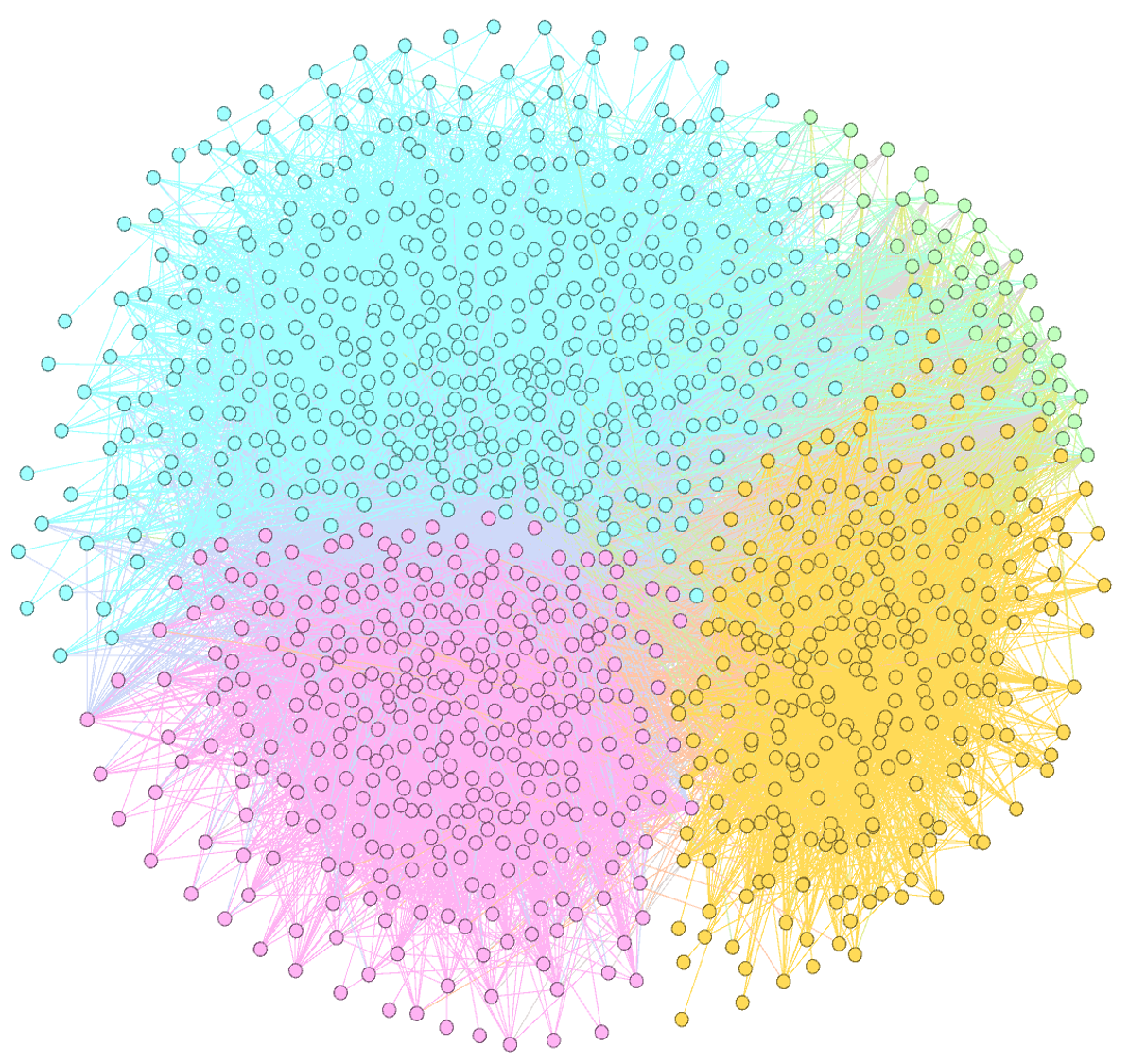}}
\caption{The topological displays for Caltech (Fig.~\ref{fig:real:a}), Football (Fig.~\ref{fig:real:b}), Blog (Fig.~\ref{fig:real:c}), and Simmons (Fig.~\ref{fig:real:d}), respectively, from the results of our SCOREH+ algorithms. }
\label{fig:real_topo} 
\end{figure}

It becomes more work to visualize the structural quality for larger-scale networks. In that case, for Caltech, Football, Blog, and Simmons, we present the topological plots in Fig. \ref{fig:real_topo} by our SCOREH+ algorithm. The Football network \citep{girvan2002community} contained the relationships of American football games between Division IA colleges during the regular season in Fall 2000. The Blog is a directed network of hyperlinks between weblogs on US politics, recorded in 2005 by Adamic and Glance \citep{adamic2005political}. In our experiment, we treat it as an undirected network. Simmons and Caltech \citep{traud2011comparing} are parts of the Facebook friendship networks, recorded in 2005. 

SCOREH+ can detect communities from the above four larger networks. Moreover, the communities show clear clustering properties where the nodes of the same color are close to each other, while the nodes are sparsely connected between different communities. For example, the Caltech social network has eight large clusters and hundreds of small clusters. The political Blog network has two communities in reality, but some ``blogs" are neutral (green on nodes). It is also reasonable and natural to split this network into three clusters.

\subsection{Analyses: Synthetic Networks} \label{subsec:synthetic}
In this subsection, we use the networks generated from the LFR criteria \citep{lancichinetti2008benchmark}. The parameters $\tau_1$ and $\tau_2$ are fixed to be $1.0$ and $1.5$, respectively, for all networks. The number of nodes ranges from $150$ to $10,000$, and the key parameter $\mu$ (mixing parameter) from $0.15$ to $0.85$. Since $\mu$ determines the network noise, we only compare the results with $0.35 \leq \mu \leq 0.65$. When $\mu$ is too small ($\mu \leq 0.25$), the network will be too easy to discover for all the algorithms, while it will be too hard when $\mu$ is too large ($\mu > 0.65$).

We report the results when the numbers of nodes are $5,000$ and $10,000$ with $\mu$ varying from $0.15$ to $0.85$, and when $\mu=0.35,0.55$ with $N$ chosen from $150$ to $10,000$. The detailed tables and extensive figures are attached to Appendix \ref{app:Q}, Appendix \ref{app:NMI}, and Appendix \ref{app:figs}, respectively.

\subsubsection{Comparisons with respect to the number of nodes}
\begin{figure}[ht]
\centering
\begin{subfigure}
\centering
\includegraphics[width=0.43\textwidth]{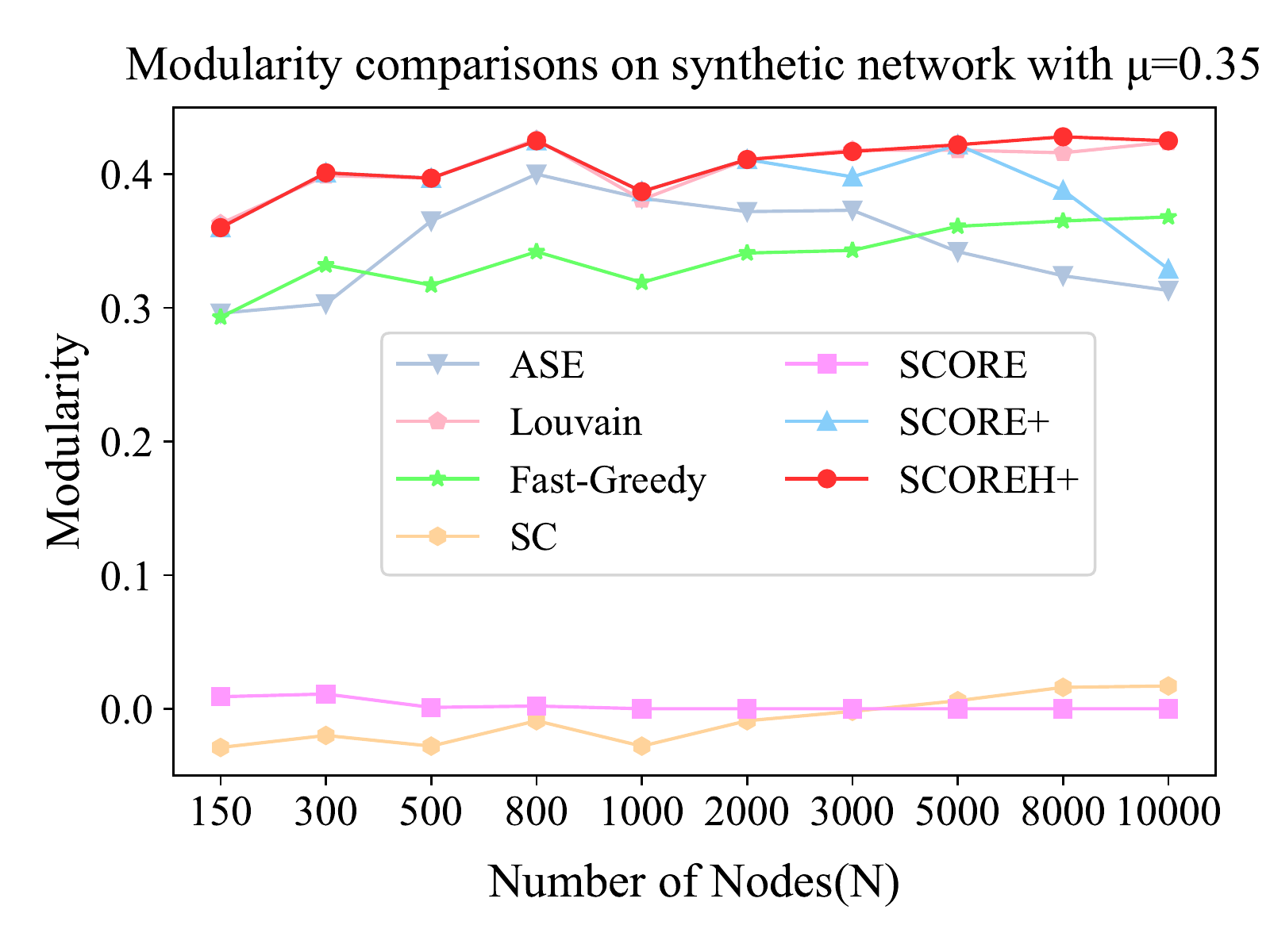}
\end{subfigure}
~
\begin{subfigure}
\centering
\includegraphics[width=0.43\textwidth]{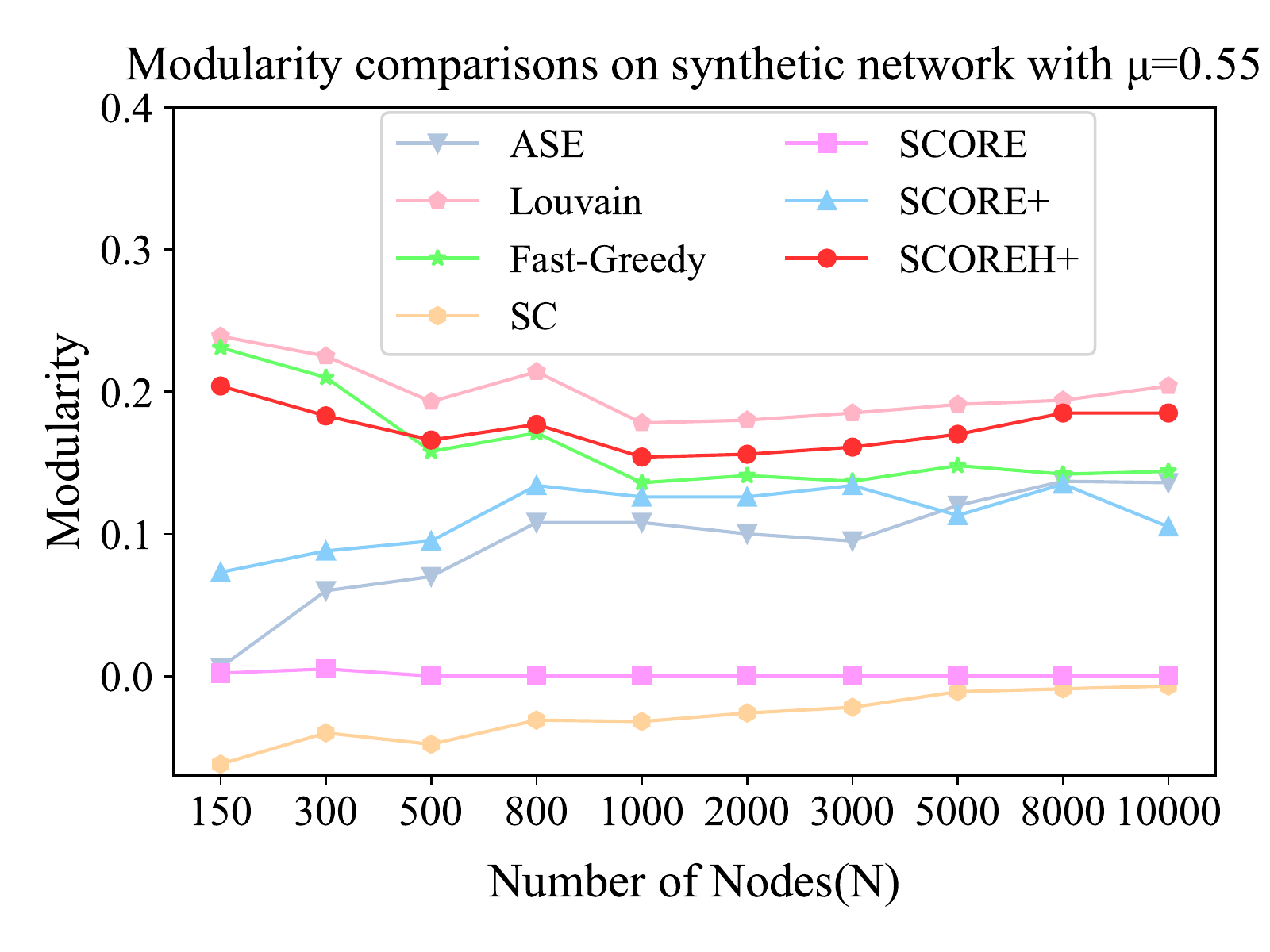}
\end{subfigure}

\begin{subfigure}
\centering
\includegraphics[width=0.43\textwidth]{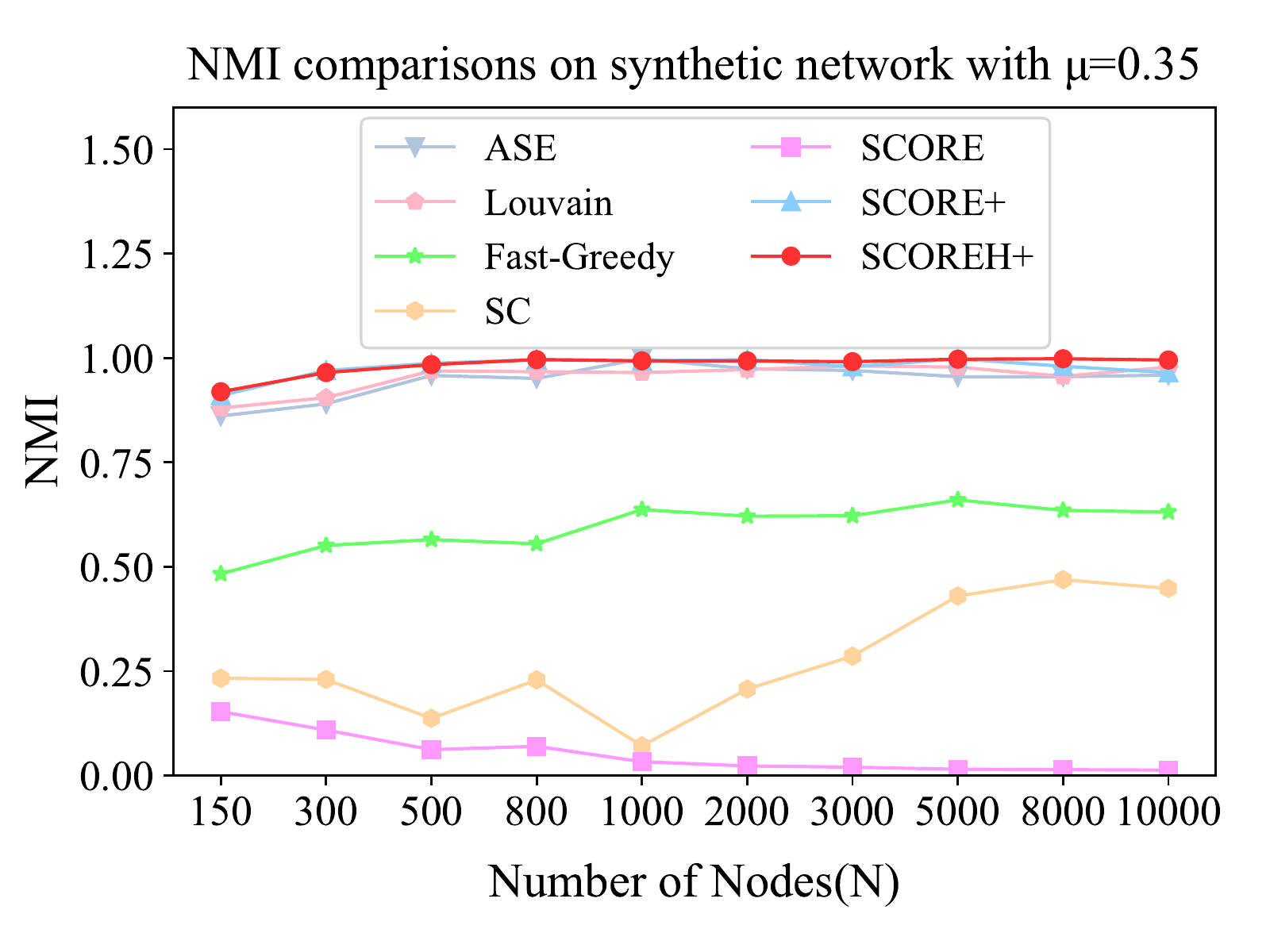}
\end{subfigure}
~
\begin{subfigure}
\centering
\includegraphics[width=0.43\textwidth]{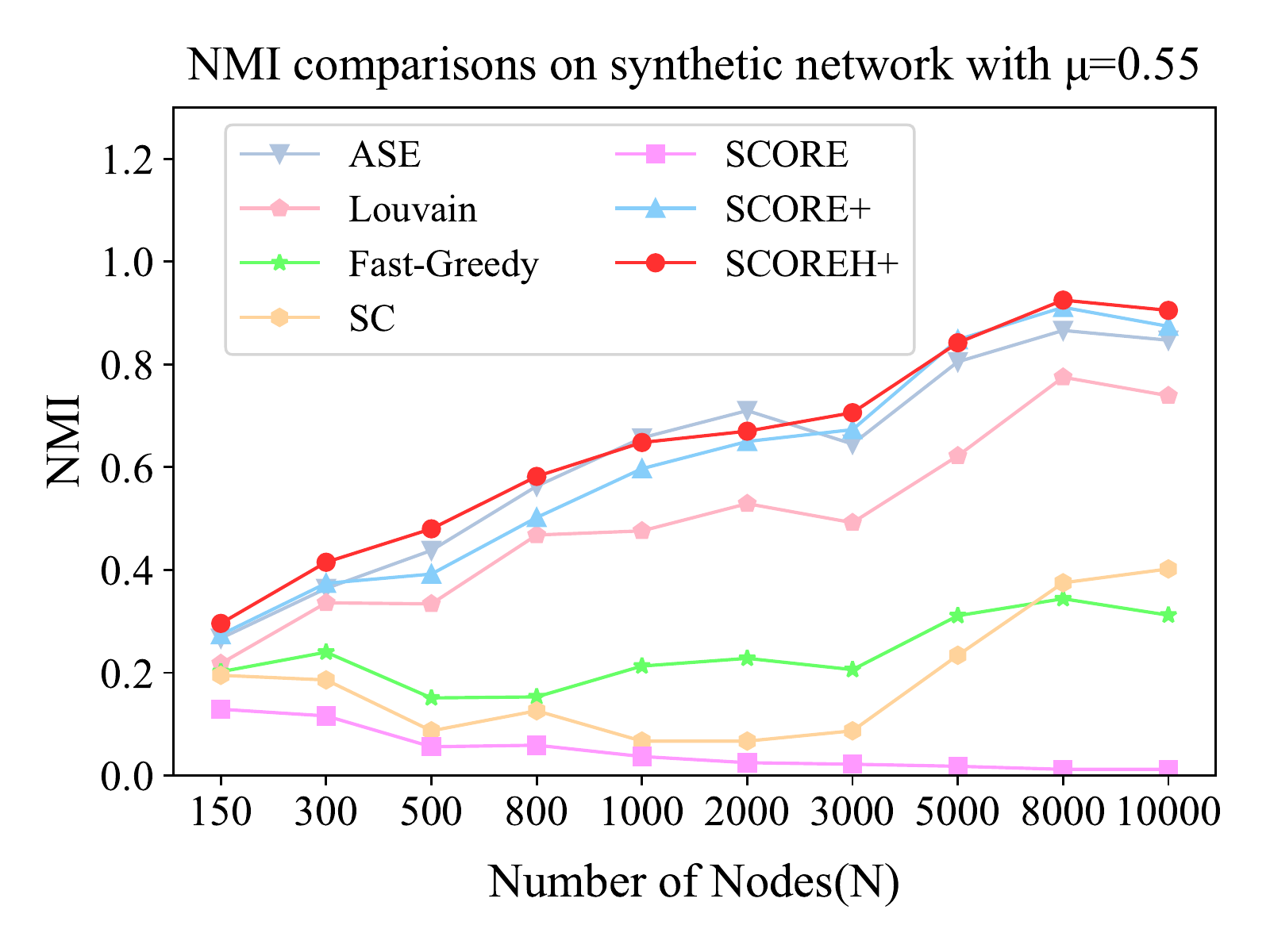}
\end{subfigure}

\caption{The comparison plots on LFR datasets with various numbers of nodes ($N$). The top two plots show the comparisons of modularity measure with respect to the various number of nodes when the mixing parameter $\mu$ is fixed. The bottom two plots are comparisons regarding NMI.
\label{LFR-N}}
\end{figure}

We fix the mixing parameter $\mu$ and compare the performance acquired from ASE, Louvain, fast-Greedy, SC, SCORE, SCORE+, and our SCOREH+. As we can observe from the modularity and NMI comparisons (Fig. \ref{LFR-N}), when $\mu$ are relatively small, SCORE+ and SCOREH+ can obtain excellent community structure, especially for small networks. However, as $\mu$ grows, the superiority of SCOREH+ is obvious. The reason is that our SCOREH+ can preserve more local information, and this property makes a difference when the network is noisy.

\subsubsection{Comparisons with respect to mixing parameter \texorpdfstring{$\mu$}{L}}
\begin{figure}[ht]
\centering
\begin{subfigure}
\centering
\includegraphics[width=0.43\textwidth]{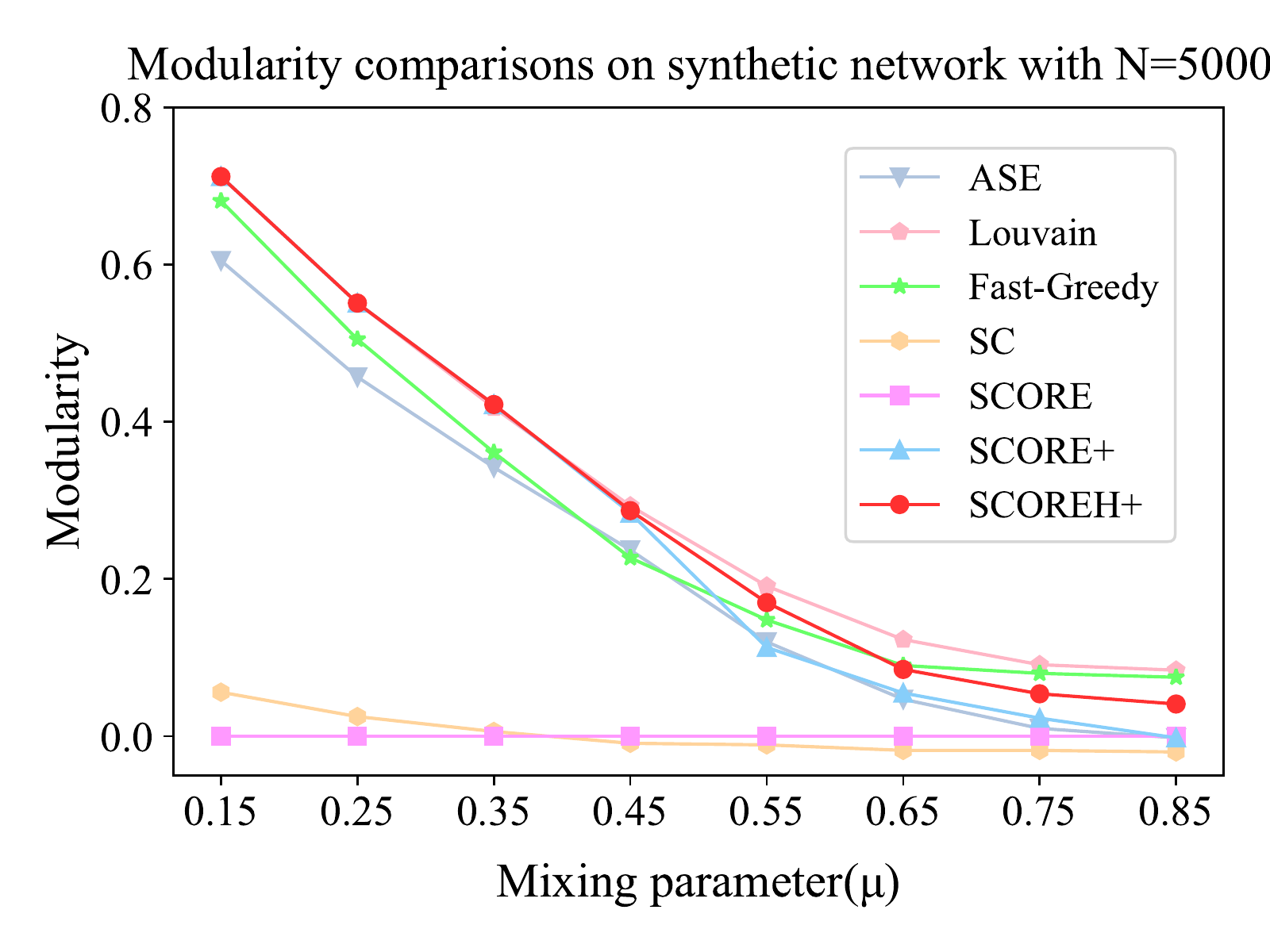}
\end{subfigure}
~
\begin{subfigure}
\centering
\includegraphics[width=0.43\textwidth]{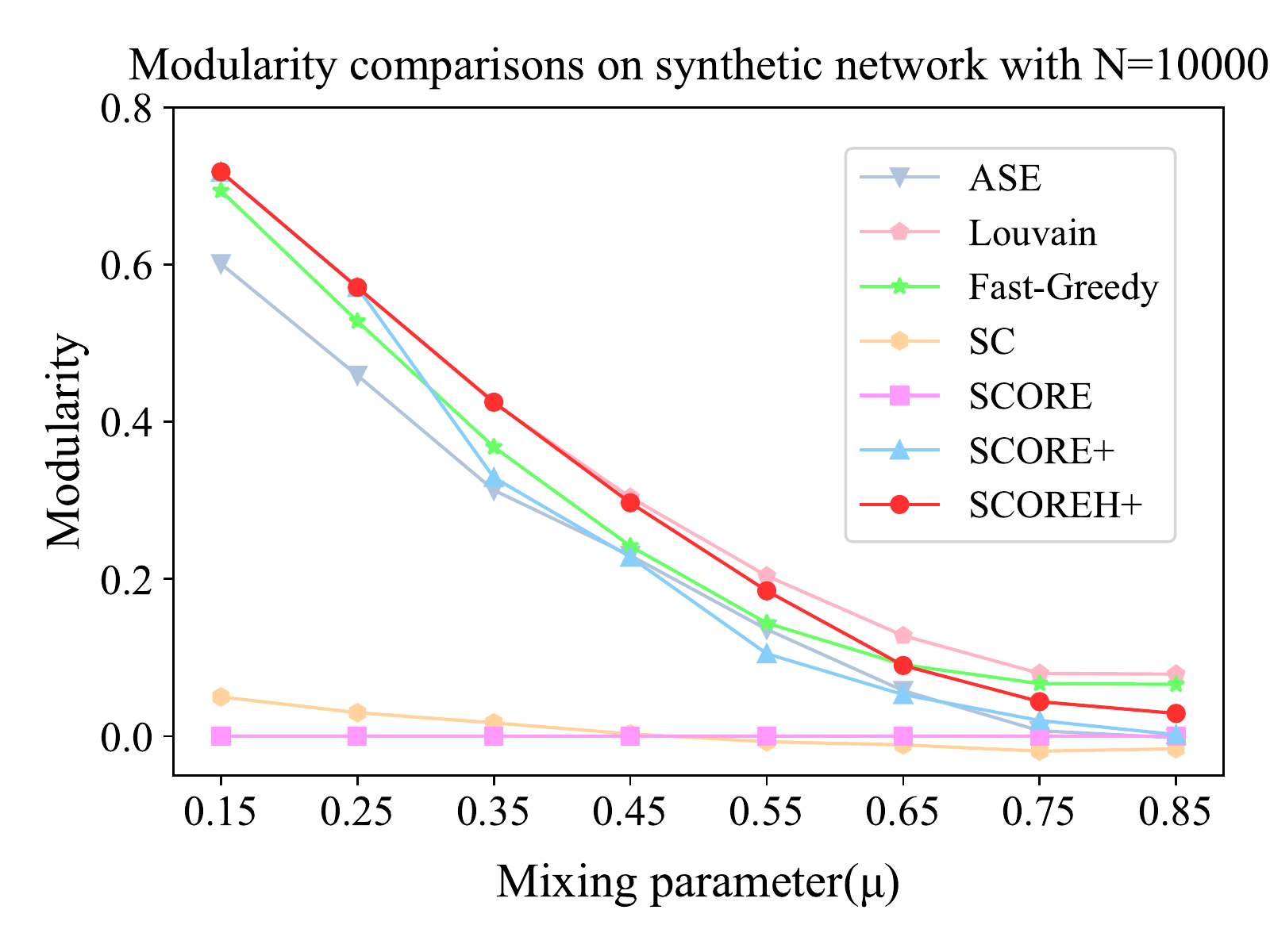}
\end{subfigure}

\begin{subfigure}
\centering
\includegraphics[width=0.43\textwidth]{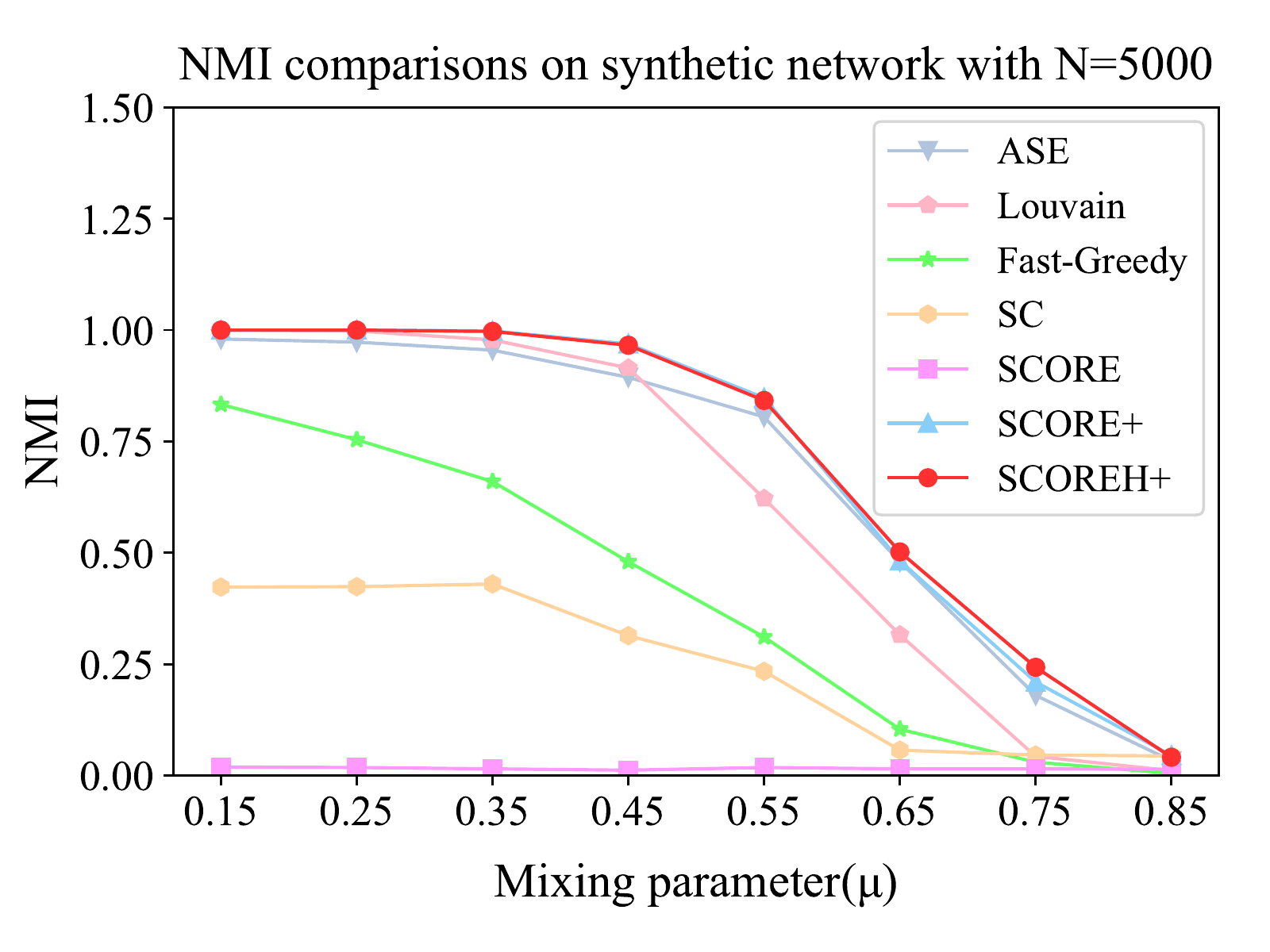}
\end{subfigure}
~
\begin{subfigure}
\centering
\includegraphics[width=0.43\textwidth]{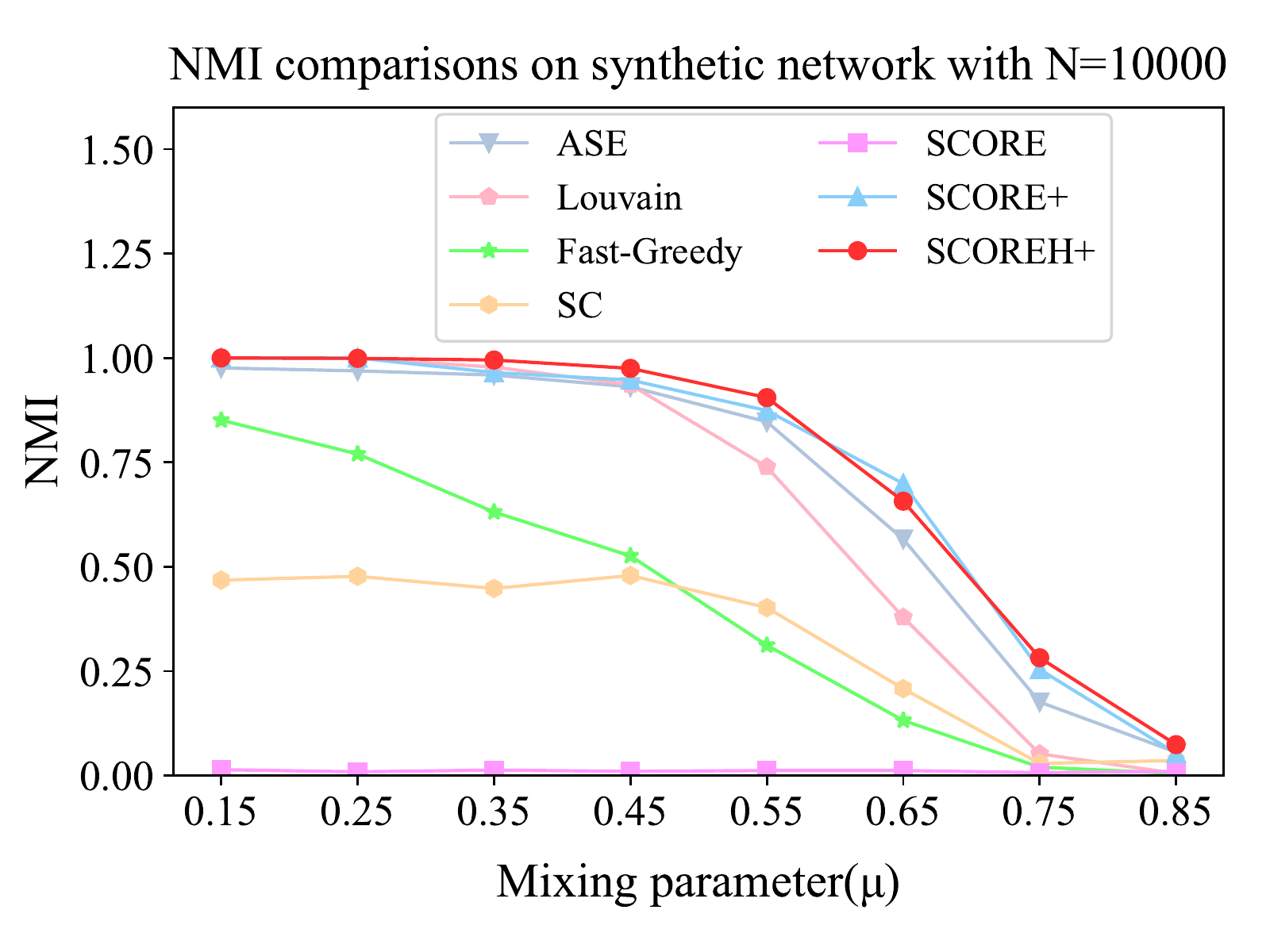}

\end{subfigure}

\caption{The comparison plots on LFR datasets with various mixing parameters $\mu$. The top two plots show the comparisons of modularity measures with respect to the mixing parameters. The bottom two plots are comparisons regarding NMI.}
\label{LFR-mu} 
\end{figure}

In this subsection, we compare the performance of each algorithm when the mixing parameter $\mu$ is fixed. Next, we show how mixing parameters affect the results on the same scale as networks. The modularity and NMI comparisons (Fig. \ref{LFR-mu}) show that $\mu$ greatly affects the performance of algorithms. This is reasonable since it determines the difficulty of a network. When $\mu$ is very small, every algorithm can detect nearly perfect communities, while a large $\mu$ can result in a modularity metric with approximately $0$, representing a nearly random community discovery.

\section{Conclusion and Future Work} 
\label{sec:conclusion}
We proposed an improved algorithm based on SCORE+ for detecting communities in complex networks. The RBF implementation and the choice of the optimal shaping parameter cast the node vector into an approximation domain while retaining high-order information. This approach proves particularly robust in noisy networks, surpassing the performance of conventional algorithms.
A careful selection of radial basis functions (RBFs) and shaping parameters is of greatest significance in order to attain a successful result. Furthermore, numerical results show that the optimal parameters frequently exist within a limited range. This observation provides a valuable guideline for adjusting RBF settings in similar large systems, promoting greater efficiency and optimal resutls.

In the future, there are various enhancements that we intend to implement. Firstly, the cost of finding the eigenvalues and eigenvectors of a large sparse matrix is very high. In the case of a large network, it is necessary to use an iterative approach, such as the Lanczos algorithm, to solve eigenvalue problems. Additionally, when implementing the Katz index for large sparse networks, applying iterative methods, for instance, the Krylov iterative methods is advantageous. Lastly, we would focus on developing a correlation between specific metrics and the optimal RBF shaping parameter. This correlation could aid in the iterative optimization of the algorithm without the need to compute the final results.
%


%
\bibliography{main}

\newpage
\onecolumn

\appendix
\section{An Example on the Toy Example in Figure \ref{flowchart}}
\label{app:example}
To help the readers better understand our framework, we present step-by-step computations on the toy example illustrated in Figure \ref{flowchart}. In this undirected toy network $G=(\{1,2,3,4,5,6\}, \{(1,2),(2,3),(2,4),(3,4),(3,5),(5,6)\})$, we assume the number of clusters is $k=2$.

\smallskip
\begin{itemize}[itemindent=2em]
    \item[\textbf{Step 1:}] {\em Sample an affinity matrix $\mathbf{A}$ from the original (undirected) network $G$.}\\
    The condition number of matrix $\mathbf{A}$ is $c_A = 3.0896$.
    \[\mathbf{A}=
    \begin{bmatrix}
    0 & 1 & 0 & 0 & 0 & 0 \\
    1 & 0 & 1 & 1 & 0 & 0 \\
    0 & 1 & 0 & 1 & 1 & 0 \\
    0 & 1 & 1 & 0 & 0 & 0 \\
    0 & 0 & 1 & 0 & 0 & 1 \\
    0 & 0 & 0 & 0 & 1 & 0 
    \end{bmatrix}, \quad \mathbf{B}=
    \begin{bmatrix}
    0 & 0.404 & 0 & 0 & 0 & 0 \\
    0.404 & 0 & 0.404 & 0.027 & 0 & 0 \\
    0 & 0.404 & 0 & 0.404 & 0.027 & 0 \\
    0 & 0.027 & 0.404 & 0 & 0 & 0 \\
    0 & 0 & 0.027 & 0 & 0 & 0.404 \\
    0 & 0 & 0 & 0 & 0.404 & 0 
    \end{bmatrix}\quad
    \]
    \smallskip
    \item[\textbf{Step 2:}] {\em Form the RBF matrix $\mathbf{B}$}.\\
    Apply Gaussian RBF with shaping parameter $c = 0.2$ and form the RBF matrix $\mathbf{B}$, using Eq.(\ref{distancematrix}). The condition number decreases to $c_B = 1.486$.
    \smallskip
    \item[\textbf{Step 3:}] {\em Compute the high-order proximity matrix $\mathbf{K}$.}\\
    Compute $\mathbf{K}$ from the RBF matrix $\mathbf{B}$ using Katz index ($\beta=0.0025$). The high-order matrix preserving the more indirect neighbor information is $\mathbf{K}$. The condition number of $\mathbf{K}$ is $c_K = 1.488$
    \[
    \mathbf{K}=
    \begin{bmatrix}
    $1.02e-06$ & 0.001 & $1.02e-06$ & $6.81e-08$ & $6.76e-11$ & $6.83e-14$ \\
    0.001 & $2.04e-06$ & 0.001 & $6.74e-05$ & $6.70e-08$ & $6.76e-11$ \\
    $1.02e-06$ & 0.001 & $2.04e-06$ & 0.001 & $6.64e-05$ & $6.70e-08$ \\
    $6.81e-08$ & $6.74e-05$ & 0.001 & $1.02e-06$ & $6.70e-08$ & $6.76e-11$ \\
    $6.76e-11$ & $6.70e-08$ & $6.64e-05$ & $6.70e-08$ & $1.02e-06$ & 0.001 \\
    $6.83e-14$ & $6.76e-11$ & $6.70e-08$ & $6.76e-11$ & 0.001 & $1.02e-06$ 
    \end{bmatrix}
    \]
    \smallskip
    \item[\textbf{Step 4:}] {\em Construct a degree matrix $\mathbf{D}$ from $\mathbf{K}$}.
        \[
        \mathbf{D}=\begin{bmatrix}
        28.627 & 0 & 0 & 0 & 0 & 0 \\ 
        0 & 20.865 & 0 & 0 & 0 & 0 \\
        0 & 0 & 20.865 & 0 & 0 & 0 \\
        0 & 0 & 0 & 27.877 & 0 & 0 \\
        0 & 0 & 0 & 0 & 27.889 & 0 \\
        0 & 0 & 0 & 0 & 0 & 28.639 
        \end{bmatrix}
        \]
    \smallskip
    \item[\textbf{Step 5:}] {\em Form a normalized Laplacian matrix $\mathbf{L}_\sigma$}\\
    $\mathbf{L}_\sigma$ is computed using Eq.~(\ref{equ:L}) with $\sigma=0.1$.
        \[
        \mathbf{L}_\sigma=
        \begin{bmatrix}
        0.0008 & 0.6028 & 0.0006 & $5.4e-05$ & $5.4e-08$ & $5.59e-11$ \\
        0.6028 & 0.0008 & 0.4394 & 0.0392 & $3.9e-05$ & $4.0e-08$ \\
        0.0006 & 0.4394 & 0.0008 & 0.5871 & 0.0386 & $4.0e-05$ \\
        $5.4e-05$ & 0.0392 & 0.5871 & 0.0007 & $5.2e-05$ & $5.4e-08$ \\
        $5.4e-08$ & $3.9e-05$ & 0.0386 & $5.2e-05$ & 0.0007 & 0.8061 \\
        $5.5e-11$ & $4.0e-08$ & $4.0e-05$ & $5.4e-08$ & 0.8061 & 0.0008 
        \end{bmatrix}
        \]
    \smallskip
    \item[\textbf{Step 6:}] {\em Compute $k+1=3$ eigenvalues $\bm{\mathbf{\hat{\lambda}}}$ and their corresponding three eigenvectors $\bm{\mathbf{\hat{\Theta}}}$.}
    \[
    \bm{\mathbf{\hat{\lambda}}}= \begin{bmatrix} 
    -0.8421 \quad
    0.8040 \quad
    0.8774
    \end{bmatrix}
    ,\quad
        \bm{\mathbf{\hat{\Theta}}}=\begin{bmatrix}
    0.3837 & 0.1053 & 0.3913 \\
    -0.5371 & 0.1402 & 0.5684 \\
    0.5349 & 0.1044 & 0.5611 \\
    -0.3475 & 0.0831 & 0.4012 \\
    -0.2866 & -0.6884 & 0.1608 \\
    0.2741 & -0.6909 & 0.1479 \\
    \end{bmatrix}
    \]
    \smallskip
    \item[\textbf{Step 7:}] {\em Decide if the network is of weak signal or strong signal.}\\
    From the Step 6, it is clear that Eq. (\ref{equ:t}) is not satisfied since the smallest eigenvalue $\bm{\mathbf{\hat{\lambda}}}_3$ is negative. Therefore, we preserve the largest two eigenvalues and their corresponding eigenvectors (last two columns).
    \smallskip
    \item[\textbf{Step 8:}] {\em Construct new feature matrix $\mathbf{F}$.}\\
    Normalize the eigenvectors with the largest eigenvector (last column of $\bm{\mathbf{\hat{\Theta}}}$). Construct a new feature matrix $\mathbf{F}$. Note that the largest eigenvector is discarded because we used it to normalize the others. It should contain all $1$'s after normalization by itself, which has no information for clustering. 
    \[\mathbf{F}=
    \begin{bmatrix} 
    0.2692\\
    0.2467\\
    0.1861\\ 
    0.2072\\ 
    -4.2802\\ 
    -4.6704\\ 
    \end{bmatrix}
    \]
    \item[\textbf{Step 9:}] {\em Compute the communities.}\\
    Apply k-means on $\mathbf{F}$ and get the final node labels $\bm{\mathbf{\hat{y}}}$.
    \[
    \bm{\mathbf{\hat{y}}}^T=[0 \quad 0 \quad 0 \quad 0 \quad 1 \quad 1]
    \]
    
    \item[\textbf{Step 10:}] {\em Evaluate the results.}\\ Evaluate the community structure using Modularity described in Section \ref{subsec:q}, the metric value is $0.208$.
\end{itemize}

\medskip
\noindent
\textbf{Remark}: In this toy-network $G$, nodes $1,2,3,4$ are divided into the same community and nodes $5,6$ are group to the other one. This result is intuitive by the topological structure illustrated in Figure \ref{flowchart}.

\section{Omitted Results in Subsection \ref{subsec:clusters}}

\subsection{Eigenvalue Ratios of Real-World Networks}
\label{app:eigen}

\begin{table}[H] 
 \centering 
 \footnotesize 
 \caption{Type of networks from the original affinity matrix and the eigenvalue ratios} \label{app:t:aff-delta} 
 \setlength{\parskip}{0.5\baselineskip} 
  \resizebox{75mm}{20mm}{
 \begin{tabular}{clccccc}
 \hline {No.} & {Dataset} & {Type} & {$\delta_{k+1}$} & {$\delta_{k+2}$} & {$\delta_{k+3}$}\\ \hline
$1$  & Les~Mis{\'e}rable & Weak   & $0.032$ & $0.066$ & $0.103$ \\
$2$  & Caltech       & Weak   & $0.078$ & $0.098$ & $0.132$ \\
$3$  & Dolphins      & Strong & $0.186$ & $0.143$ & $0.165$ \\
$4$  & Football      & Strong & $0.187$ & $0.071$ & $0.218$ \\
$5$  & Karate        & Strong & $0.414$ & $0.208$ & $0.356$ \\
$6$  & Polbooks      & Strong & $0.503$ & $0.045$ & $0.143$ \\
$7$  & Blog         & Strong & $0.6$   & $0.162$ & $0.085$ \\
$8$  & Simmons       & Weak   & $0.08$  & $0.1$   & $0.058$ \\
$9$  & Ukfaculty     & Strong & $0.314$ & $0.117$ & $0.118$ \\
$10$ & Github        &  Strong      & $0.256$      &  $0.112$     &  $ 0.063 $   \\
$11$ & Facebook      &  Strong      &  $0.128 $    &  $ 0.279$    & $0.106$ \\
 \hline 
 \end{tabular} 
 }
 \end{table}

\begin{table}[H] 
 \centering 
 \footnotesize 
 \caption{Type of networks from the high-order matrix and the eigenvalue ratios} \label{app:t:hp-delta} 
 \setlength{\parskip}{0.5\baselineskip} 
  \resizebox{75mm}{20mm}{
 \begin{tabular}{clccccc}
 \hline {No.} & {Dataset} & {Type} & {$\delta_{k+1}$} & {$\delta_{k+2}$} & {$\delta_{k+3}$}\\ \hline
$1$  & Les~Mis{\'e}rable & Strong & $0.172$ & $ 0.04 $ & $0.231$ \\
$2$  & Caltech       & Weak   & $0.079$ & $0.025 $& $0.075$ \\
$3$  & Dolphins      & Strong & $0.486$ & $0.187$ & $0.076$ \\
$4$  & Football      & Weak   & $0.019$ & $0.098$ & $0.101$ \\
$5$  & Karate        & Strong & $0.257$ & $0.426$ &$ 0.164$ \\
$6$  & Polbooks      & Strong & $0.818$ & $0.092$ & $0.148$ \\
$7$  & Blog          & Strong & $0.369$ & $0.198$ & $0.041$ \\
$8$  & Simmons       & Strong & $0.242$ & $0.328$ & $0.205$ \\
$9$  & Ukfaculty     & Strong & $0.397$ & $0.109$ & $0.143$ \\
$10$ & Github        & Strong &  $0.224$  & $ 0.124$  &  $0.031$ \\
$11$ & Facebook      &  Strong  &  $0.115$  & $0.365$  & $0.216$ \\
 \hline 
 \end{tabular} 
 }
 \end{table}

\subsection{Extensive Experiments on Weak Signal Networks}
\label{app:weak}

\begin{table}[H]
\centering 
 \footnotesize 
 \caption{Extensive Experiments on Weak Signal Networks} \label{t:weak} 
  \resizebox{70mm}{10mm}{
\begin{tabular}{clcccc}
\hline
Dataset & Metric & $k$     & $k+1$   & $k+2$   & $k+3$   \\ \hline
 & NMI    & $0.63$  & $0.646$ & $0.647$ & $0.646$ \\ \cline{2-6} 
\multirow{-2}{*}{Caltech}       & Q      & $0.36$  & $0.4$   & $0.394$ & $0.391$ \\ \hline
 & NMI    & $0.934$ & $0.958$ & $0.957$ & $0.957$ \\ \cline{2-6} 
\multirow{-2}{*}{Football}      & Q      & $0.623$ & $0.62$  & $0.62$  & $0.62$  \\ \hline
\end{tabular}
}
\end{table}

\section{Comparisons of Running Time on Real-World Networks}
\label{app:time}
 \begin{table}[H] 
 \centering 
 \footnotesize 
 \caption{Comparisons of running time(s) on real-world networks} \label{t:r:time} 
 \setlength{\parskip}{0.5\baselineskip} 
  \resizebox{140mm}{25mm}{
 \begin{tabular}{clccccccc}
 \hline {No.} & {Dataset} & ASE & Louvain & Fast-Greedy & {SC} & {SCORE} & {SCORE+} & {SCOREH+}  \\ \hline
$1$   & Les Miserable & 0.034 & $0.03$   & $0.012$       & $0.03$   & $0.048$ & $0.024$  & $0.033$   \\
$2$   & Caltech & 0.208         & $0.096$   & $1.225 $      & $0.087$  & $0.097$ & $0.096$  & $0.178$   \\
$3$   & Dolphins & 0.024      & $0.012$   & $0.006 $      & $0.013$  & $0.01$  & $0.017$  & $0.013$   \\
$4$   & Football & 0.028       & $0.015$   & $0.025$       & $0.024$  & $0.011$ & $0.019$  & $0.026$   \\
$5$   & Karate & 0.011         & $0.011$   & $0.013$       & $0.01$   & $0.01$  & $0.011$  & $0.011$   \\
$6$   & Polbooks & 0.015       & $0.012$   & $0.018 $      & $0.016$  & $0.008$ & $0.01$   & $0.014$   \\
$7 $  & Blog & 0.408           & $0.231$   & $0.995$       & $0.057$  & $0.037$ & $0.113$  & $0.448 $  \\
$8$   & Simmons & 0.256        & $0.212 $  & $0.243$        & $0.109$  & $0.121$ & $0.144 $ & $0.365$   \\
$9$   & UKfaculty & 0.017      & $0.015$   & $0.018 $      & $0.013$  & $0.01$  & $0.011 $ & $0.018$   \\
$10$  & github & 22.6         & $12.45$   & $18.6$         & $15.2$ & $15.4$   & $20.7$    & $28.8$     \\
$11$  & facebook & 21.78       & $13.19$    & $29.9$         & $22.5$ & $22.3$   & $29.2$    & $33.7$    \\  
 \hline 
 \end{tabular} 
 }
 \end{table}

\section{Additional Analyses for Section \ref{sec:real-network-analyses}}
\label{app:dolphins-polbooks}
\subsection{Analysis of Dolphins Network}
The Dolphins network contained an undirected social network in a community living off Doubtful Sound, New Zealand. This network is constructed from frequent associations between $62$ nodes (dolphins) \citep{lusseau2003bottlenose}. It has two communities.

\begin{figure*}
\centering
\subfigure[]{
\label{fig:dol:a} 
\includegraphics[width=0.23\textwidth]{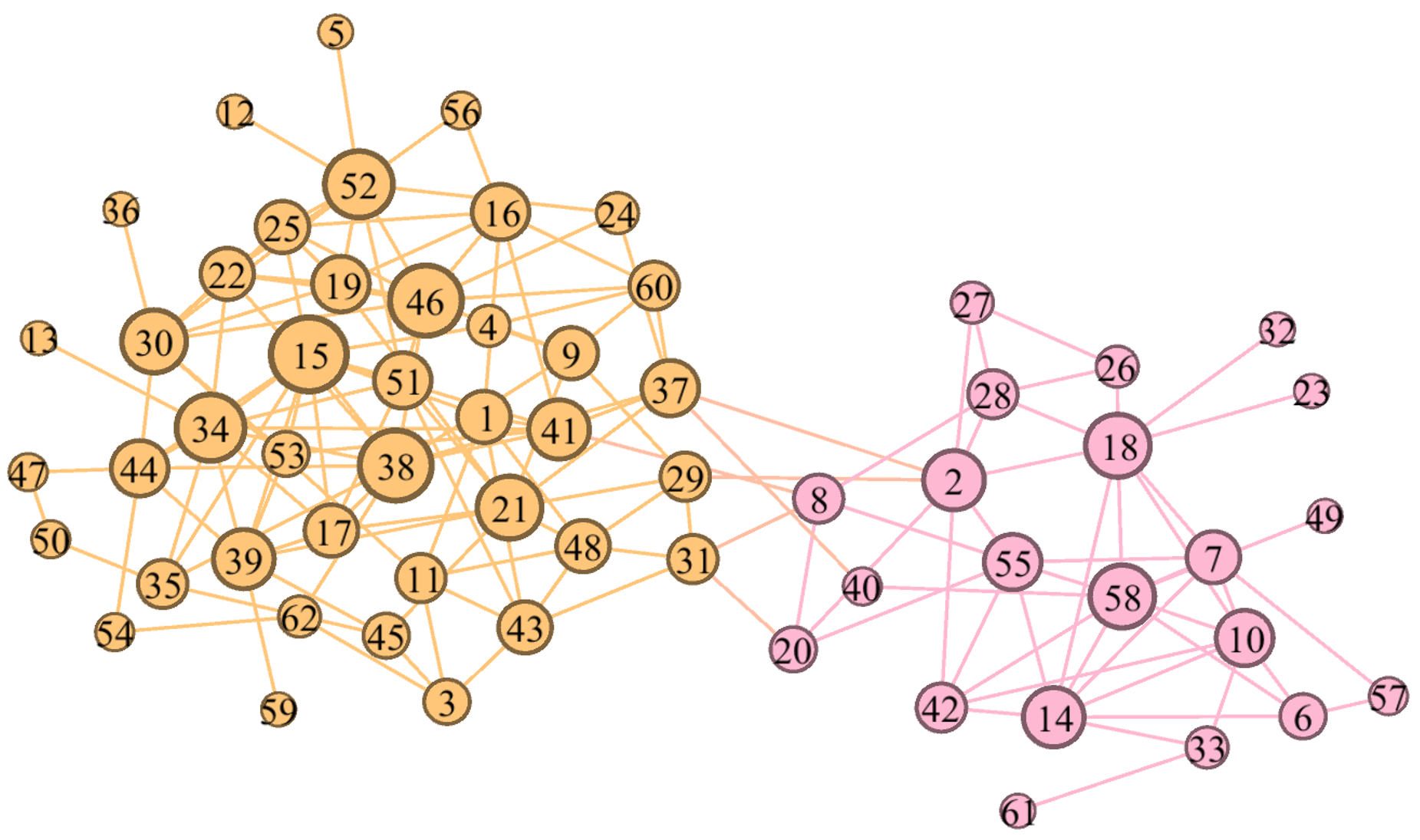}}
\subfigure[]{
\label{fig:dol:b} 
\includegraphics[width=0.23\textwidth]{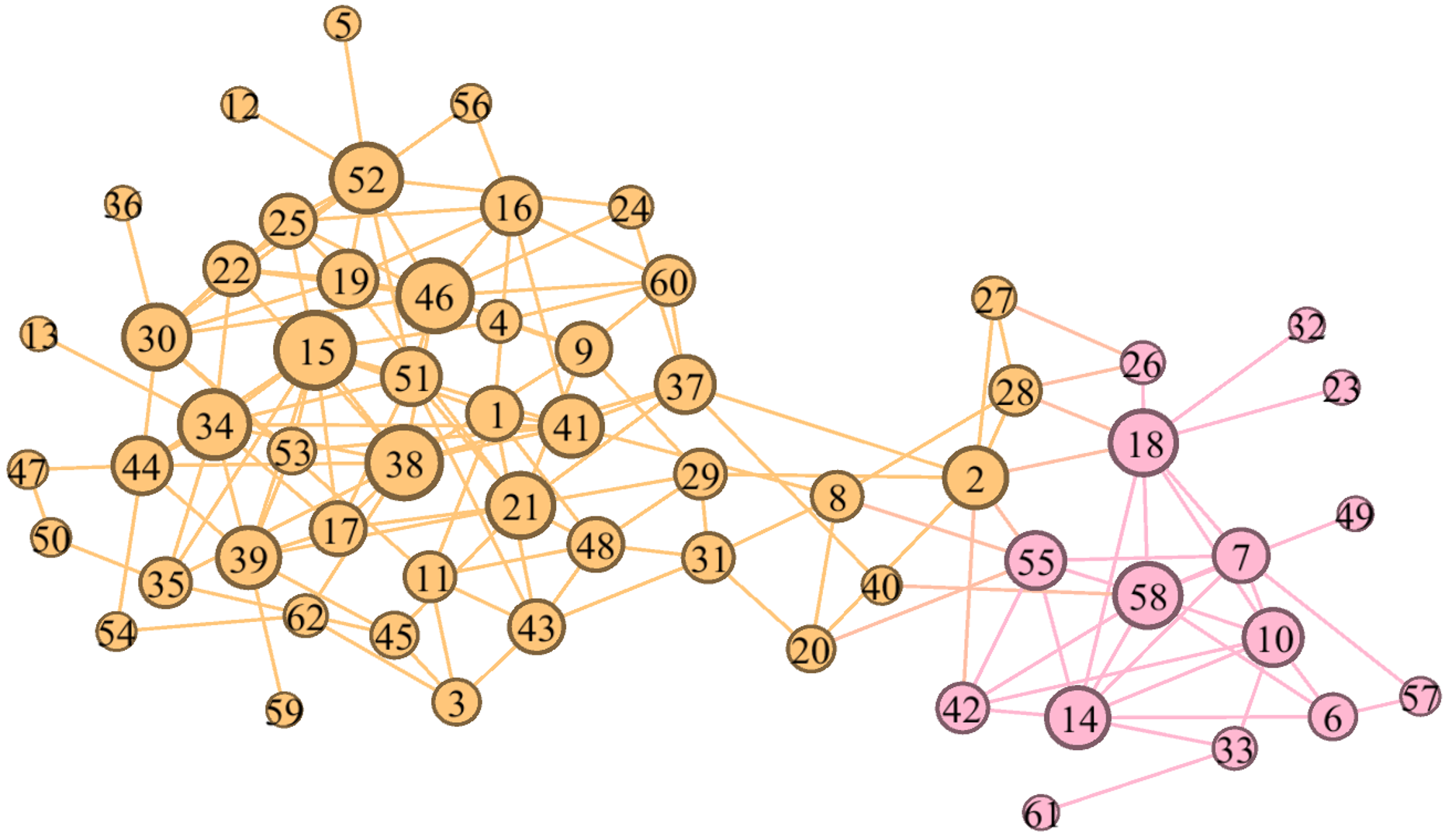}}
\subfigure[]{
\label{fig:dol:c} 
\includegraphics[width=0.23\textwidth]{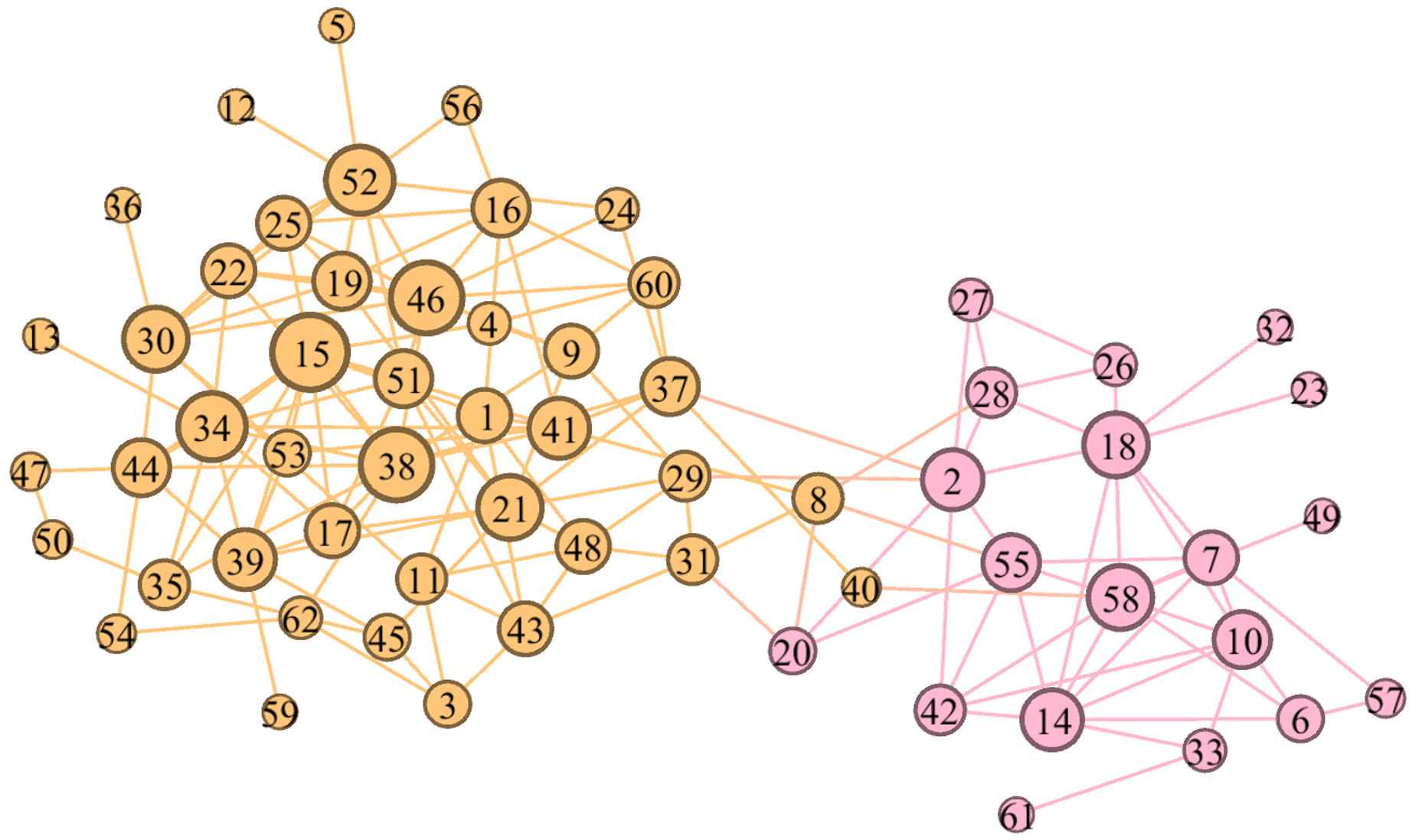}}
\subfigure[]{
\label{fig:dol:d} 
\includegraphics[width=0.23\textwidth]{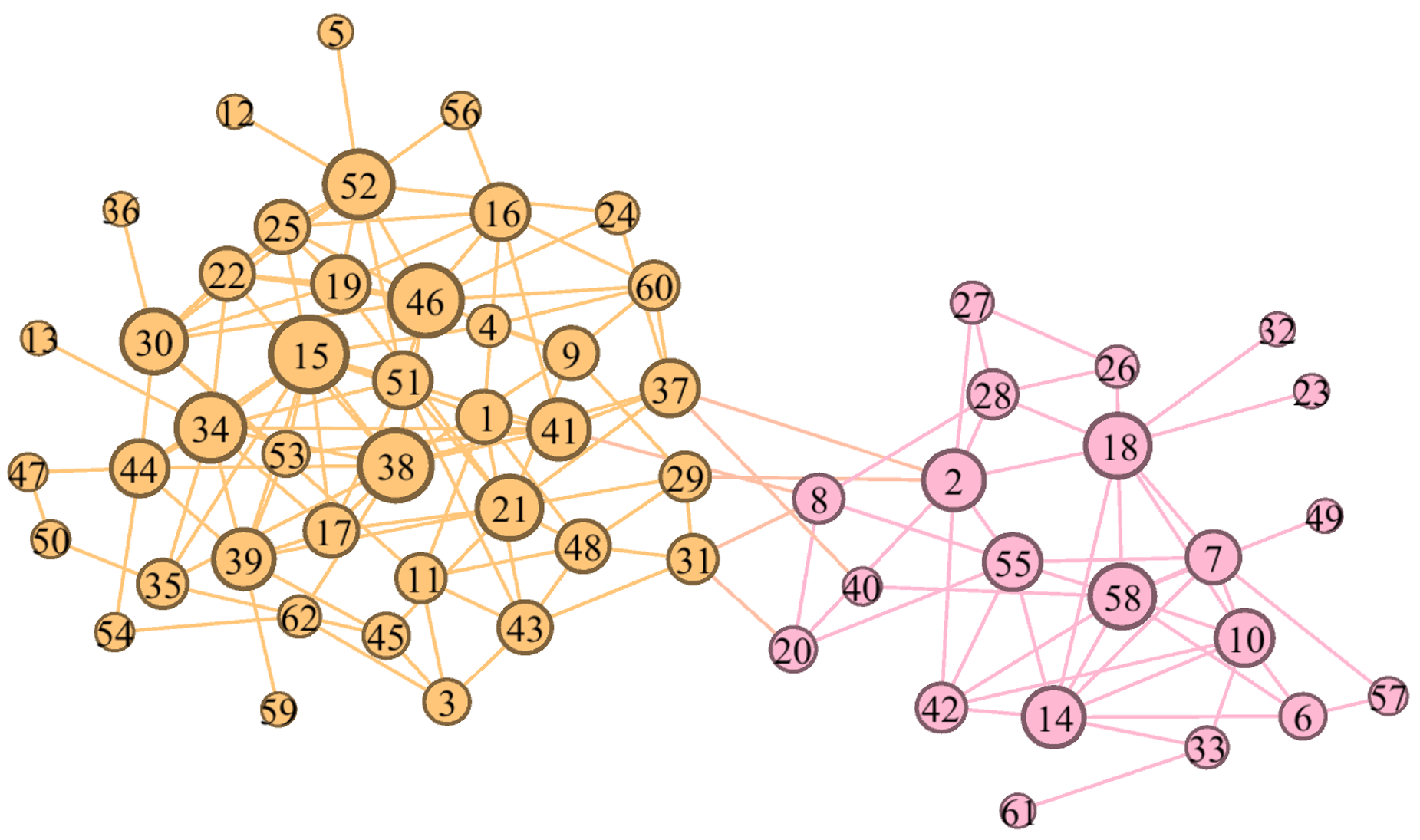}}
\caption{The topological displays for the Dolphins network from SCORE, SCORE+, and our SCOREH+ algorithms (\ref{fig:dol:a} is from the ground-truth of the network). }
\label{fig:dolphins} 
\end{figure*}
For the Dolphins network, the results demonstrate that SCORE misclassified the numbers $2$, $8$, $20$, $27$, $28$, and $40$ nodes, and SCORE+ made mistakes on the numbers $8$ and $40$ nodes. Refer to Fig. \ref{fig:dolphins}, and our SCOREH+ perfectly acquired the community structure.

\subsection{Analysis of Polbooks Network}

The Polbooks network contained books on American politics, published around the 2004 presidential election and sold by Amazon.com. This data set was not published, but we can access it on V. Krebs' website \footnote{http://www.orgnet.com/divided.html}. In this network, the edges between books represent that they were sold together. Typically, this network has two communities. However, the network structure is difficult to discover since ``neutral" or ``moderate" people can buy books promoting both parties. The experiments by detecting communities from the constructed network also demonstrate this viewpoint.

\begin{figure*}
\centering
\subfigure[]{
\label{fig:polbooks:a} 
\includegraphics[width=0.23\textwidth]{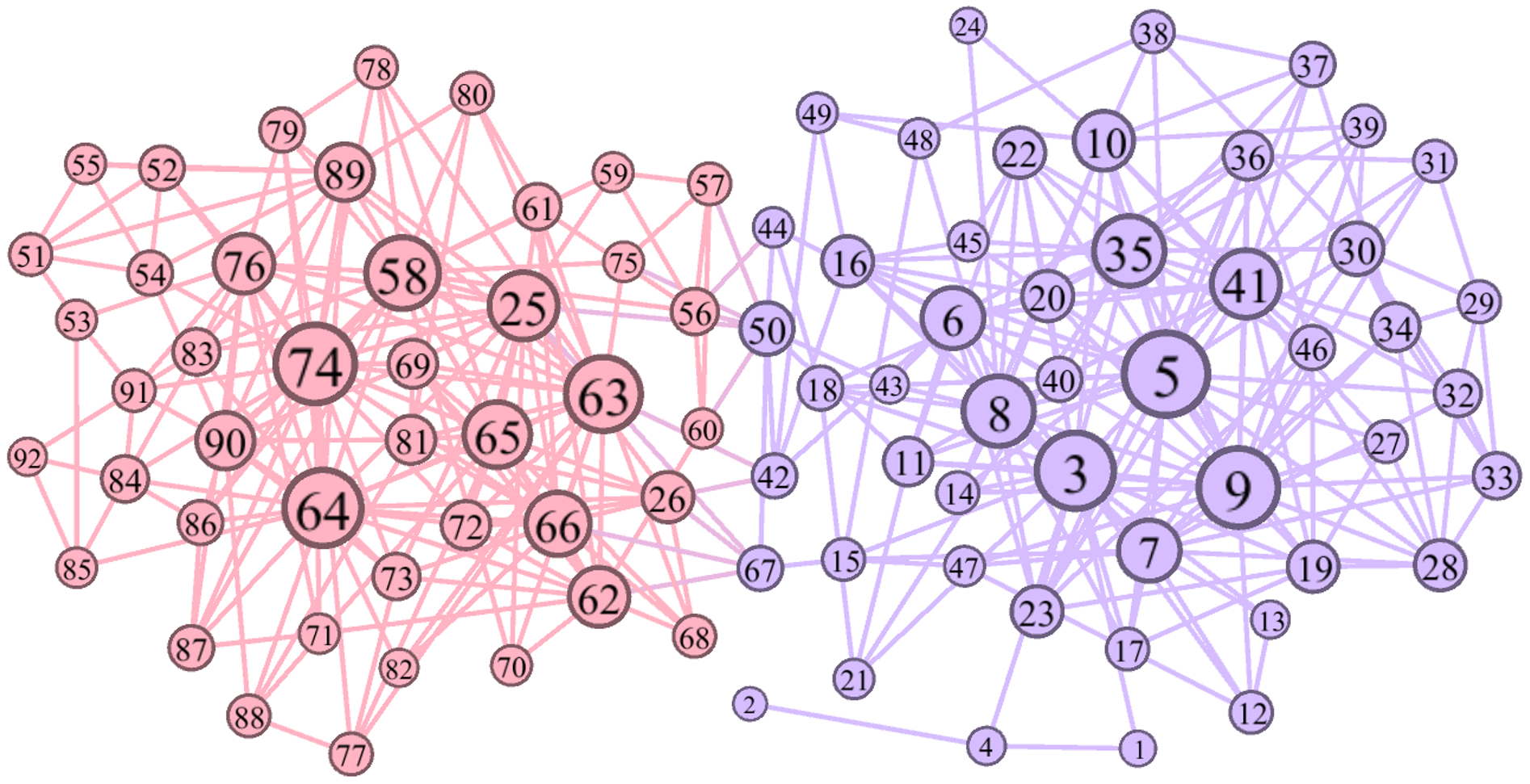}}
\subfigure[]{
\label{fig:polbooks:b} 
\includegraphics[width=0.23\textwidth]{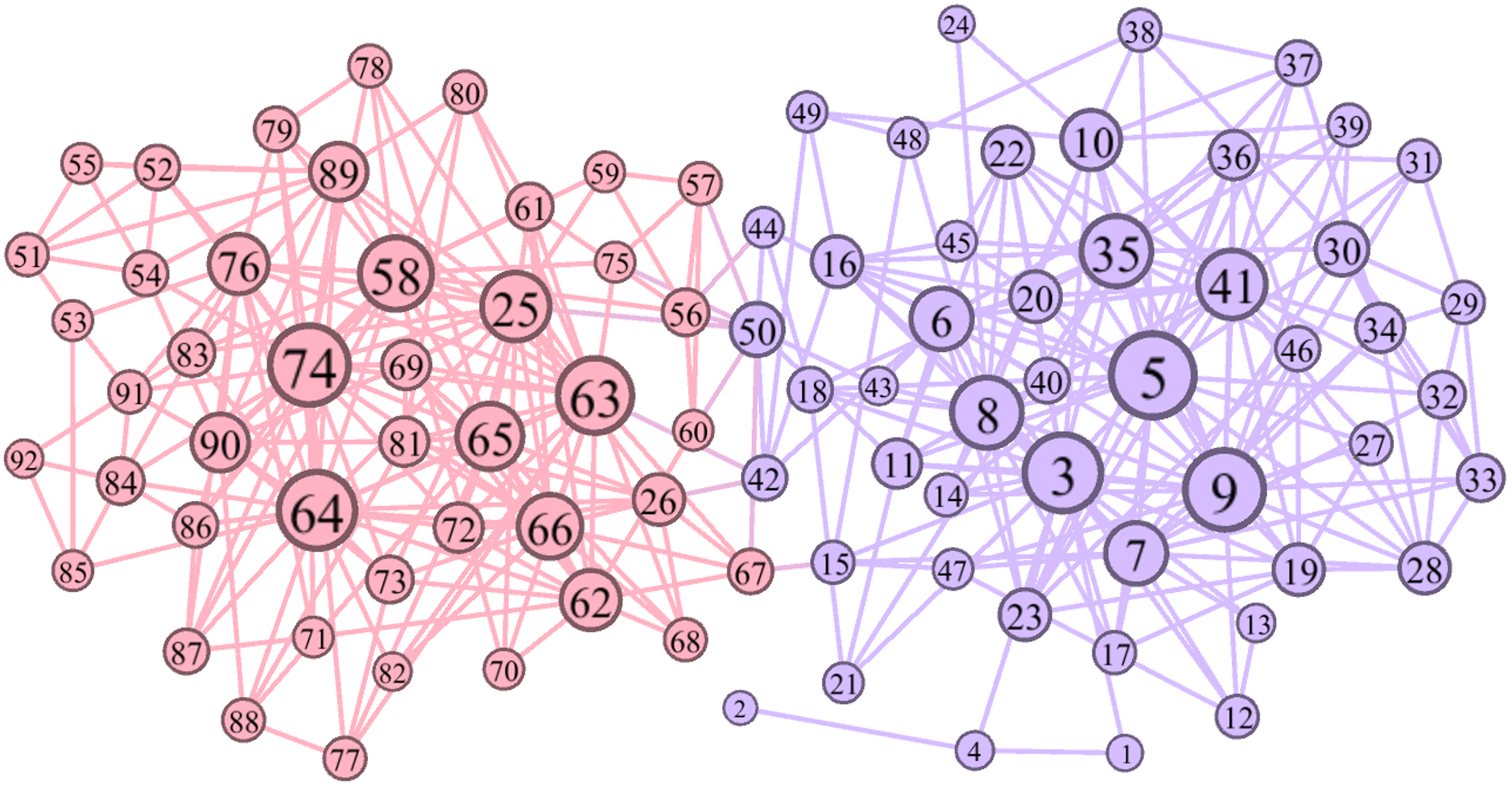}}
\subfigure[]{
\label{fig:polbooks:c} 
\includegraphics[width=0.23\textwidth]{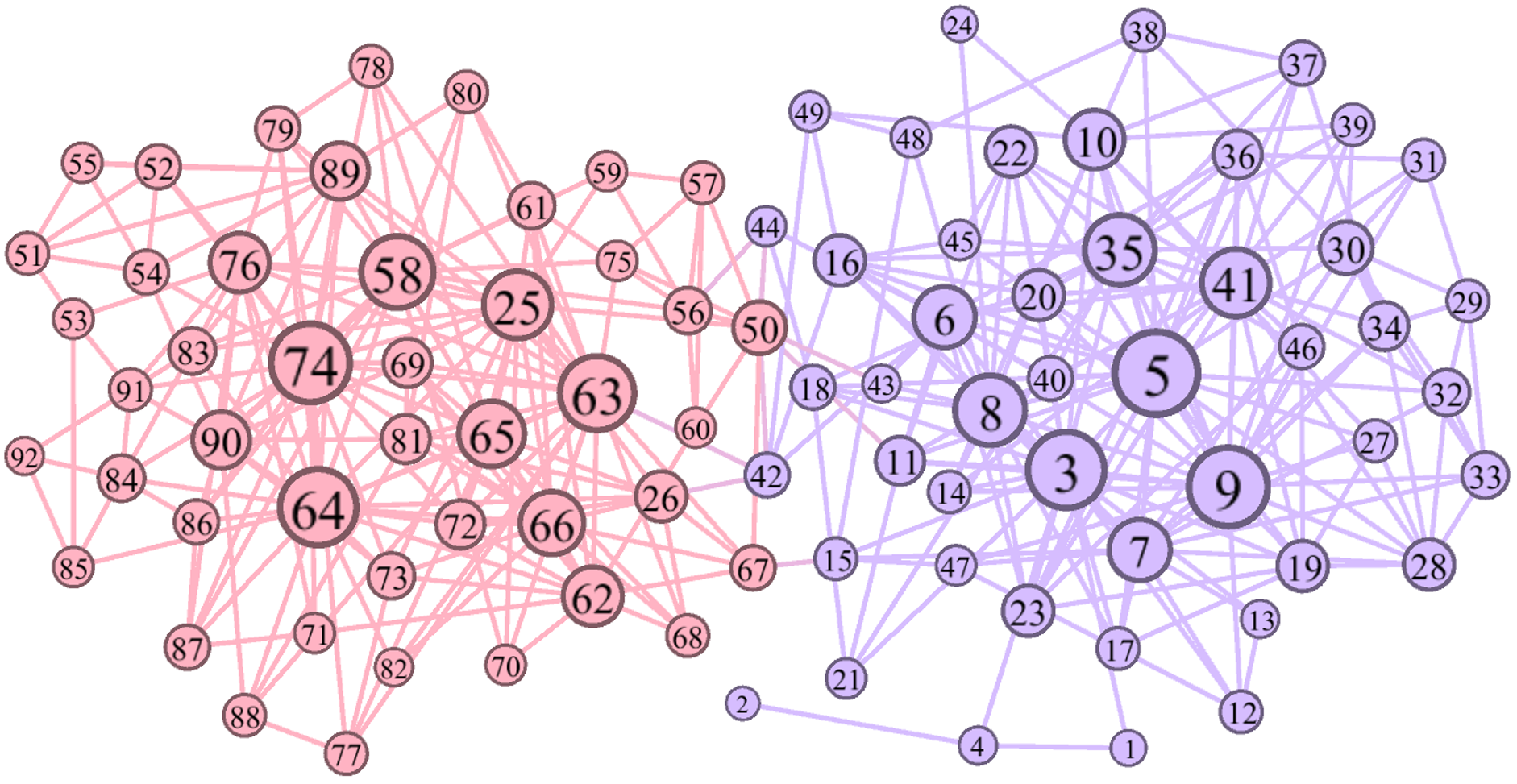}}
\subfigure[]{
\label{fig:polbooks:d} 
\includegraphics[width=0.23\textwidth]{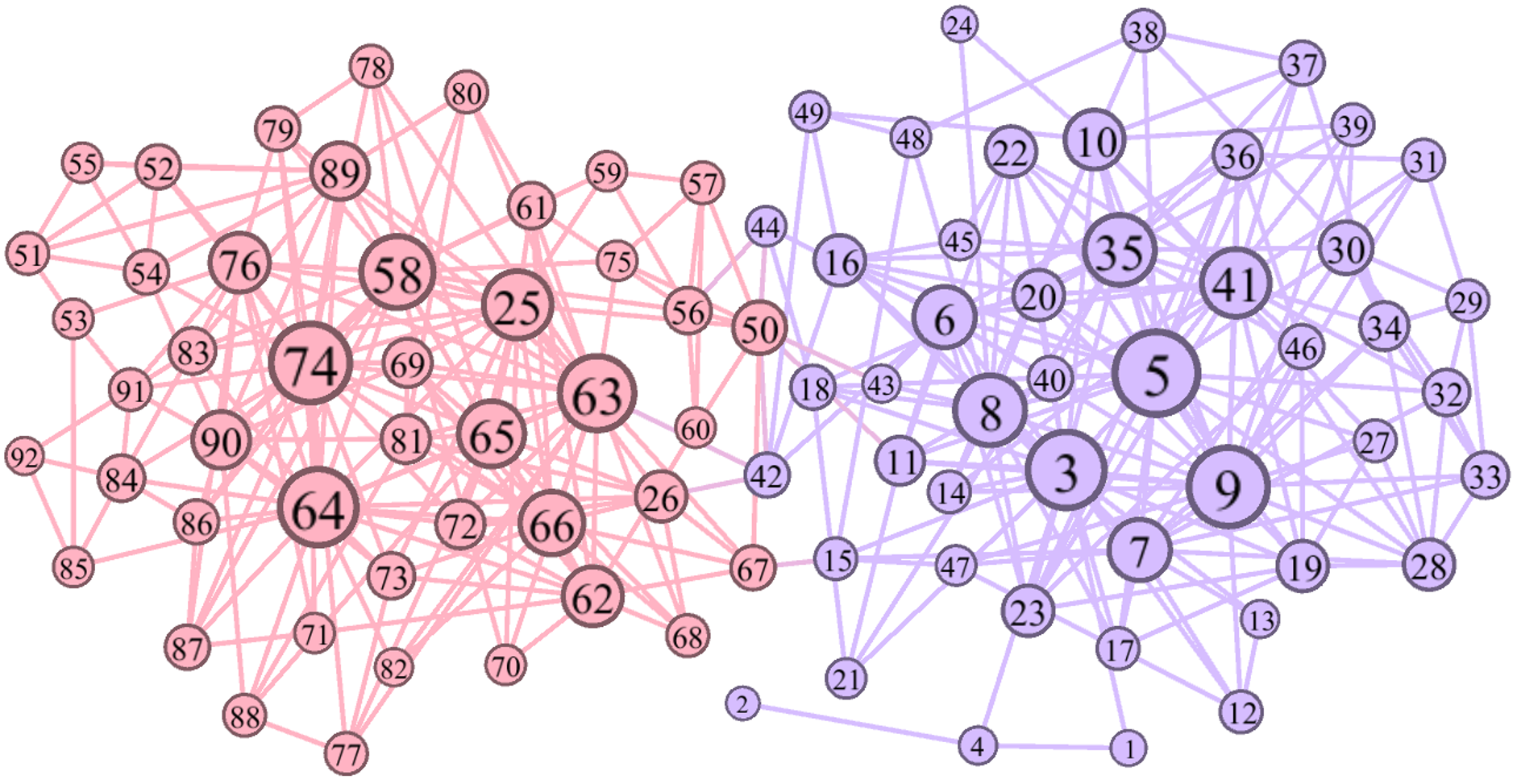}}
\caption{The topological displays for the Polbooks network from SCORE, SCORE+, and our SCOREH+ algorithms (\ref{fig:polbooks:a} is from the ground-truth of the network). The Polbooks network, like Dolphins and Karate, has two communities.}
\label{fig:polbooks} 
\end{figure*}

As shown in Fig. \ref{fig:polbooks}, both of the SCORE+ and SCOREH+ algorithms made mistakes on the $50$ and $67$ nodes, whereas the SCORE misclassified only one node: the $66$ node. However, it is this small one-node difference that improved the NMI of the SCORE algorithm on this network, from $0.87$ to $0.924$.

\section{Additional Results for Section \ref{subsec:synthetic}}
\subsection{Modularity Tables}
\label{app:Q}
\begin{table}[H] 
 \centering 
 \footnotesize 
 \caption{Numerical results on synthetic networks with N=2,000 and N=5,000 (Modularity)} \label{tab:syn-q1} 
 \setlength{\parskip}{0.5\baselineskip} 
  \resizebox{180mm}{14mm}{
\begin{tabular}{ccccccccc|ccccccc}
 \hline
{\multirow{2}{*}{No.}} & {\multirow{2}{*}{$\mu$}} & & \multicolumn{4}{c}{N=2,000  } & & &\multicolumn{4}{c}{N=5,000  }\\
\cline{3-16}  &  & {ASE} & {Louvain} & {Fast-Greedy} & {SC} & {SCORE} & {SCORE+} & {SCOREH+} & {ASE} & {Louvain} & {Fast-Greedy} & {SC} & {SCORE} & {SCORE+} & {SCOREH+}\\\hline
$1$ & $0.15$ & $0.694$ & $\textbf{0.698}$ & $0.675$ &  $0.049(0.006)$ & $0.001(0.0)$ & $\textbf{0.698(0.0)}$ & $\textbf{0.698(0.0)}$ & $0.605$ & $\textbf{0.712}$ & $0.681$ & $0.056(0.001)$ & $0.0(0.0)$ & $\textbf{0.712(0.0)}$ & $\textbf{0.712(0.0)}$\\
$2$ & $0.25$ & $0.546$ & $0.55$ & $0.503$ &  $0.013(0.002)$ & $0.0(0.0)$ & $\textbf{0.55(0.0)}$ & $\textbf{0.55(0.0)}$ & $0.457$ & $\textbf{0.551}$ & $0.505$ & $0.025(0.001)$ & $0.0(0.0)$ & $\textbf{0.551(0.0)}$ & $\textbf{0.551(0.0)}$\\
$3$ & $0.35$ & $0.372$ & $\textbf{0.411}$ & $0.341$ &  $-0.009(0.003)$ & $0.0(0.0)$ & $\textbf{0.411(0.0)}$ & $\textbf{0.411(0.0)}$ & $0.342$ & $0.418$ & $0.361$ &  $0.006(0.001)$ & $0.0(0.0)$ & $\textbf{0.422(0.0)}$ & $\textbf{0.422(0.0)}$ \\
$4$ & $0.45$ & $0.215$ & $\textbf{0.284}$ & $0.21$ &  $-0.019(0.001)$ & $0.0(0.0)$ & $0.269(0.0)$ & $0.279(0.0)$ &  $0.237$ & $\textbf{0.293}$ & $0.227$ & $-0.009(0.001)$ & $0.0(0.0)$ & $0.284(0.0)$ & $0.287(0.002)$\\
$5$ & $0.55$ & $0.1$ & $\textbf{0.18}$ & $0.141$ &  $-0.026(0.001)$ & $0.0(0.0)$ &  $0.126(0.008)$ & $0.156(0.001)$ & $0.12$ & $\textbf{0.191}$ & $0.148$ & $-0.011(0.001)$ & $0.0(0.0)$ & $0.113(0.008)$ & $0.17(0.001)$\\
$6$ & $0.65$ & $0.045$ & $\textbf{0.129}$ & $0.109$ & $-0.027(0.0)$ & $-0.0(0.0)$ & $0.044(0.003)$ & $0.095(0.001)$ & $0.047$ & $\textbf{0.123}$ & $0.09$ & $-0.018(0.0)$ & $0.0(0.0)$ & $0.055(0.005)$ & $0.085(0.002)$ \\
$7$ & $0.75$ & $0.011$ & $\textbf{0.108}$ & $0.095$ &  $-0.03(0.0)$ & $-0.0(0.0)$ & $0.016(0.003)$ & $0.069(0.001)$ & $0.01$ & $\textbf{0.091}$ & $0.08$ & $-0.018(0.0)$ & $0.0(0.0)$ & $0.023(0.001)$ & $0.054(0.0)$\\
$8$ & $0.85$ & $0.001$ &$\textbf{0.105}$ & $0.096$ &  $-0.026(0.0)$ & $0.0(0.0)$ & $-0.004(0.002)$ & $0.059(0.001)$ & $-0.002$ & $\textbf{0.084}$ & $0.075$ & $-0.02(0.0)$ & $-0.0(0.0)$ & $-0.002(0.001)$ & $0.041(0.0)$\\
 \hline 
 \end{tabular} 
}
 \end{table}

\begin{table}[H] 
 \centering 
 \footnotesize 
 \caption{Numerical results on synthetic networks with N=8,000 and N=10,000 (Modularity)} \label{tab:syn-q2} 
 \setlength{\parskip}{0.5\baselineskip} 
  \resizebox{180mm}{14mm}{
\begin{tabular}{ccccccccc|ccccccc}
 \hline
{\multirow{2}{*}{No.}} & {\multirow{2}{*}{$\mu$}} & & \multicolumn{4}{c}{N=8,000  } & & &\multicolumn{4}{c}{N=10,000  }\\
\cline{3-16}  &  & {ASE} & {Louvain} & {Fast-Greedy} & {SC} & {SCORE} & {SCORE+} & {SCOREH+} & {ASE} & {Louvain} & {Fast-Greedy} & {SC} & {SCORE} & {SCORE+} & {SCOREH+}\\\hline
$1$ & $0.15$ & $0.608$ &$\textbf{0.71}$ & $0.686$ & $0.046(0.002)$ & $0.0(0.0)$ & $\textbf{0.71(0.0)}$ & $\textbf{0.71(0.0)}$ & $0.601$ & $\textbf{0.718}$ & $0.694$ & $0.05(0.001)$ & $0.0(0.0)$ & $\textbf{0.718(0.0)}$ & $\textbf{0.718(0.0)}$ \\
$2$ & $0.25$ & $0.445 $ & $0.562$ & $0.531$ & $0.031(0.0)$ & $0.0(0.0)$ & $0.495(0.0)$ & $\textbf{0.563(0.0)}$  & $0.459$ & $\textbf{0.571}$ & $0.528$ & $0.03(0.001)$ & $0.0(0.0)$ & $\textbf{0.571(0.0)}$ & $\textbf{0.571(0.0)}$ \\
$3$ & $0.35$ & $0.324$ & $0.416$ & $0.365$ & $0.016(0.001)$ & $0.0(0.0)$ & $0.388(0.0)$ & $\textbf{0.428(0.0)}$  & $0.313$ & $0.424$ & $0.368$ &$0.017(0.001)$ & $0.0(0.0)$ & $0.329(0.0)$ & $\textbf{0.425(0.0)}$ \\
$4$ & $0.45$ & $0.233$ &$\textbf{0.303}$ & $0.24$ & $0.004(0.0)$ & $0.0(0.0)$ & $0.249(0.001)$ & $0.302(0.003)$ &  $0.23$ & $\textbf{0.304}$ & $0.242$ & $0.003(0.001)$ & $0.0(0.0)$ & $0.228(0.0)$ & $0.297(0.0)$\\
$5$ & $0.55$ & $0.137$ & $\textbf{0.194}$ & $0.142$ & $-0.009(0.0)$ & $-0.0(0.0)$ & $0.135(0.0)$ & $0.185(0.002)$ & $0.136$ & $\textbf{0.204}$ & $0.144$ &$-0.007(0.0)$ & $0.0(0.0)$ & $0.105(0.003)$ & $0.185(0.003)$ \\
$6$ & $0.65$ & $0.047$&$\textbf{0.118}$ & $0.09$ & $-0.014(0.0)$ & $0.0(0.0)$ & $0.047(0.004)$ & $0.082(0.003)$ & $0.058$ &$\textbf{0.128}$ & $0.091$ &$-0.011(0.0)$ & $0.0(0.0)$ & $0.053(0.003)$ & $0.09(0.001)$ \\
$7$ & $0.75$ & $0.002$ &$\textbf{0.081}$ & $0.071$ & $-0.02(0.0)$ & $0.0(0.0)$ & $0.018(0.001)$ & $0.045(0.0)$  & $0.007$ &$\textbf{0.08}$ & $0.067$ & $-0.019(0.0)$ & $0.0(0.0)$ & $0.02(0.0)$ & $0.044(0.0)$ \\
$8$ & $0.85$ & $0$ & $\textbf{0.077}$ & $0.068$ & $-0.016(0.0)$ & $0.0(0.0)$ & $0.006(0.0)$ & $0.036(0.0)$ & $-0.002$ & $\textbf{0.079}$ & $0.066$ & $-0.016(0.0)$ & $-0.0(0.0)$ & $0.002(0.0)$ & $0.029(0.0)$ \\

 \hline 
 \end{tabular} 
 }
 \end{table}

\subsection{NMI Tables}
\label{app:NMI}

\begin{table}[H] 
 \centering 
 \footnotesize 
 \caption{Numerical results on synthetic networks with N=2,000 and N=5,000 (NMI)} \label{tab:syn-nmi1} 
 \setlength{\parskip}{0.5\baselineskip} 
  \resizebox{180mm}{14mm}{
 \begin{tabular}{ccccccccc|ccccccc}
 \hline
{\multirow{2}{*}{No.}} & {\multirow{2}{*}{$\mu$}} & & \multicolumn{4}{c}{N=2,000  } & & &\multicolumn{4}{c}{N=5,000  }\\
\cline{3-16}  &  & {ASE} & {Louvain} & {Fast-Greedy} & {SC} & {SCORE} & {SCORE+} & {SCOREH+} & {ASE} & {Louvain} & {Fast-Greedy} & {SC} & {SCORE} & {SCORE+} & {SCOREH+}\\\hline
$1$ & $0.15$ & $\textbf{1.0}$ & $\textbf{1.0}$ & $0.903$ & $0.407(0.013)$ & $0.029(0.001)$ & $\textbf{1.0(0.0)}$ & $\textbf{1.0(0.0)}$ & $0.98$ & $\textbf{1.0}$ & $0.883$ & $0.423(0.005)$ & $0.019(0.002)$ & $\textbf{1.0(0.0)}$ & $\textbf{1.0(0.0)}$\\
$2$ & $0.25$ & $\textbf{1.0}$ & $\textbf{1.0}$ & $0.79$ & $0.296(0.005)$ & $0.024(0.001)$ & $\textbf{1.0(0.0)}$ & $\textbf{1.0(0.0)}$ & $0.973$ & $0.998$ & $0.754$ & $0.424(0.007)$ & $0.018(0.001)$ & $\textbf{1.0(0.0)}$ & $\textbf{1.0(0.0)}$\\
$3$ & $0.35$ & $0.974$ &$0.972$ & $0.621$ & $0.207(0.005)$ & $0.023(0.001)$ & $\textbf{0.996(0.0)}$ & $0.993(0.0)$ & $0.955$ & $0.978$ & $0.66$ & $0.43(0.006)$ & $0.015(0.001)$ & $\textbf{0.998(0.0)}$ & $0.997(0.0)$ \\
$4$ & $0.45$ & $0.879$ &$0.892$ & $0.41$ & $0.169(0.006)$ & $0.028(0.001)$ & $0.952(0.001)$ & $\textbf{0.953(0.0)}$ & $0.894$ &$0.915$ & $0.48$ &$0.314(0.008)$ & $0.012(0.001)$ & $\textbf{0.969(0.0)}$ & $0.966(0.003)$ \\
$5$ & $0.55$ & $\textbf{0.71}$ &$0.529$ & $0.228$ & $0.067(0.004)$ & $0.025(0.001)$ & $0.65(0.002)$ & $0.67(0.004)$ & $0.805$ &$0.622$ & $0.311$ & $0.234(0.006)$ & $0.018(0.001)$ & $\textbf{0.848(0.003)}$ & $0.842(0.006)$\\
$6$ & $0.65$ & $\textbf{0.42}$ &$0.195$ & $0.088$ & $0.056(0.002)$ & $0.025(0.0)$ & $0.294(0.009)$ & $0.36(0.003)$ & $0.48$ & $0.316$ & $0.104$ & $0.057(0.002)$ & $0.015(0.001)$ & $0.482(0.002)$ & $\textbf{0.502(0.007)}$\\
$7$ & $0.75$ &$\textbf{0.132}$& $0.035$ & $0.013$ & $0.044(0.002)$ & $0.026(0.002)$ & $0.104(0.006)$ & $0.127(0.005)$& $0.18$ &$0.043$ & $0.03$ &$0.046(0.001)$ & $0.015(0.0)$ & $0.21(0.004)$ & $\textbf{0.243(0.002)}$ \\
$8$ & $0.85$ & $0.052$ &$0.017$ & $0.013$ & $0.07(0.004)$ & $0.032(0.001)$ & $\textbf{0.072(0.002)}$ & $0.064(0.003)$ & $0.033$&$0.011$ & $0.005$ & $0.044(0.002)$ & $0.014(0.001)$ & $\textbf{0.045(0.002)}$ & $0.041(0.003)$\\

 \hline 
 \end{tabular} 
 }
 \end{table}

\begin{table}[H] 
 \centering 
 \footnotesize 
 \caption{Numerical results on synthetic networks with N=8,000 and N=10,000 (NMI)} \label{tab:syn-nmi2} 
 \setlength{\parskip}{0.5\baselineskip} 
  \resizebox{180mm}{14mm}{
\begin{tabular}{ccccccccc|ccccccc}
 \hline
{\multirow{2}{*}{No.}} & {\multirow{2}{*}{$\mu$}} & & \multicolumn{4}{c}{N=8,000  } & & &\multicolumn{4}{c}{N=10,000  }\\
\cline{3-16}  &  & {ASE} & {Louvain} & {Fast-Greedy} & {SC} & {SCORE} & {SCORE+} & {SCOREH+} & {ASE} & {Louvain} & {Fast-Greedy} & {SC} & {SCORE} & {SCORE+} & {SCOREH+}\\\hline

$1$ & $0.15$ & $0.978$&$\textbf{1.0}$ & $0.872$ & $0.462(0.007)$ & $0.011(0.001)$ & $\textbf{1.0(0.0)}$ & $\textbf{1.0(0.0)}$ & $0.976$ &$\textbf{1.0}$ & $0.851$ & $0.468(0.002)$ & $0.014(0.001)$ & $\textbf{1.0(0.0)}$ & $\textbf{1.0(0.0)}$ \\
$2$ & $0.25$ & $0.971$ &$0.985$ & $0.794$ & $0.439(0.008)$ & $0.011(0.0)$ & $0.979(0.0)$ & $\textbf{0.999(0.0)}$ & $0.969$ & $0.999$ & $0.77$ &  $0.477(0.005)$ & $0.009(0.0)$ & $\textbf{1.0(0.0)}$ & $0.999(0.0)$\\
$3$ & $0.35$ & $0.955$&$0.956$ & $0.635$ &  $0.469(0.002)$ & $0.014(0.001)$ & $0.98(0.0)$ & $\textbf{0.998(0.0)}$& $0.959$& $0.978$ & $0.631$ & $0.448(0.003)$ & $0.013(0.001)$ & $0.965(0.0)$ & $\textbf{0.995(0.0)}$ \\
$4$ & $0.45$ &$0.943$& $0.898$ & $0.508$ &  $0.477(0.002)$ & $0.013(0.001)$ & $0.959(0.001)$ & $\textbf{0.978(0.002)}$ &$0.931$& $0.936$ & $0.526$ & 
 $0.479(0.001)$ & $0.01(0.0)$ & $0.947(0.001)$ & $\textbf{0.975(0.001)}$ \\
$5$ & $0.55$ & $0.866$ &$0.775$ & $0.344$ &  $0.375(0.002)$ & $0.012(0.001)$ & $0.911(0.0)$ & $\textbf{0.925(0.001)}$ & $0.847$ &$0.739$ & $0.312$ &  $0.402(0.001)$ & $0.012(0.0)$ & $0.874(0.001)$ & $\textbf{0.905(0.001)}$ \\
$6$ & $0.65$ & $0.483$ &$0.333$ & $0.118$ & $0.125(0.004)$ & $0.013(0.001)$ & $\textbf{0.613(0.001)}$ & $0.605(0.006)$  &$0.566$& $0.379$ & $0.132$ & $0.208(0.011)$ & $0.012(0.0)$ & $\textbf{0.699(0.001)}$ & $0.657(0.005)$ \\
$7$ & $0.75$ &$0.123$ &$0.025$ & $0.035$ & $0.026(0.001)$ & $0.01(0.0)$ & $0.186(0.002)$ & $\textbf{0.22(0.001)}$ & $0.176$&$0.052$ & $0.02$ & $0.029(0.001)$ & $0.007(0.0)$ & $0.254(0.001)$ & $\textbf{0.282(0.002)}$ \\
$8$ & $0.85$ & $0.055$&$0.015$ & $0.016$ &  $0.047(0.001)$ & $0.012(0.001)$ & $0.122(0.002)$ & $\textbf{0.129(0.002)}$ & $0.056$&$0.005$ & $0.005$ & $0.036(0.001)$ & $0.01(0.001)$ & $0.053(0.001)$ & $\textbf{0.074(0.002)}$ \\

 \hline 
 \end{tabular} 
 }
 \end{table}
 
\newpage
\subsection{Additional Figures for Section \ref{subsec:synthetic}}
\label{app:figs}

\begin{figure}[ht]
\centering
\subfigure[]{
\label{fig1:subfig:a} 
\includegraphics[width=0.43\textwidth]{figures/figure/3a.pdf}}
\subfigure[]{
\label{fig1:subfig:b} 
\includegraphics[width=0.43\textwidth]{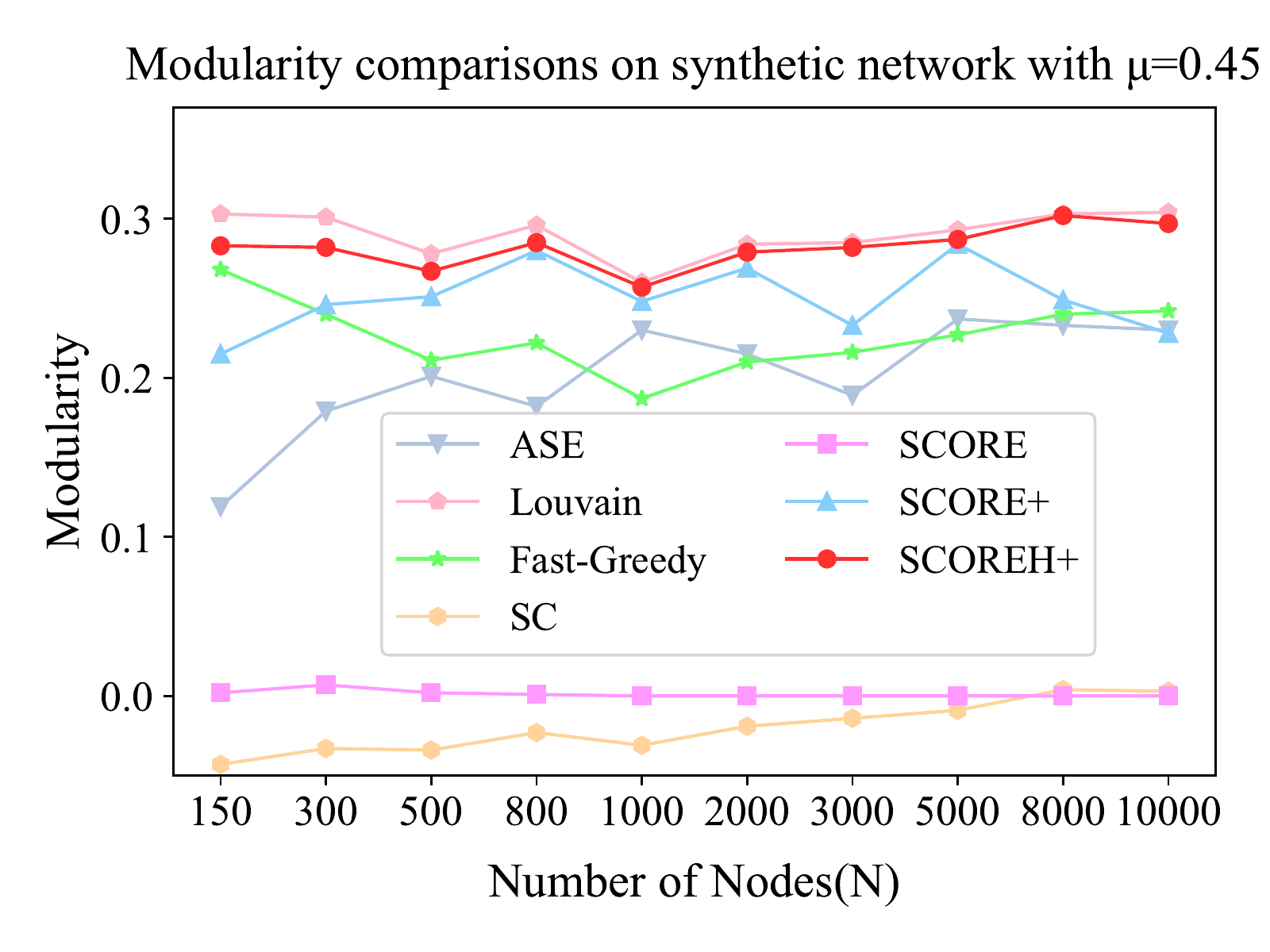}}
\subfigure[]{
\label{fig1:subfig:c} 
\includegraphics[width=0.43\textwidth]{figures/figure/3c.pdf}}
\subfigure[]{
\label{fig1:subfig:d} 
\includegraphics[width=0.43\textwidth]{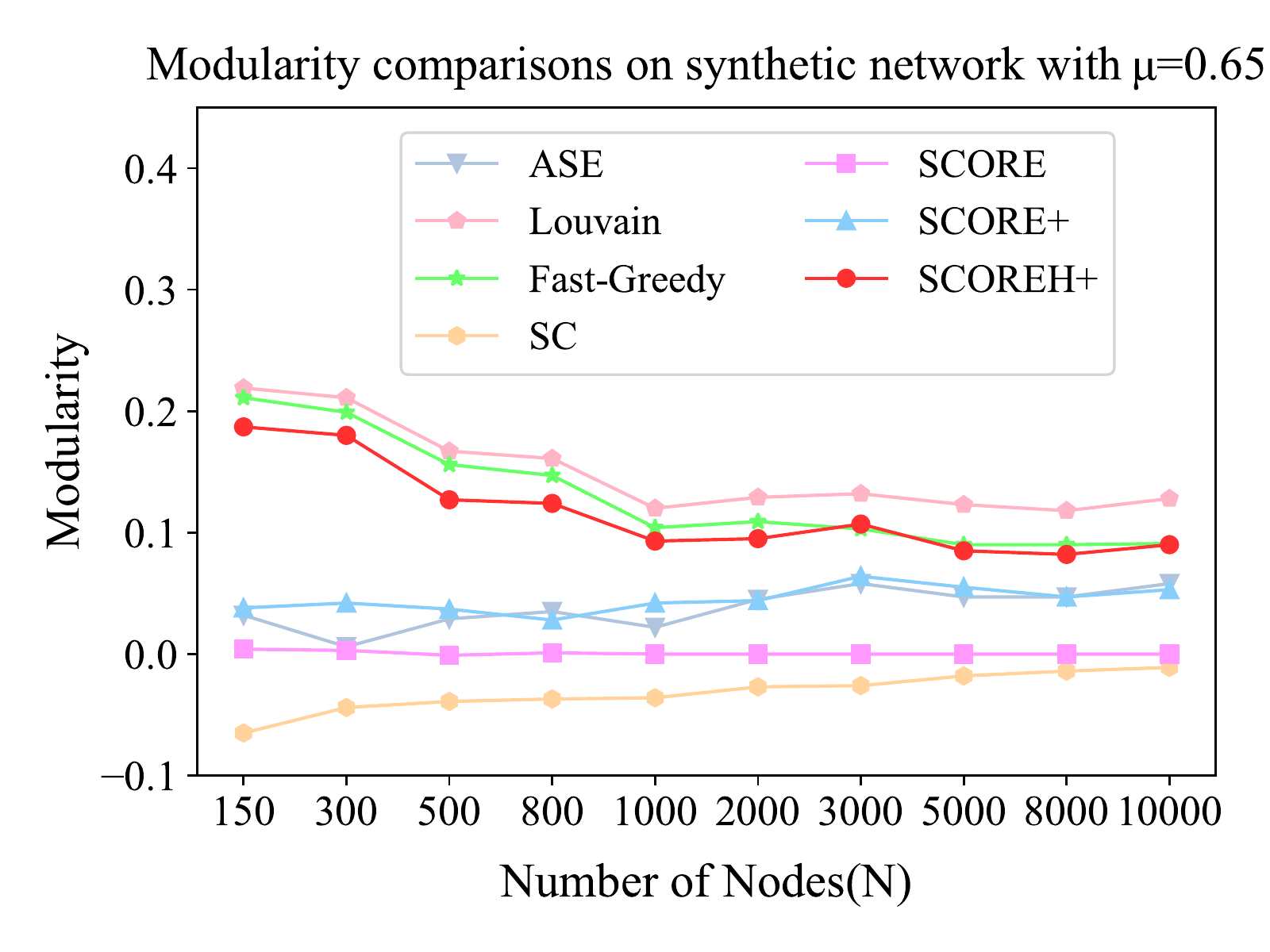}}
\caption{The comparison plots of Modularity on LFR datasets with different N.
\label{app:LFR-Q-N}}
\end{figure}

\begin{figure}[ht]
\centering
\subfigure[]{
\label{fig11:subfig:a} 
\includegraphics[width=0.43\textwidth]{figures/figure/5a.pdf}}
\subfigure[]{
\label{fig11:subfig:b} 
\includegraphics[width=0.43\textwidth]{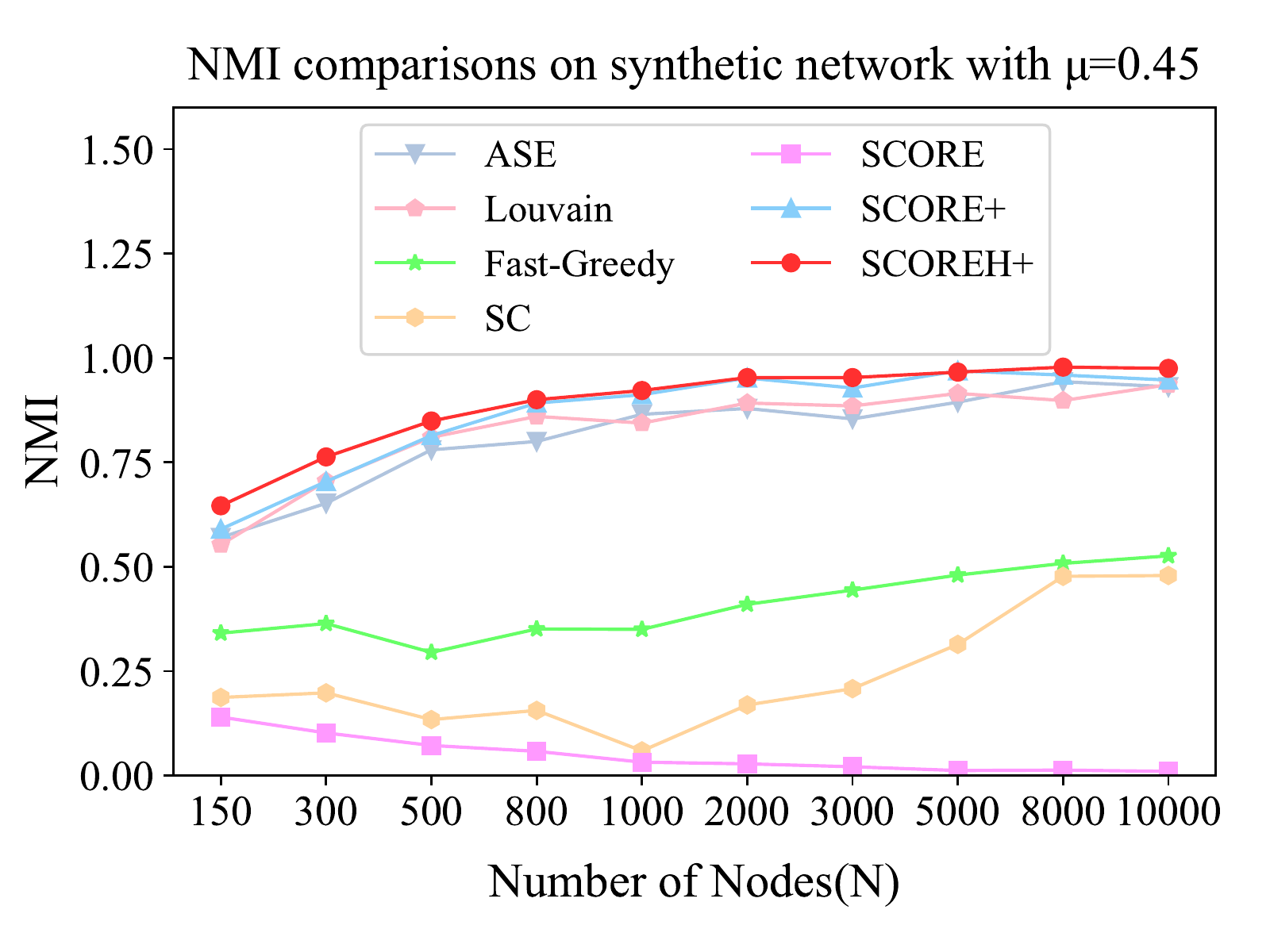}}
\subfigure[]{
\label{fig11:subfig:c} 
\includegraphics[width=0.43\textwidth]{figures/figure/5c.pdf}}
\subfigure[]{
\label{fig11:subfig:d} 
\includegraphics[width=0.43\textwidth]{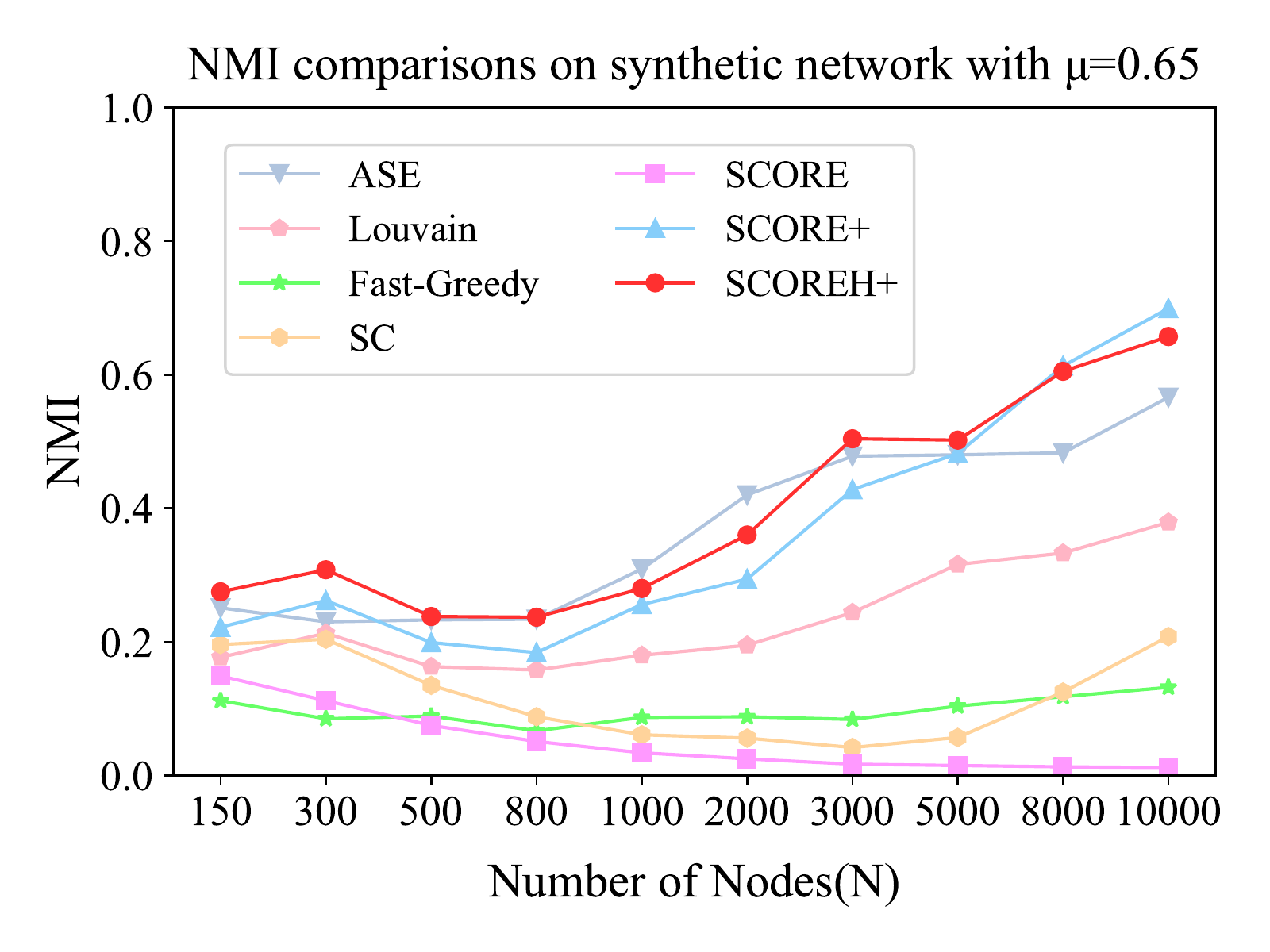}}
\caption{The comparison plots of NMI on LFR datasets with different N. \label{app:LFR-NMI-N}}
\end{figure}

\begin{figure}[ht]
\centering
\subfigure[]{
\label{fig12:subfig:a} 
\includegraphics[width=0.43\textwidth]{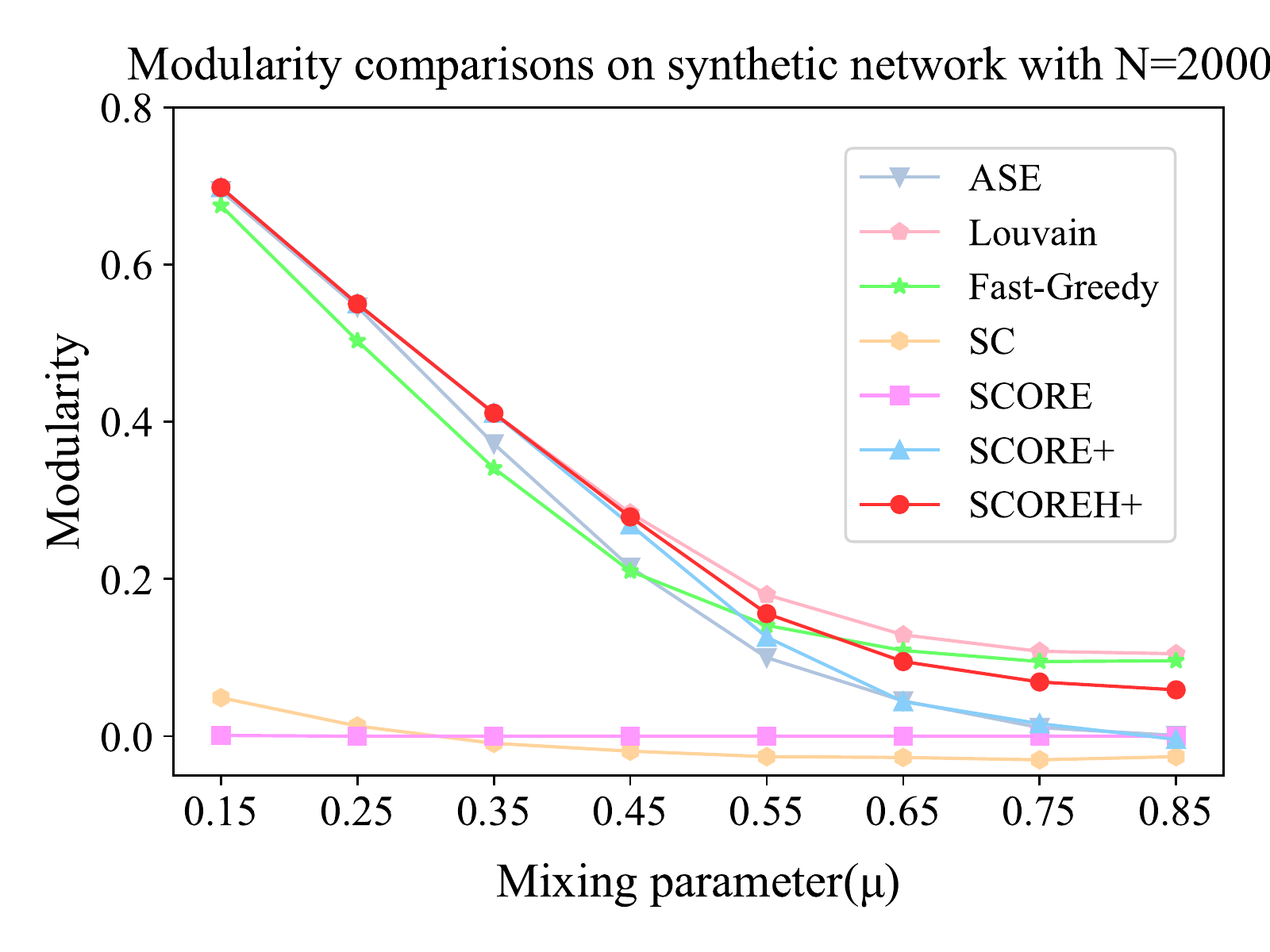}}
\subfigure[]{
\label{fig12:subfig:b} 
\includegraphics[width=0.43\textwidth]{figures/figure/4b.pdf}}
\subfigure[]{
\label{fig12:subfig:c} 
\includegraphics[width=0.43\textwidth]{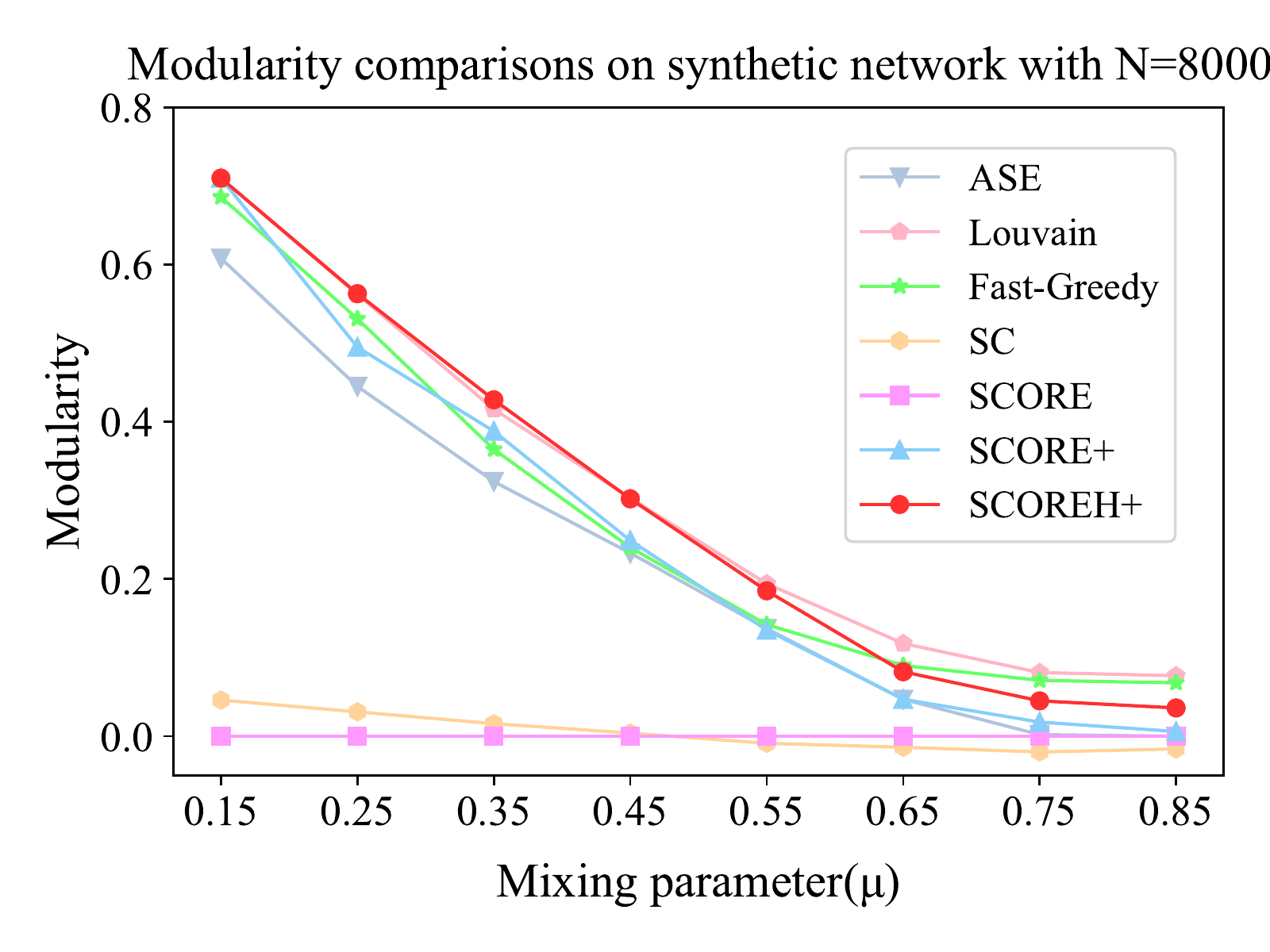}}
\subfigure[]{
\label{fig12:subfig:d} 
\includegraphics[width=0.43\textwidth]{figures/figure/4d.pdf}}
\caption{The comparison plots of Modularity on LFR datasets with different $\mu$.}
\label{app:LFR-Q-mu} 
\end{figure}

\begin{figure}[ht]
\centering
\subfigure[]{
\includegraphics[width=0.43\textwidth]{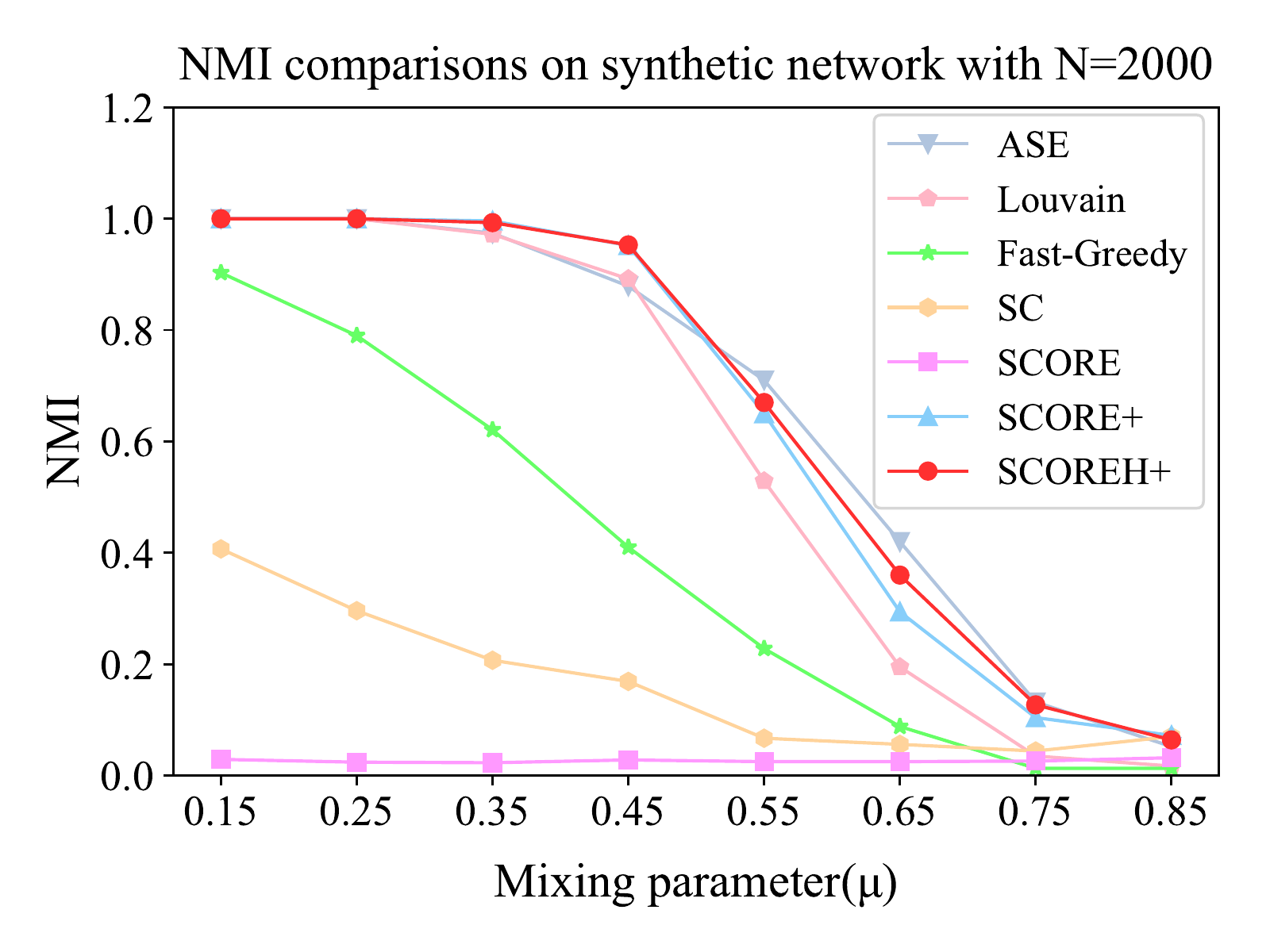}}
\subfigure[]{
\includegraphics[width=0.43\textwidth]{figures/figure/6b.pdf}}
\subfigure[]{
\includegraphics[width=0.43\textwidth]{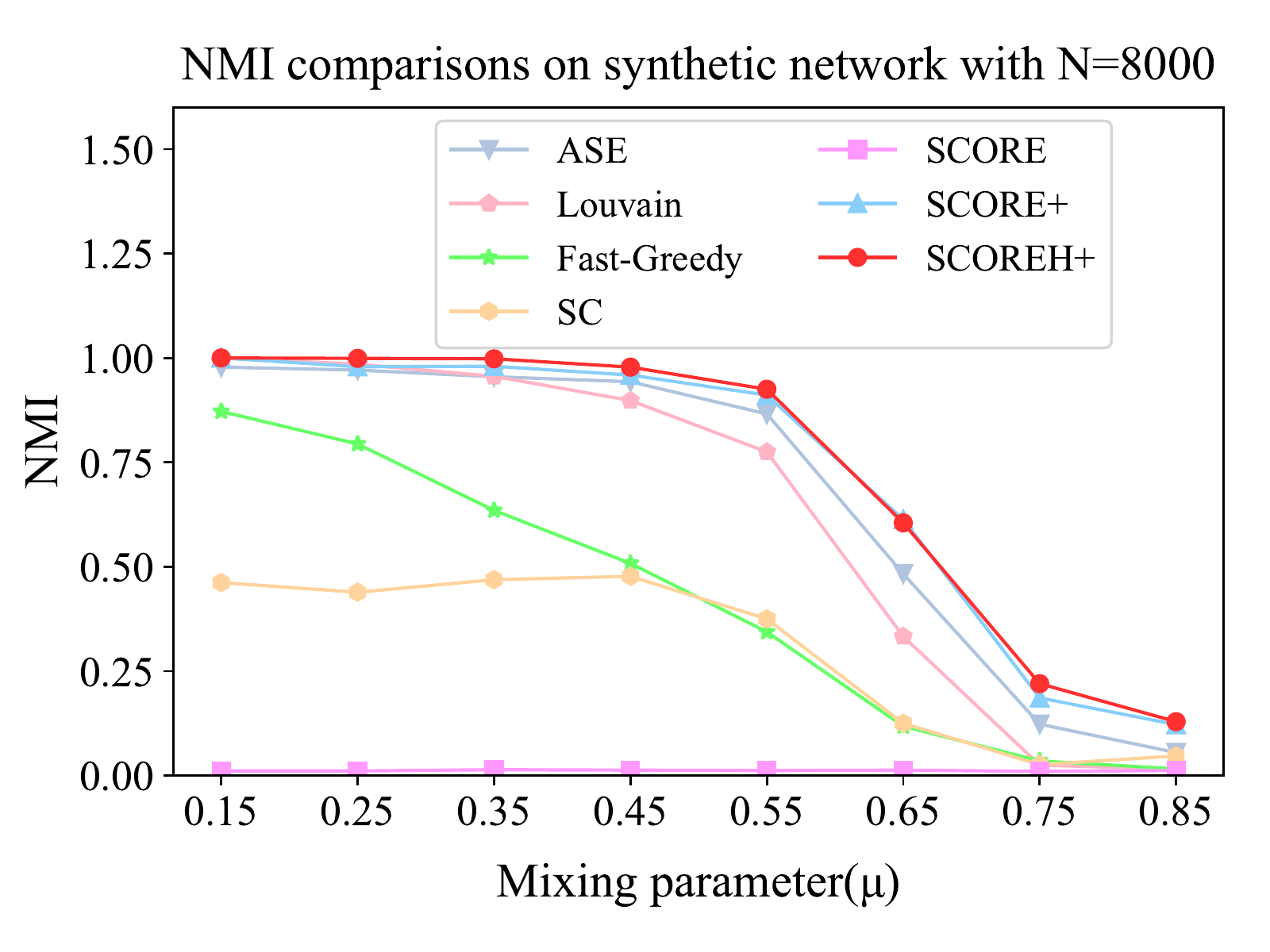}}
\subfigure[]{
\includegraphics[width=0.43\textwidth]{figures/figure/6d.pdf}}
\caption{The comparison plots of NMI on LFR datasets with different $\mu$.}
\label{app:LFR-NMI-mu} 
\end{figure}

\end{document}